\documentclass[%
 pdftex,
 letterpaper,
 aps,
 prc,
 floatfix,
 twocolumn,
 superscriptaddress,
 nofootinbib,
 dvipsnames,
 amssymb,
 amsmath,
 amsfonts
]{revtex4-2}

\usepackage{tikz}
\usepackage{graphicx}
\usepackage{cmap} 
\usepackage{bm}
\usepackage{xcolor}
\usepackage{hyperref}
\hypersetup{%
colorlinks=true,%
citecolor=Blue,%
filecolor=Black,%
linkcolor=Blue,
urlcolor=Magenta 
}
\ifpdf
\usepackage{microtype}
\fi

\graphicspath{{./}}

\newlength{\lminus}
\newcommand{\ud}  {\mathrm{d}}
\newcommand{\gev} {~\mathrm{GeV}}

\newcommand{\gevsq}{~\mathrm{GeV}^2}

\newcommand{\ceps}{\varepsilon}
\newcommand{\bra}[1]{\left\langle{#1}\left|}
\newcommand{\ket}[1]{\right|{#1}\right\rangle}

\newcommand{\eq}[1]{Eq.~(\ref{#1})}
\newcommand{\Eqs}[2]{Eq.~(\ref{#1}) and (\ref{#2})}

\AtBeginDocument{%
    \newwrite\bibnotes
    \def\bibnotesext{Notes.bib}
    \immediate\openout\bibnotes=\jobname\bibnotesext
    \immediate\write\bibnotes{@CONTROL{REVTEX42Control}}
    \immediate\write\bibnotes{@CONTROL{%
    apsrev42Control,author="08",editor="0",pages="0",title="0",year="1"}}
     \if@filesw
     \immediate\write\@auxout{\string\citation{apsrev42Control}}%
    \fi
}%

\begin{document}

\title{%
Hybrid model of proton structure functions
}

\author{S.~A.~Kulagin}
\email[]{kulagin.physics@gmail.com}
\affiliation{%
Institute for Nuclear Research of the Russian Academy of Sciences,
Moscow 117312, Russia}
\author{V.~V.~Barinov}
\email[]{barinov.vvl@gmail.com}
\affiliation{%
Physics Department, Lomonosov Moscow State University, Leninskie Gory, Moscow 119991, Russia}
\affiliation{%
Institute for Nuclear Research of the Russian Academy of Sciences,
Moscow 117312, Russia}

\begin{abstract}
\noindent
We develop a ``hybrid''  model of the proton inelastic structure functions applicable
in a wide region of invariant mass of produced states $W$ and invariant momentum transfer $Q$
including deep inelastic scattering (DIS),
nucleon resonance production
as well as the region close to inelastic threshold.
DIS is described in terms of the parton distributions
together with higher-twist corrections from an available global QCD fit.
The resonant part is addressed in terms of the Breit-Wiegner contributions from five states including
the $\Delta(1232)$ resonance,
the $N(1440)$ Roper resonance,
and three effective resonances describing the second and third resonance regions.
The couplings of the nucleon resonances to photon are described in terms of helicity amplitudes.
The nonresonant background is addressed in terms of
DIS structure functions smoothly extrapolated
to low-$W$ and low-$Q$ values with the proper behavior at the real photon limit $Q^2=0$ as well as  near the inelastic threshold.
We independently treat the transverse $F_T$ and the longitudinal $F_L$ structure function
and fix the model parameters from a global analysis of the world hydrogen
electroproduction and photoproduction cross-section data.
We demonstrate a very good performance of the model by comparing our predictions with data on differential cross sections
and the structure functions $F_2$ and $R=F_L/F_T$.
\end{abstract}
\preprint{INR-TH-2021-002}

\maketitle

\section{\label{sec:intro}Introduction}

The rate of various high-energy processes
is driven by the parton distribution functions (PDFs),
which are universal (process-independent) characteristics of the
target at high invariant momentum transfer $Q$ and
are normally determined from global QCD analyses of high-energy data
(for a recent review see~\cite{Accardi:2016ndt,Gao:2017yyd}).
As the values of $Q$ and $W$ decrease
various subleading effects, such as higher-order  perturbative QCD  corrections in the strong coupling constant,
higher-twist power corrections and/or target mass corrections become increasingly important~\cite{Accardi:2016ndt}.
In a strong coupling regime at a low scale $Q\lesssim 1\gev$ available methods of perturbative QCD,
the twist expansion,
and the methods based on the renormalization group equations are not applicable and the validity of the partonic picture becomes controversial.

In a low-$Q$ and also low-$W$ region, $W<2\gev$,
the lepton-nucleon inelastic scattering is dominated by production of nucleon resonance states.
Unlike deep inelastic scattering (DIS), which is described in terms of the partonic picture and PDFs,
the resonance region is usually addressed in terms of hadronic degrees of freedom.
Starting from Ref.~\cite{Walker:1968xu},
a number of phenomenological analyses are available~\cite{Drechsel:1998hk,Christy:2007ve,HillerBlin:2019hhz},
in which the resonance production is considered as a superposition of the Breit-Wiegner poles
and the photon-resonance couplings are described in terms of helicity amplitudes
(for a recent review see~\cite{Aznauryan:2011qj}).
The resonant inelastic scattering is accompanied by non-resonant background processes
whose rate rises with $W$.
In the region of $W>2\gev$ they  dominate the cross sections.
In available analyses the background contributions are often treated empirically in a limited region of $W$ and $Q$~\cite{Walker:1968xu,Drechsel:1998hk,Christy:2007ve}.
Also other approaches are discussed such as the phenomenology based on a Regge-dual model~\cite{Fiore:2003dg} and neural network analyses~\cite{Brown:2021upr}.

In this paper we develop a combined (hybrid) approach to the proton inelastic structure functions (SFs)
spanning both the resonant production and the DIS region.
We address both the transverse ($F_T$) and longitudinal ($F_L$) SFs,
which form a complete set of functions required to compute the spin-independent cross sections.
In this work we are motivated by the fact that
a quantitative model of this kind,
thoroughly verified with experimental data,
could be applied in various studies,
in which the integration over the full spectrum of intermediate states
and momentum transfer is required.
In this context
we mention the studies of the Bloom-Gilman quark-hadron duality~\cite{Bloom:1970xb},
evaluation of the photon content in the proton~\cite{Manohar:2016nzj,Gao:2017yyd},
the studies of nuclear effects in a resonance-DIS transition region~\cite{Kulagin:2018mxb},
and calculation of the total neutrino cross sections~\cite{Kulagin:2016bdo}.
It should be also remarked that in the energy region of modern long-baseline neutrino experiments
the neutrino-nuclear interaction is dominated by excitation
of nucleon resonances together with low-$W$ DIS and a reliable modeling of
corresponding cross sections is needed to reduce systematic uncertainties and for a correct
interpretation of experimental results~\cite{DUNE:2015lol}.

It is well known that the experimental spectrum of inelastic electron-nucleon scattering for $W<2\gev$ and $Q<2\gev$
has three pronounced resonance structures.
The first resonance region is due to excitation of $\Delta(1232)$ resonance state,
while the other two involve a superposition of a number of individual resonance states.
In this paper the resonance contribution to SFs is addressed in terms of five Breit-Wiegner resonances:
the $\Delta(1232)$ resonance state,
the $N(1440)$ Roper resonance state,
and
three more heavy resonances describing the second and the third resonance regions.
We treat the resonances heavier than the Roper state
as \emph{effective} Breit-Wiegner resonances which incorporate
contributions from a number of excited nucleon states.
Nevertheless, the effective resonances are assigned particle quantum numbers,
such as spin and mass, and their couplings to photons are described in terms of helicity amplitudes.
This choice on the number of effective resonances is motivated by
a balance between the accuracy of description of cross-section data
and the overall stability of our analysis of world cross-section data.

The resonance contributions vanish at sufficiently high values of $W$
and in this region the cross sections are dominated by nonresonant background processes
which eventually turn into DIS at high values of invariant momentum transfer $Q$.
Motivated by this we model the nonresonant background in the resonance region in terms of
the DIS structure functions properly continued
into a low-$W$ and low-$Q^2$ region down to photoproduction limit $Q^2=0$.
To this end we develop an extrapolation method allowing to smoothly match
the DIS SFs at a scale $Q=Q_0$.
This scale defines the onset of a low-$Q$ region and its value
is optimized from analysis of cross-section data.
Our extrapolation method ensures that $F_T/Q^2$ matches the photoproduction cross section in the $Q^2\to0$ limit and also provides vanishing $R=F_L/F_T$ in this limit.

Inelastic scattering off the nucleon is characterized by the presence of the pion production threshold
at $W=M+m_\pi$ with $M$ and $m_\pi$ the proton and the pion mass, respectively.
The resonant contributions explicitly respect the inelastic threshold behavior, as discussed below in Sec.~\ref{sec:R}.
Our method to compute the background contributions also ensures that they vanish
smoothly at the inelastic production threshold.
We also note in this context that the inelastic threshold effect
should impact DIS SF studies at high Bjorken $x\approx 1$ in terms of PDFs.
The threshold effect is also relevant for calculation of nuclear corrections in this region~\cite{Kulagin:2018mxb}.


The parameters of our model, such as the resonance masses, widths,
parameters of helicity amplitudes,
the transition scale $Q_0$,
as well as the parameters responsible for extrapolation into a low-$W$ and low-$Q$ region,
are determined from a global fit to the world data
on the hydrogen electroproduction differential cross section
and the total photoproduction cross section off hydrogen.
We recall that our model smoothly matches the DIS region and
we use the results of a global QCD analysis from Refs.~\cite{Alekhin:2007fh,Alekhin:2008ua}
in order to compute the background contributions in the resonance region.
For this reason we do not need to refit DIS data and we focus on the studies of
the resonance and DIS transition region.
After fixing the model parameters in a fit to cross-section data,
we verify our predictions on $F_2$ and $R$ by comparing with available measurements.

The article is organized as follows.
In Sec.~\ref{sec:frame} we outline the basic theory framework to be used in this study.
In Sec.~\ref{sec:dis} we summarize basic information on DIS SFs required in our analysis,
while in Sec.~\ref{sec:R} we address the derivation of
the resonance SFs in terms of the Breit-Wiegner poles and helicity amplitudes.
In Sec.~\ref{sec:model} we discuss in detail our model of the resonance and background contributions.
In Sec.~\ref{sec:fit} we describe the details of data analysis.
Our results and observations are discussed in Sec.~\ref{sec:discus} and we summarize in Sec.~\ref{sec:sum}.
Supplemental Material~\cite{supplement} provides a detailed comparison of our predictions
with cross-section data from various experiments used in our analysis.


\section{\label{sec:frame}Framework}

The scattering of charged leptons by hadrons in the leading order in the electromagnetic coupling constant $\alpha=e^2/(4\pi)$ is determined by the standard one-photon exchange process. In inclusive scattering, the final hadronic state is not  detected and the differential cross section is given by the hadronic tensor $W_{\mu\nu}$ (see, e.g., Ref.~\cite{Ioffe:1985ep}):
\begin{align}\label{eq:Wmunu}
W_{\mu\nu}(p,q) &=  \frac{1}{8\pi}  \sum_{\lambda,n}
(2\pi)^4 \delta(p+q-p_n) \notag \\ & \quad \times
\bra{P,\lambda}J_\mu^{\mathrm{em}}(0)\ket{n} \bra{n}J_\nu^{\mathrm{em}}(0)\ket{P,\lambda},
\end{align}
where $J_\mu^{\rm em}$ is the electromagnetic current, the sum is taken over all final hadronic states $n$, and $p$ is the proton four-momentum and $q$ is four-momentum transfer.
We do not consider the polarization effects and explicitly average over proton polarization $\lambda$.
Only the symmetric part of the hadronic tensor contributes to the spin-averaged cross section.
Because of the current conservation, time reversal invariance, and parity conservation,
the symmetric hadronic tensor has only two independent Lorentz structures which are usually written as follows:
\begin{align}\label{eq:SF}
W_{\mu\nu}(p,q) &= \left(\frac{q_\mu q_\nu}{q^2} - g_{\mu\nu}\right) F_1
\notag \\ &  +
\frac{F_2}{p\cdot q}
   \left(p_\mu - q_\mu\frac{p\cdot q}{q^2}\right)\left(p_\nu - q_\nu\frac{p\cdot q}{q^2}\right),
\end{align}
where $F_{1,2}$ are the Lorentz-invariant dimensionless structure functions.
The structure functions depend on two independent Lorentz-invariant variables.
In the DIS region, the Bjorken
variable $x=Q^2/(2p\cdot q)$ and the four-momentum transfer squared $Q^2=-q^2$ are used.%
\footnote{%
For the particle state we use the normalization
$\langle p|p'\rangle = 2p_0(2\pi)^3 \delta(\bm{p}-\bm{p}')$,
where $p_0$ is the energy of the particle.
We also use the standard notation of the scalar product of two four-vectors $a\cdot b=a_0 b_0-\bm{a b}$.%
}
In the discussion of the resonance and transition region,
we will also consider the structure functions as a function of the invariant mass of produced states,
\begin{equation}\label{eq:w2}
    W^2 = M^2 + Q^2\left( 1/x - 1 \right).
\end{equation}

The differential cross section in terms of $x$ and $Q^2$ variables reads
\begin{align}
\label{eq:xsec:xQ}
\frac{\ud^2\sigma}{\ud x\ud Q^2} &= \frac{4\pi\alpha^2}{xQ^4}
\left[xy^2\left(1-\frac{2m_l^2}{Q^2}\right) F_1  \right.\notag\\ & \left. \quad +
\left(1-y-\frac{M^2x^2y^2}{Q^2}\right)F_2 \right],
\end{align}
where $m_l$ is the lepton mass and $y=p\cdot q/p\cdot k$ is the inelasticity parameter (here $k$ is the four-momentum of the incoming lepton).
The variable $y$ is not independent but is related to $x$ and $Q^2$ as $xy=Q^2/(2p\cdot k)$.
In the laboratory frame, the differential cross section is measured as a function of scattering angle and the outgoing lepton energy $E'$, and is
related to \eq{eq:xsec:xQ} as
\begin{equation}\label{eq:xsec:lab}
\frac{\ud^2\sigma}{\ud\Omega \ud E'} = \frac{x E'}{\pi y} \frac{\ud^2\sigma}{\ud x\ud Q^2}.
\end{equation}

The structure functions $F_{1,2}$ can be related to the virtual photon
helicity cross sections by projecting the hadronic tensor onto the photon polarization vectors of definite helicity
$\ceps_\mu^{(m)}$ with $m=\pm 1,0$.
It is convenient to chose the $z$-axis along the momentum transfer, $q_z = |\bm{q}|$. Then we have
\begin{subequations}
\label{eq:epsTL}
\begin{align}
\ceps^{(\pm 1)} &= (0,1,\pm i,0)/\sqrt2 ,
\\
\ceps^{(0)} &= (q_z, \bm{0}_\perp, q_0)/Q .
\end{align}
\end{subequations}
where $Q=\sqrt{Q^2}$. The polarization vectors $\ceps^{(+1)}$ and $\ceps^{(-1)}$
describe the transversely polarized states with photon helicity $m=+1$ and $m=-1$, while
the vector $\ceps^{(0)}$ corresponds to the longitudinally polarized virtual photon.
Note that the polarization vectors $\ceps^{(m)}$ are orthogonal to the photon momentum, $\ceps^{(m)}\cdot q=0$, and
normalized as $\ceps^{(0)}\cdot\ceps^{(0)}=1$ and ${\ceps^{(m)}}^*\cdot\ceps^{(m)}=-1$ for $m=\pm 1$,
where $\ceps^*$ is the complex conjugate.
The transverse, $F_T$, and the longitudinal, $F_L$, SFs can be found by contracting the hadronic tensor, \eq{eq:Wmunu}, with the photon polarization vectors, \eq{eq:epsTL}.
We have
\begin{align}
\label{eq:FT}
F_T &= x\sum_{m=\pm 1} {\ceps^{(m)}_\mu}^* W_{\mu\nu} \ceps^{(m)}_\nu = 2xF_1,
\\
\label{eq:FL}
F_L &= 2x\, {\ceps^{(0)}_\mu} W_{\mu\nu} \ceps^{(0)}_\nu = \gamma^2 F_2 - F_T,
\end{align}
where $\gamma^2=1+4x^2M^2/Q^2$.
The transverse and longitudinal cross sections in terms of \Eqs{eq:FT}{eq:FL} can be written as follows (see, e.g., Ref.~\cite{Ioffe:1985ep}):
\begin{equation}\label{eq:sigTL}
\sigma_{T,L} = \frac{4\pi^2\alpha F_{T,L}}{(1-x)Q^2}.
\end{equation}
Note that the definition of the virtual photon flux is somewhat uncertain.
In \eq{eq:sigTL} we assume the virtual photon flux equals the real photon one with
the condition that the mass of produced hadronic states $W$ is the same for the real and virtual photon~\cite{Ioffe:1985ep}.

Let us briefly discuss the real photon limit. To this end it is convenient to consider the SF as a function of $W^2$ and $Q^2$.
Owing to conservation of electromagnetic current, the longitudinal cross section vanishes at $Q^2=0$.
This in turn suggests $F_L$ vanishing faster than $Q^2$ as $Q^2\to 0$.
On the other hand, the cross section $\sigma_T$ for transverse virtual photons
goes to the total photoproduction cross section in this limit:
\begin{equation}\label{eq:sig:gamma}
\sigma_{\gamma}(W)=4\pi^2\alpha\lim_{Q^2\to 0} F_T(W^2,Q^2)/Q^2 .
\end{equation}
For this reason
the ratio $F_T(W^2,Q^2)/Q^2$ (as well as $F_2/Q^2$) is finite at $Q^2\to 0$
and fixed $W$.
We use \eq{eq:sig:gamma} to constrain a low-$Q$ behavior of our model in Sec.~\ref{sec:model}.

\subsection{\label{sec:dis}Deep inelastic scattering}

In QCD, a common framework to address DIS is the operator product expansion (OPE),
a procedure producing the power series in $Q^{-2}$ (twist expansion).
In the leading order of this expansion, i.e., in the leading twist (LT),
SFs factorize into  a convolution of the coefficients functions,
describing quark-gluon interaction at a hard scale $Q$,
and the parton distribution functions~\cite{Collins:1989gx}.
In the lowest order in QCD coupling constant $\alpha_s$
the coefficient functions are simply the quark charges squared,
and SFs are given in terms of PDFs according to the celebrated quark-parton model.
The gluon PDF contributes to SFs in higher orders in $\alpha_s$ through quark-gluon radiation processes.
For more detailed discussion see, e.g., Ref.~\cite{Accardi:2016ndt}.

The power corrections in $Q^{-2}$ can be of two different types:
contributions from higher-twist (HT) operators describing quark-gluon correlations
and
correction arising from a finite nucleon mass (target mass correction, or TMC).
Summarizing, we write the proton SF as follows
\begin{align}\label{eq:sf:dis}
F_i^\text{DIS}(x,Q^2) &= F_i^{\text{TMC}}(x,Q^2) 
+ H_i(x)/Q^{2}, 
\end{align}
where $i=T,L$ and the superscript TMC labels the LT SF corrected for the target mass effect
while the functions $H_i$ describe the dynamical
twist-4 contribution (for brevity, we suppress explicit notation to the twists higher than 4).

The PDFs are usually determined in a global QCD analysis of high-energy data including DIS,
muon pair production in proton-proton collisions (DY),
and $W/Z$ boson production at colliders
(for more detail we refer to Refs.~\cite{Accardi:2016ndt,Gao:2017yyd}).
In this study we use the proton PDFs from a global QCD fit of Refs.\cite{Alekhin:2007fh,Alekhin:2008ua},
which was performed to the next-to-next-to-leading-order (NNLO) approximation in the QCD coupling constant.
Although updated PDF analyses are available, such as Refs.~\cite{Alekhin:2017kpj,Alekhin:2017fpf}
as well as the other results discussed in Refs.~\cite{Accardi:2016ndt,Gao:2017yyd},
which include the most recent LHC data and thus better constrain the low-$x$ region,
we use the PDFs of Refs.~\cite{Alekhin:2007fh,Alekhin:2008ua} as the base.
We are motivated by the fact that the analysis of Ref.~\cite{Alekhin:2007fh}
applies the cuts $Q>1\gev$ and $W>1.8\gev$ and thus includes  low-$Q$ data.%
\footnote{SFs of Ref.~\cite{Alekhin:2007fh}
are available at significantly lower values of $Q^2$ owing an extrapolation procedure,
which is different from that discussed below.}
Along with PDFs the analysis of Refs.~\cite{Alekhin:2007fh,Alekhin:2008ua} provides
the determination of $H_i$ functions describing twist-4 terms.
In this study we address the resonance and DIS transition region and for this reason
a low-$Q$ PDF fit is preferred over recent PDF analyses~\cite{Alekhin:2017kpj,Accardi:2016ndt},
which utilize the more stringent cut $Q^2>2.5\gevsq$.
We also comment that the proton and the neutron SFs of Ref.~\cite{Alekhin:2007fh}
show a very good performance in analysis of
the nuclear EMC effect in Refs.~\cite{Kulagin:2004ie,Kulagin:2010gd}
as well as in the interpretation of a recent measurement of $F_2^n/F_2^p$~\cite{MARATHON:2021vqu}.

TMC is accounted for within the OPE framework of Ref.~\cite{Georgi:1976ve}.
Similar TMC was used in analyses of Refs.~\cite{Alekhin:2007fh,Alekhin:2008ua}.
It should be remarked that the TMC procedure of Ref.\cite{Georgi:1976ve}
violates the inelastic threshold behavior of the SFs, leading to nonzero values at $x\ge 1$
(see, e.g., the discussion in Ref.~\cite{Kulagin:2004ie}).
The region of large Bjorken $x$ corresponds to low values of $W$.
By increasing the value of $x$ at some point we enter the resonance region
even at high values of $Q^2$
and thus leave the region of applicability of an OPE-based description.
In our analysis below, we use DIS SFs in a safe region of $W^2>4\gevsq$ and $Q^2>2\gevsq$.
To calculate nonresonant background at low-$W$ and low-$Q$ values we use extrapolated
DIS SFs as discussed in Sec.~\ref{sec:model}.


\subsection{\label{sec:R}Excitation of nucleon resonances}

In the region of $W<2\gev$ and for a low momentum transfer $Q\lesssim 2\gev$
the inelastic cross section is dominated by excitation of nucleon resonance states.
Their contribution to the hadronic tensor \eq{eq:Wmunu} can be written as
\begin{align}
\label{eq:Wmunu:R}
W_{\mu\nu}^\text{Res} &=  \frac{1}{4} \sum_{R,\lambda,\lambda'} \delta(W^2-M_R^2)
\notag\\ & \times
\bra{P,\lambda}  J_\mu^\text{em}(0) \ket{R,\lambda'}
\bra{R,\lambda'} J_\nu^\text{em}(0) \ket{P,\lambda},
\end{align}
where the sum is taken over the nucleon resonant states and $M_R$ is the mass of the corresponding state.
We first consider narrow resonance states with zero width;
the effect of a finite width will be discussed below.
In \eq{eq:Wmunu:R} we also explicitly sum over the polarization $\lambda'$ of an intermediate state
and average over the proton polarization $\lambda$.

We discuss in detail the derivation of the structure functions $F_T$ and $F_L$
from hadronic tensor (\ref{eq:Wmunu:R}) in terms of helicity amplitudes
thus updating existing studies~\cite{HillerBlin:2019hhz,Christy:2007ve}.
$F_T$ and $F_L$ are given by contracting
the hadronic tensor (\ref{eq:Wmunu:R}) with the corresponding photon polarization vector
by \Eqs{eq:FT}{eq:FL}.
The matrix elements of electromagnetic current can be described in terms of helicity
amplitudes~\cite{Bjorken:1966ij,Walker:1968xu,Aznauryan:2011qj}:
\begin{align}\label{eq:ha:R}
\bra{R,h'}\ceps^{(m)}\cdot J^\text{em}(0)\ket{P,h} = H^m_{h'h},
\end{align}
where $h'$ and $h$ are the helicities of the resonance state $R$ and the proton, respectively
(for brevity, we suppress explicit notation of the dependence of helicity amplitudes on the resonance $R$).
Because helicity is conserved,  $h'=h+m$. In the parity transformation the helicity changes its sign,
and, because of symmetry under parity transformation,
we have the relation $H^{-m}_{-h'-h}=H^m_{h'h}$ (see also Ref.~\cite{Walker:1968xu}).
We consider the helicity amplitudes in the center-of-mass frame (c.m.)
and chose the spin quantization axis along the photon momentum.
Then the proton helicity $h$ has the sign opposite to its polarization,
as the proton momentum balances the photon momentum,
and the helicity $h'$ of the resonance state $R$  corresponds to its polarization.

Consider \eq{eq:ha:R} for the proton polarization $\lambda=1/2$ and respectively $h=-1/2$.
The helicity amplitudes can be related to the standard electrocouplings $A_{1/2}$, $A_{3/2}$,
and $S_{1/2}$, which are commonly used to describe electroexcitation of the resonance states~(see, e.g., Ref.\cite{Aznauryan:2011qj}):
\begin{align}
\label{eq:hpl:R}
H^{+1}_{\frac12, -\frac12} &= c_R A_{1/2}(Q^2),\\
\label{eq:hmn:R}
H^{-1}_{-\frac32, -\frac12} &= c_R A_{3/2}(Q^2),\\
\label{eq:h0:R}
H^{0}_{-\frac12, -\frac12} &= c_R  S_{1/2}(Q^2) (Q/|\bm q|_\textsc{cm}),
\end{align}
where $|\bm q|_\textsc{cm}$ is the photon momentum in the c.m. frame.%
\footnote{%
Note that in the c.m. frame $|\bm q|_\textsc{cm}$ is also the magnitude of the proton momentum.
For completeness, $|\bm q|_\textsc{cm}^2=E_\textsc{cm}^2-M^2$, where  $E_\textsc{cm}=(W^2+Q^2+M^2)/(2W)$ is the proton c.m. energy.}
The normalization factor $c_R$ can be determined by requiring the
electromagnetic decay width $\Gamma_R^\gamma(R\to P\gamma)$ to be~\cite{Tanabashi:2018oca,Aznauryan:2011qj}
\begin{align}\label{eq:Rwidth}
\Gamma_R^\gamma = \frac{2 K_R^2 M}{\pi(2S_R+1)M_R} &
\left( |A_{1/2}^R(0)|^2 + 
       |A_{3/2}^R(0)|^2 \right),
\end{align}
where we assume averaging over the resonance polarization and summing over photon polarization,
$S_R$ is the resonance spin, and
$K_R=(M_R^2-M^2)/(2M_R)$ is the energy of a real photon in the c.m. frame needed to produce the state with the mass $M_R$.
Using \Eqs{eq:hpl:R}{eq:hmn:R} we have
\begin{equation}\label{eq:cR}
	c_R^2 = \frac{M(M_R^2-M^2)}{\pi\alpha}.
\end{equation}
Note that the electromagnetic current in \eq{eq:Wmunu} is normalized such that the electric charge $e=1$.
By definition the electric charge is absorbed in the amplitudes $A_{1/2}$, $A_{3/2}$, $S_{1/2}$, and for that reason
we have $\alpha$ in the denominator in \eq{eq:cR}.

We now apply the results of the present discussion to compute the resonant contribution to $F_T$ and $F_L$ in terms of the amplitudes $A_{1/2}$, $A_{3/2}$ and $S_{1/2}$. From \Eqs{eq:FT}{eq:FL} we have
\begin{align}
	\label{eq:FT:R0}
	F_T^{\text{Res}} &= \frac{xM}{2\pi \alpha} \sum_{R}
	\delta(W^2-M_R^2)(M_R^2-M^2)
\notag\\ & \quad \times
	\left( |A_{1/2}^R(Q^2)|^2 + |A_{3/2}^R(Q^2)|^2 \right),
	\\
	\label{eq:FL:R0}
	F_L^{\text{Res}} &= \frac{xM}{\pi \alpha}  \sum_{R}
	\delta(W^2-M_R^2)(M_R^2-M^2)
\notag\\ & \quad \times
	(Q^2/|\bm q|^2_\text{\sc cm}) |S_{1/2}^R(Q^2)|^2,
\end{align}
where $x$ is the Bjorken variable and the sum is taken over the resonance states.

Let us now discuss the effect of finite resonance width in \eq{eq:Wmunu:R} in some more detail.
Following a traditional approach, we replace $\delta(W^2-M_R^2)$ with the standard Breit-Wiegner factor:
\begin{equation}\label{eq:delta_to_BW}
	\delta(W^2-M_R^2) \to \frac{1}{\pi}
	 \frac{M_R\Gamma_R}{(W^2-M_R^2)^2+M_R^2\Gamma_R^2},
\end{equation}
where $\Gamma_R$ is the resonance total width.

The $\Delta(1232)$ resonance width is entirely due to the $\pi N$ decay channel.
For heavier resonance states there are also other decay modes.
In our analysis we will assume that any resonance state $R$ in \Eqs{eq:FT:R0}{eq:FL:R0} decays either in $\pi N$, $\eta N$, or $2\pi N$ channels:
\begin{equation}\label{eq:gamR}
	\Gamma_R = \beta_R^\pi \Gamma_R^\pi +
				\beta_R^\eta \Gamma_R^\eta +
				\beta_R^{2\pi} \Gamma_R^{2\pi},
\end{equation}
with $\beta_R^{\pi,\eta,2\pi}$  the corresponding branching fractions.

Generally, the resonances can be excited off the resonance pole,  $W^2\not=M_R^2$,
and in \eq{eq:Wmunu:R} one has to consider off-mass-shell effects on the resonance parameters.
In particular, the resonance width becomes a function of running mass, $\Gamma_R=\Gamma_R(W)$.
Indeed, near the inelastic threshold $W_\text{th}=M+m_\pi$ the cross section should vanish that in turn requires vanishing resonance width. On the other hand, $\Gamma_R(W)$ increases with $W$ as the phase space available for the resonance decay increases.
In order to account for this effect, we parametrize the energy dependence of $\Gamma_R(W)$  following Ref.\cite{Walker:1968xu}:
\begin{equation}\label{eq:gampi}
	\Gamma_R^\pi = \Gamma_R^0 \left(\frac{p_\pi(W)}{p_\pi(M_R)}\right)^{2L+1}
	\left(\frac{p_\pi(M_R)^2+X_R^2}{p_\pi(W)^2+X_R^2}\right)^L ,
\end{equation}
where $\Gamma_R^0$ is the intrinsic resonance width,
$p_\pi(W)$ is the meson c.m. momentum in the decay $R\to \pi P$ of the resonance with mass $W$,
$L$ is angular momentum of the resonance, and $X_R$ is a phenomenological parameter (damping factor).
The parametrization of the $\eta N$ decay mode is similar to \eq{eq:gampi} with $p_\eta$ the $\eta$ meson c.m. momentum.
For the $2\pi$ decay mode we use~\cite{Aznauryan:2011qj}
\begin{equation}\label{eq:gam2pi}
	\Gamma_R^{2\pi} = \Gamma_R^0 
	\left(\frac{p_{2\pi}(W)}{p_{2\pi}(M_R)}\right)^{\!\!2L+4}
	\!\!\left(\frac{p_{2\pi}(M_R)^2+X_R^2}{p_{2\pi}(W)^2+X_R^2}\right)^{\!\!L+2} ,
\end{equation}
where $p_{2\pi}$ is effective two-pion momentum in c.m. frame which is computed similarly to $p_{\pi}$ but replacing $m_\pi$ with $2m_\pi$.
Apparently, $p_{2\pi}=0$ below the $2\pi$ production threshold and $p_\eta=0$ below the $\eta$ meson production threshold.

Also the $\gamma P R$ vertex, or helicity amplitudes, acquire $W$ dependence in the resonance off-pole region.
We phenomenologically account for the off-shell effect following Ref.\cite{Walker:1968xu} with the factor $f_R^\gamma$:
\begin{equation}\label{eq:fgamma}
	f_R^\gamma(W) = \frac{K^2}{K_R^2} \frac{K_R^2+X_R^2}{K^2+X_R^2},
\end{equation}
where  $K=K(W)=(W^2-M^2)/(2W)$ is the equivalent photon c.m. momentum, $K_R=K(M_R)$ and $X_R$ is the same damping parameter as in \Eqs{eq:gampi}{eq:gam2pi}.
At the resonance pole $f_R^\gamma(M_R)=1$.

Summarizing, we have for the resonant contribution to $F_T$ and $F_L$:
\begin{align}
\label{eq:FT:R}
F_T^{\text{Res}} &= \frac{xM}{\pi^2\alpha} \sum_{R}
    \frac{M_R^2\Gamma_R K_R f_R^{\gamma}(W)}
         {(W^2-M_R^2)^2+M_R^2\Gamma_R^2}
\notag\\ & \quad \times
          \left( |A_{1/2}^R(Q^2)|^2 + |A_{3/2}^R(Q^2)|^2 \right),
\\
\label{eq:FL:R}
F_L^{\text{Res}} &= \frac{2xM}{\pi^2 \alpha} \sum_{R}
    \frac{M_R^2 \Gamma_R K_R f_R^{\gamma}(W)}
         {(W^2-M_R^2)^2+M_R^2\Gamma_R^2}
\notag\\ & \quad \times
         (Q^2/|\bm q|^2_\text{\sc cm}) |S_{1/2}^R(Q^2)|^2,
\end{align}
where we sum over the resonance states and $\Gamma_R=\Gamma_R(W)$ is the total resonance width by \eq{eq:gamR}.
For completeness we also present the corresponding contributions to the virtual photon cross section $\sigma_T$ and $\sigma_L$ by \eq{eq:sigTL}. Using \Eqs{eq:FT:R}{eq:FL:R} and also the relation $Q^2(1-x)=x(W^2-M^2)$ we have
\begin{align}
\label{eq:sigT:R}
\sigma_T^{\text{Res}} &= \frac{2M}{W} \sum_{R}
\frac{M_R^2\Gamma_R (K_R/K) f_R^{\gamma}(W)}
{(W^2-M_R^2)^2+M_R^2\Gamma_R^2}
\notag \\ & \quad \times
\left( |A_{1/2}^R(Q^2)|^2 + |A_{3/2}^R(Q^2)|^2 \right),
\\
\label{eq:sigL:R}
\sigma_L^{\text{Res}} &= \frac{4M}{W} \sum_{R}
\frac{M_R^2 \Gamma_R (K_R/K) f_R^{\gamma}(W)}
{(W^2-M_R^2)^2+M_R^2\Gamma_R^2}
\notag\\ & \quad \times
(Q^2/|\bm q|^2_\text{\sc cm}) |S_{1/2}^R(Q^2)|^2.
\end{align}
The explicit parametrization of the $Q^2$ dependence of the helicity amplitudes
entering \eq{eq:FT:R} to (\ref{eq:sigL:R}) is discussed in Sec.~\ref{sec:model}.

It should be commented that the off-shell continuation of amplitudes for a particle with finite width is not unique.
In order to illustrate this statement, we first observe that
in the right side of \Eqs{eq:FT:R0}{eq:FL:R0} $M_R$ can be replaced with $W$ and then the factor $W^2-M^2$ can be taken out of the sum over the resonance states.
On the other hand, this operation does not commute with \eq{eq:delta_to_BW}
and the result would depend on the order of these two operations.
We will proceed with \Eqs{eq:FT:R}{eq:FL:R} and fix phenomenological parameters there from a fit to the cross section data described in Sec.~\ref{sec:fit}.

\section{\label{sec:model}Description of the model}

The full structure functions include contributions from both,
the resonance states discussed in Sec.~\ref{sec:R},
and non-resonantly produced continuum states  [background (BG) contributions]:
\begin{equation}\label{eq:sffull}
	F_i = F_i^\text{Res} + F_i^\text{BG},
\end{equation}
where $i=T,L$. In what follows it will be convenient to consider the structure functions as a function of $Q^2$ and $W^2$.
From \eq{eq:w2} the Bjorken variable $x=Q^2/(Q^2+W^2-M^2)$.
While the resonance part dominates in the region of low $W$, BG contribution rises with $W$ and prevails for $W>2\gev$.
If $W$ and $Q$ are sufficiently high then BG contributions are driven by DIS,
$F_i^\text{BG}\to F_i^\text{DIS}$.
We will use this simple observation to also model BG contributions in the whole region of $W$ and $Q$
with suitable extrapolation of the DIS structure functions outside the region of their applicability.

\subsection{Resonance contributions}
\label{sec:model:r}

We apply \Eqs{eq:FT:R}{eq:FL:R} to compute the resonant contribution to the transverse and the longitudinal SFs.
Note that the unpolarized scattering is not sensitive to individual amplitudes $A_{1/2}$ and $A_{3/2}$ and only their quadrature sum is relevant.
For this reason, for each of the resonance state in \Eqs{eq:FT:R}{eq:FL:R}, we discuss the average amplitude $A(Q^2)$ defined as
\begin{equation}\label{eq:Aav}
	|A(Q^2)|^2 = |A_{1/2}(Q^2)|^2 + |A_{3/2}(Q^2)|^2.
\end{equation}

In order to describe the observed resonant inclusive spectra,
we include five resonant contributions in \Eqs{eq:FT:R}{eq:FL:R}.
The first resonance region is described by a well separated $\Delta(1232)$ resonance state.
The second and third resonance regions are described in terms of the $N(1440)$ Roper resonance
and three more heavier states $R_1$, $R_2$, and $R_3$.
As outlined in Sec.~\ref{sec:intro}, we treat these states as effective Breit-Wiegner resonances incorporating contributions from a number of individual excited-nucleon states.
The relevant resonance parameters are listed in Table~\ref{tab:respars}.
The parameter values  are determined from a combined fit to
the hydrogen inclusive electroproduction differential cross-section data
and photoproduction cross-section data,
as described in Sec.~\ref{sec:fit}.

\begingroup
\squeezetable
\begin{table}[htb]
\caption{The best fit values for the mass $M_R$, the intrinsic width $\Gamma_R$, the angular momentum $L$, the damping parameter $X_R$,
and the decay branching fractions $\beta$ for each of the resonant state.
The dimensional parameters are in GeV units.
The estimate of the fit parameter uncertainty is given in parentheses in percent units.
\label{tab:respars}}
\begin{ruledtabular}
\begin{tabular*}{\textwidth}{l@{\extracolsep{\fill}}lllllll}
{}             & $M_R$        & $\Gamma_R$   &$L$& $X_R$        & $\beta_{1\pi}$ & $\beta_{2\pi}$ & $\beta_{\eta}$\\
\hline
$\Delta(1232)$ & 1.2270(0.02) & 0.1128(0.48) & 1 & 0.0554(1.07) & 1.00 & 0.00 & 0.00 \\
$N(1440)$      & 1.4487(0.34) & 0.4022(3.34) & 1 & 0.1125(3.85) & 0.65 & 0.35 & 0.00 \\
$R_1$          & 1.5123(0.02) & 0.0945(1.83) & 2 & 0.4959(4.81) & 0.75 & 0.25 & 0.00 \\
$R_2$          & 1.5764(0.16) & 0.5005(1.76) & 0 & 0.3097(2.12) & 0.15 & 0.85 & 0.00 \\
$R_3$          & 1.7002(0.03) & 0.1177(1.66) & 2 & 0.2583(10.8) & 0.15 & 0.60 & 0.25 \\
\end{tabular*}
\end{ruledtabular}
\end{table}
\endgroup

To parametrize $Q^2$ dependence of the transverse and the longitudinal amplitudes in \Eqs{eq:FT:R}{eq:FL:R} we use the following model:
\begin{align}
\label{eq:AMPL:T}
A(Q^2) &=
\left(a_1 + a_2 Q^2\right)/\left(1 + a_3 Q^2\right)^{a_4}
\\
\label{eq:AMPL:L}
S_{1/2}(Q^2) &= \left(c_1 + c_2 Q^2\right)\exp\left(-c_3 Q^2\right)
\end{align}
The determination of parameter values in \Eqs{eq:AMPL:T}{eq:AMPL:L}, as well as other model parameters entering \Eqs{eq:FT:R}{eq:FL:R}, is discussed in Sec.~\ref{sec:fit}.

\subsection{\label{sec:model:bg}Background contributions}

In order to model BG contributions we first consider the extrapolation of DIS structure functions to low $Q^2$.
Note that the framework of Sec.~\ref{sec:dis} applies for $Q>Q_0$ with the scale $Q_0\sim 1\gev$.
Going into the region $Q<Q_0$,
we consider extrapolation of DIS $F_T$ and $F_L$ from the scale $Q_0^2$ down to $Q^2=0$,
taking into account the real photon limit [see the discussion after \eq{eq:sigTL}].
Let us first discuss the function $F_T$ which vanishes as $Q^2$ at $Q^2\to 0$.
From \eq{eq:sig:gamma} the ratio $F_T/Q^2$ in the limit $Q^2\to 0$ is given
by the total photoproduction cross section $\sigma_\gamma(W)$.
Taking this into account, we consider the following model for $0\le Q^2 \le Q_0^2$:
\begin{equation}\label{eq:ext:FT1}
F_{T}^{\text{Ext}}(W^2,t) = f_0 t + f_1 t^m + f_2 t^n ,
\end{equation}
where for $Q^2$ we use a more handy notation $t$, and $f_0$, $f_1$ and $f_2$ are the functions of $W$, and we assume $m>1$ and $n>1$.
Taking the limit $t\to 0$ and using \eq{eq:sig:gamma}  we have
\begin{equation}\label{eq:ext:f0}
f_0(W)=\sigma_\gamma(W)/(4\pi^2\alpha).
\end{equation}
%
The functions $f_1$ and $f_2$ are determined by requiring the smoothness of the extrapolation function by \eq{eq:ext:FT1} at $t=t_0=Q_0^2$;
i.e., we require the continuity of the function and its first derivative at the DIS matching point.
We have
\begin{align}
\label{eq:ext:FT:f1}
f_1 &= t_0^{-m} \left( nF_{T}^\text{DIS} - t_0\partial_t F_{T}^\text{DIS}
\right. \notag\\ &\quad \left.
{}- (n-1)f_0t_0\right)/(n-m),
\\
\label{eq:ext:FT:f2}
f_2 &= t_0^{-n} \left( mF_{T}^\text{DIS} - t_0\partial_t F_{T}^\text{DIS}
 \right. \notag\\ &\quad \left.
{}- (m-1)f_0t_0\right)/(m-n),
\end{align}
where
$F_{T}^{\text{DIS}}$ and its derivative $\partial_t F_{T}^{\text{DIS}}$ are computed at $t=t_0$ for given $W^2$.
The exponents $m$ and $n$ controlling the transition to the low-$t$ region are adjusted from the data analysis in Sec.~\ref{sec:fit}.
Note that both \Eqs{eq:ext:FT:f1}{eq:ext:FT:f2} have a pole at $n=m$.
However, \eq{eq:ext:FT1} is finite in the limit $n\to m$.
Taking this limit we see that a low-$t$ behavior is given by a combination of $t^m$ and $t^m\ln t$ terms, and \eq{eq:ext:FT1} can be written as follows:
\begin{align}\label{eq:ext:FT2}
F_{T}^{\text{Ext}} &= f_0 t +
\left({t}/{t_0}\right)^m
\left[ F_{T}^{\text{DIS}} - f_0t_0 
+ \left(mF_{T}^{\text{DIS}}
\right.\right. \notag\\ & \quad \left.\left.
{} - t_0\partial_t F_{T}^{\text{DIS}} - (m-1)f_0t_0\right)\ln\left(t_0/t\right)
\right].
\end{align}
In practice this is an important case, preferred by data as described in Sec.~\ref{sec:fit}.

To extrapolate the longitudinal SF in the region $0\le t \le t_0$ we use a model similar to \eq{eq:ext:FT1} with $f_0=0$ as the longitudinal cross section vanishes for real photon:
\begin{equation}\label{eq:ext:FL1}
F_L^{\text{Ext}}(W^2,t) = f'_1 t^{m'} + f'_2 t^{n'} .
\end{equation}
The functions $f'_{1,2}(W)$ are fixed by requiring smoothness of the function \eq{eq:ext:FL1} at $t=t_0$,
similarly to the $F_T$ case.
On the functions $f'_1$ and $f'_2$ we obtain the equations similar to \Eqs{eq:ext:FT:f1}{eq:ext:FT:f2} with $F_T$ replaced with $F_L$ and $f_0=0$.
The case $n'=m'$, which is preferred by our analysis in Sec.~\ref{sec:fit}, reads
\begin{align}\label{eq:ext:FL2}
	F_L^{\text{Ext}} &=
	\left({t}/{t_0}\right)^{m'}
	\left[ F_L^{\text{DIS}}  +
	\left(m'F_L^{\text{DIS}} \right.\right.\notag\\ & \quad \left.\left.
	{} - t_0\partial_t F_L^{\text{DIS}}\right)\ln\left(t_0/t\right)
	\right].
\end{align}

Note that in global QCD fits the PDFs are parameterized in the full region of the variable $x$ ($0<x<1$)
and for this reason
the DIS structure functions by \eq{eq:sf:dis} can be computed in the full region of $W$.
However, for $W<2\gev$, and therefore in the region of large $x$,
the structure functions from global PDF fits are not directly constrained by data
because low-$W$ data are explicitly removed from the fits
(for instance, the fit of Refs.~\cite{Alekhin:2007fh,Alekhin:2017kpj} applies the cut $W>1.8\gev$).
It should be also recalled, that
the target mass correction of Ref.\cite{Georgi:1976ve}
generates unphysical contributions at $x\ge1$.
Taking this into account, we address the low-$W$ region in our model by introducing a correction factor
$B$ as follows:
\begin{equation}\label{eq:BG}
F_{i}^{\text{BG}} 
= B_{i}(W^2)
	\begin{cases}
	F_{i}^{\text{DIS}}(W^2,Q^2)\ \text{if}\ Q^2 \geq Q_0^2,
\\
	F_{i}^{\text{Ext}}(W^2,Q^2)\ \text{if}\ Q^2 < Q_0^2,
	\end{cases}
\end{equation}
where $i=T,L$ and $F_{i}^{\text{BG}}$ is the corresponding background structure function,
and $F_T^\text{Ext}$ and $F_L^\text{Ext}$ are given  by \Eqs{eq:ext:FT1}{eq:ext:FL1}.
The factors $B_T$ and $B_L$ are responsible for extrapolation to the low-$W$ region and in this study we assume them to be the functions of $W$ only.
The $B_{T,L}$ functions are positively defined and required to vanish at the pion production threshold
$W\to W_\text{th}=M + m_{\pi}$.
On the other hand, they rise with $W$ and $B_{T,L} \to 1$ above the resonance region.
In order to respect these requirements, we use the following model:
\begin{equation}
\label{eq:B}
	B = 1 - \exp\left(-b_1(W^2-W^2_\text{th})^{b_2}\right),
\end{equation}
where for simplicity we suppress the explicit subscript $i=T,L$ for $B$.
The parameters $b_1$ and $b_2$, which are assumed to be positive,
are adjusted from a fit to cross-section data in Sec.~\ref{sec:fit}.
We independently treat these parameters for $F_T$ and $F_L$.

\subsection{\label{sec:photo}Real photon limit}

In the limit of $Q^2=0$ the longitudinal cross section vanishes and
the photoproduction cross section $\sigma_{\gamma}$ is given by \eq{eq:sig:gamma}.
Note that $\sigma_{\gamma}$ receives contributions from the resonance production
process as well as from a nonresonant background scattering:
\begin{equation}
\label{eq:photocs}
\sigma_{\gamma}(s) = \sigma_{\gamma}^\text{Res}(s) + \sigma_{\gamma}^\text{BG}(s),
\end{equation}
where $s=W^2=M^2+2ME_\gamma$ and $E_\gamma$ is the photon energy in the target rest frame.
The resonant part is given by the $Q^2\to0$ limit of \eq{eq:sigT:R} and we have
\begin{equation}
\label{eq:photocs:R}
\sigma_{\gamma}^\text{Res}(s) = \frac{2M}{W} \sum_{R}\frac{M_R^2 \Gamma_R  (K_R/K) f_R^{\gamma}(W)}{(s-M_R^2)^2+M_R^2\Gamma_R^2} |A_R(0)|^2 ,
\end{equation}
where the notations are similar to those in  \eq{eq:sigT:R}.

Above the resonance region, $W>2\gev$, the total photoproduction cross section is dominated
by nonresonant processes.
At high energy, $s>10\gevsq$, available photoproduction data can be
described to a high accuracy in terms of a Regge model fit~\cite{Cudell:1999tx} whose best fit result is
\begin{equation}\label{eq:photocs:reggefit}
\sigma_\gamma^\text{Regge}(s)=0.0598 s^{0.0933} + 0.1164 s^{-0.357}\ \text{mb}.
\end{equation}
%
Using this result
we model the background cross section $\sigma_{\gamma}^\text{BG}$ in the full region of $s$ by
applying a correction function $B_T$ by \eq{eq:B} to \eq{eq:photocs:reggefit}:
\begin{equation}\label{eq:photocs:BG}
\sigma_{\gamma}^\text{BG}(s)=B_T(s) \sigma_\gamma^\text{Regge}(s).
\end{equation}
Note that in this analysis we assume the correction function by \eq{eq:B} to be independent of $Q^2$.
For this reason the same function $B_T(s)$ can be applied to both,
the photoproduction cross section and the structure function $F_T$.
We further check this assumption in our analysis of data in Sec.~\ref{sec:fit}.

\section{\label{sec:fit}Data sets and fit}

\begingroup
\begin{table*}[htb]
\vspace{-5ex}
\caption{\label{tab:csdata}Hydrogen electroproduction cross-section data sets used in our analysis.
Listed are the experiments with corresponding number of data points (NDP) and kinematics coverage.
The values of $Q^2$ and $W^2$ are in $\gevsq$ units.
The cut $W^2>1.16\gevsq$ was applied.
The ``DIS" label indicates data which are mostly in the DIS region while the ``RES" label is
for data samples which are mostly in the resonance region.
The last two columns are the values of $\chi^2$ normalized per NDP computed, respectively,
in our model and in the model of Ref.~\cite{Christy:2007ve}, for comparison
(the symbol ``N/A" indicates that the model of Ref.~\cite{Christy:2007ve} is not applicable for kinematics reason).
}
\begin{ruledtabular}
\begin{tabular*}{\textwidth}{l@{\extracolsep{\fill}}rcccccc}
Data set
\footnote{Here we show the primary source of data in our analysis which is not always the full reference to the corresponding experiment.}
 &  NDP & $Q^2_\text{min}$ &  $Q^2_\text{max}$  & $W^2_\text{min}$  & $W^2_\text{max}$          & $\chi^2$ & $\chi_{\text{CB}}^2$\\
	\hline
SLAC-E49a     (DIS)~\cite{Whitlow:1990dr}  &   117   &  0.586  &  8.067  &  3.130  &  27.19 & 0.55 &  N/A \\
SLAC-E49b     (DIS)~\cite{Whitlow:1990dr}  &   208   &  0.663  &  20.08  &  3.010  &  27.51 & 1.32 &  N/A \\
SLAC-E61      (DIS)~\cite{Whitlow:1990dr}  &    32   &  0.581  &  1.738  &  3.210  &  16.00 & 0.44 &  N/A \\
SLAC-E87      (DIS)~\cite{Whitlow:1990dr}  &   109   &  3.959  &  20.41  &  3.280  &  17.18 & 0.57 &  N/A \\
SLAC-E89a     (DIS)~\cite{Whitlow:1990dr}  &    77   &  3.645  &  30.31  &  3.300  &  20.43 & 0.60 &  N/A \\
SLAC-E89b     (DIS)~\cite{Whitlow:1990dr}  &   118   &  0.887  &  19.18  &  3.100  &  27.75 & 0.70 &  N/A \\
SLAC-E004     (DIS)~\cite{ResDatabase}     &   198   &  0.249  &  20.07  &  3.561  &  26.84 & 0.44 &  N/A \\
SLAC-E49a6    (RES)~\cite{ResDatabase}     &   460   &  0.146  &  3.708  &  1.177  &  3.992 & 0.72 &  1.16\\
SLAC-E49a10   (RES)~\cite{ResDatabase}     &   541   &  0.445  &  8.593  &  1.171  &  4.000 & 0.84 &  1.04\\
SLAC-E49b     (RES)~\cite{ResDatabase}     &   366   &  1.018  &  16.74  &  1.153  &  3.992 & 0.81 &  1.15\\
SLAC-E61      (RES)~\cite{ResDatabase}     &  1075   &  0.061  &  1.839  &  1.160  &  4.000 & 1.20 &  1.76\\
SLAC-E87      (RES)~\cite{ResDatabase}     &    22   &  1.821  &  20.54  &  3.183  &  3.988 & 0.25 &  N/A\\
SLAC-E89a\footnote{Listed as SLAC-E891 in \cite{ResDatabase}.}
              (RES)~\cite{ResDatabase}     &    90   &  7.124  &  32.39  &  1.156  &  4.000 & 0.14 &  N/A\\
SLAC-E89b\footnote{Listed as SLAC-E8920 in \cite{ResDatabase}.}
              (RES)~\cite{ResDatabase}     &   492   &  0.395  &  20.66  &  1.197  &  3.984 & 1.12 &  N/A\\
SLAC-E133     (RES)~\cite{ResDatabase}     &   178   &  2.287  &  9.914  &  1.153  &  3.037 & 3.19 &  5.04\\
SLAC-E140     (RES)~\cite{ResDatabase}     &    87   &  0.717  &  20.41  &  3.010  &  3.950 & 1.46 &  N/A\\
SLAC-E140X    (RES)~\cite{ResDatabase}     &   153   &  1.118  &  8.871  &  1.200  &  3.720 & 2.88 &  3.27\\
SLAC-NE11     (RES)~\cite{Stuart:1996zs}   &   113   &  1.606  &  6.855  &  1.164  &  1.788 & 2.27 &  5.78\\
SLAC-Onen1half (RES)~\cite{ResDatabase}     &   745   &  0.011  &  0.263  &  1.153  &  4.000 & 6.18 &  7.00\\
Jlab-CLAS E1\footnote{The JLab-CLAS data sets E1, E2, E3, E4, E5 correspond to the beam energies 1.515, 2.567, 4.056, 4.247, 4.462~GeV, respectively.}
              (RES)~\cite{Osipenko:2003bu,Osipenko:2003ua,Osipenko:pc,CLASDatabase}   &   509   &  0.225 &  0.925 &  1.162 &  2.544 & 1.15 & 19.5\\
Jlab-CLAS E2  (RES)~\cite{Osipenko:2003bu,Osipenko:2003ua,Osipenko:pc,CLASDatabase}   &  1443   &  0.475 &  2.175 &  1.162 &  3.987 & 1.44 & 11.3\\
Jlab-CLAS E3  (RES)~\cite{Osipenko:2003bu,Osipenko:2003ua,Osipenko:pc,CLASDatabase}   &  2484   &  1.325 &  4.175 &  1.162 &  5.537 & 1.04 & 2.73\\
Jlab-CLAS E4  (RES)~\cite{Osipenko:2003bu,Osipenko:2003ua,Osipenko:pc,CLASDatabase}   &  2637   &  1.325 &  4.425 &  1.164 &  5.643 & 0.95 & 1.93\\
Jlab-CLAS E5  (RES)~\cite{Osipenko:2003bu,Osipenko:2003ua,Osipenko:pc,CLASDatabase}   &  2681   &  1.375 &  4.725 &  1.162 &  5.971 & 0.96 & 1.51\\
JLab-E94-110  (RES)~\cite{ResDatabase}     &  1273   &  0.181  &  5.168  &  1.225 &  3.850 & 3.15 & 1.33\\
JLab-E00-116  (RES)~\cite{Malace:2009kw}   &   261   &  3.585  &  7.384  &  1.243 &  5.131 & 1.48 & 1.58\\
JLab-E00-002  (RES)~\cite{Tvaskis:2016uxm,ResDatabase}  &  1477   &  0.055  &  2.079  &  1.163 &  7.932 & 1.22 & 0.88\\
\end{tabular*}
\end{ruledtabular}
\caption{\label{tab:photodata}Hydrogen photoproduction cross-section data sets used in our analysis.
Listed are the experiments with corresponding number of data points (NDP) and kinematics coverage.
The values of $W^2$ are in $\gevsq$ units.
The last two columns are the values of $\chi^2$ normalized per NDP computed in our model
and in the model of Ref.~\cite{Christy:2007ve}, respectively
(``N/A" has the same meaning as in Table~\ref{tab:csdata}).}
\begin{ruledtabular}
\begin{tabular*}{\textwidth}{l@{\extracolsep{\fill}}rcccc}
Data set                               &  NDP &  $W^2_\text{min}$    & $W^2_\text{max}$   & $\chi^2$ & $\chi_{\text{CB}}^2$\\
\hline
Armstrong   \cite{Armstrong:1971ns}   &  159 &   1.378   &   8.790 & 2.39     & 1.34 \\
Maccormick  \cite{MacCormick:1996jz}  &  57  &   1.263   &   2.361 & 2.15     & 7.12 \\
Meyer       \cite{Meyer:1970fya}      &  18  &   3.038   &   12.61 & 0.69     & 0.54 \\
Hilpert     \cite{Hilpert:1968uzd}    &  6   &   2.121   &   9.212 & 3.09     & 1.66 \\
Dieterle    \cite{Perl:1969kf}        &  5   &   2.382   &   11.67 & 1.70     & N/A  \\
Ballam      \cite{Ballam:1971yd}      &  3   &   6.135   &   14.95 & 0.79     & N/A  \\
Bingham     \cite{Bingham:1973fu}     &  1   &   18.33   &   18.33 & 0.26     & N/A  \\
Caldwell    \cite{Caldwell:1973bu}    &  9   &   8.518   &   31.62 & 1.10     & N/A  \\
Caldwell    \cite{Caldwell:1978yb}    &  30  &   35.22   &   343.7 & 0.64     & N/A  \\
Michalowski \cite{Michalowski:1977eg} &  6   &   4.633   &   18.73 & 1.08     & N/A  \\
Alexander   \cite{Alexander:1974zc}   &  1   &   14.95   &   14.95 & 0.003     & N/A  \\
Aid         \cite{Aid:1995bz}         &  2   &   39999   &   43681 & 0.27     & N/A  \\
Vereshkov   \cite{Vereshkov:2003cp}   &  4   &   2065    &   17822 & 0.24     & N/A  \\
GRAAL       \cite{Bartalini:2008zza}  &  62  &   1.950   &   3.564 & 9.92     & 6.50 \\
\end{tabular*}
\end{ruledtabular}
\end{table*}
\endgroup

\begin{figure*}[htb]
\centering
\includegraphics[width=0.5\textwidth]{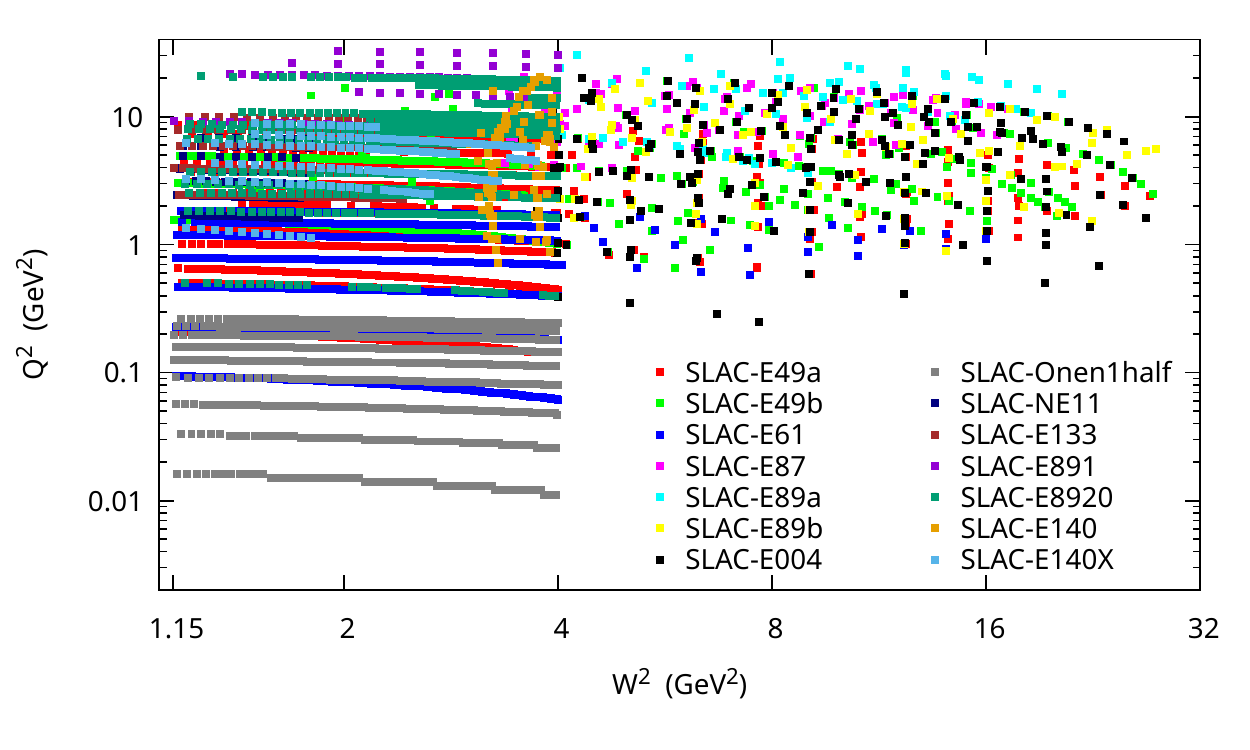}%
\includegraphics[width=0.5\textwidth]{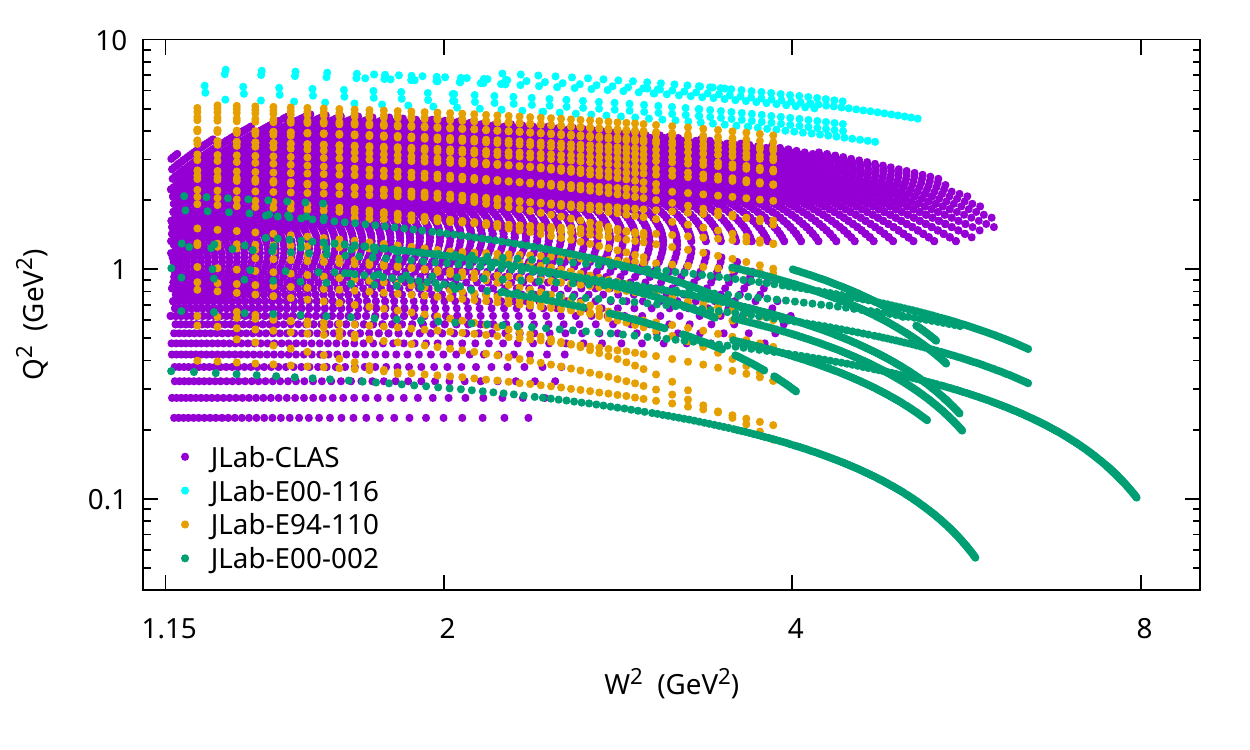}
\caption{\label{fig:csdata}
The $(W^2,Q^2)$ space populated by data points of the electroproduction data sets in Table~\ref{tab:csdata}.
The left (right) panel is for SLAC (JLab) experiments.
The color code and legend for each of the data set are shown in the panels.}
\end{figure*}

We adjust parameters of our model by fitting to world data on the
differential cross section of inelastic electron scattering
together with photoproduction cross section data off a hydrogen target.
The data sets used in our analysis are listed in Table~\ref{tab:csdata} and Table~\ref{tab:photodata}.
The electron beam energy of SLAC experiments~\cite{Poucher:1973rg,Bodek:1979rx,Atwood:1976ys,Mestayer:1982ba,Stein:1975yy,Whitlow:1990dr,Whitlow:1991uw,Dasu:1993vk,Tao:1995uh,Abe:1998ym,Stuart:1996zs} 
span the region from $2$ to $20$~GeV and the data cover a wide kinematical region including the resonance production and DIS,
while the beam energy of JLab experiments~\cite{Osipenko:2003bu,Osipenko:2003ua,Osipenko:pc,Liang:2004tj,Malace:2009kw,Tvaskis:2016uxm} 
was in the
range from $1.15$ to $5.5$~GeV and their data cover the resonance and transition region.
The kinematics as well as statistics coverage of electron cross section data sets are illustrated in the $(W^2,Q^2)$ plane in Fig.~\ref{fig:csdata}.
The photoproduction cross section data at $Q^2=0$ are listed in Table~\ref{tab:photodata}.

Note that our hybrid model is designed to smoothly match the DIS region.
For this reason we do not include in the fit DIS data sets from
NMC~\cite{Arneodo:1996qe},
BCDMS~\cite{Benvenuti:1989rh},
H1~\cite{Adloff:2000qk},
and ZEUS~\cite{Chekanov:2001qu},
which were included in a global QCD analysis of Ref.~\cite{Alekhin:2007fh} whose PDFs and HT terms are used in our study.

The parameters of our model are adjusted from minimization of the $\chi^2$ function
\begin{equation}\label{eq:chi2}
	\chi^2 = \sum_i \left(v_i^\text{exp} - v_i^\text{model}\right)^2/\sigma_i^2,
\end{equation}
where the sum runs over the cross section data points used in our fit,
and $v_i^\text{exp}$ and $v_i^\text{model}$ are the corresponding experimental and model values,
and $\sigma_i^2$ is the experimental uncertainty squared,
for which we take the quadrature sum of statistical and systematic error.
We use data from Tables~\ref{tab:csdata} and \ref{tab:photodata} as provided with no attempt to adjust overall normalization of particular data sets.

We use the {\sc minuit} program~\cite{James:1975dr} to minimize $\chi^2$ and
thus determine the parameters of our model.
In preliminary trials we had 68 free parameters including
the masses, widths, orbital momenta, damping parameters, and the branching fractions of resonance states in Table~\ref{tab:respars};
the parameters $a_i$ and $c_i$ describe the resonance amplitudes by  \Eqs{eq:AMPL:T}{eq:AMPL:L},
the parameters $b_i$ are for the nonresonant background in \eq{eq:B} for transverse and longitudinal channels,
and the scale parameter $Q_0$
as well as the parameters $m$ and $n$ drive the $Q^2$ dependence of extrapolated structure functions in \Eqs{eq:ext:FT2}{eq:ext:FL2}.

An iterative procedure was applied to find the best fit model parameters.
We first addressed the total photoproduction cross section data at $Q^2=0$ in Table~\ref{tab:photodata}.
This allowed us to determine some of the model parameters,
in particular the resonance masses and widths, the normalization of resonance
helicity amplitudes $|A(0)|=a_1$, and the parameters of the $B_T$ function.
Those parameters were then used as input for a further combined fit including both
the photoproduction and electroproduction cross-section data from
Tables \ref{tab:csdata} and \ref{tab:photodata}.
By studying the photoproduction cross section we also tried to optimize
the angular momenta of effective resonances together with their branching fractions,
which then were fixed to the values listed in Table~\ref{tab:respars}.

In a combined fit to electroproduction and photoproduction data we had a number of sequential iterations.
In initial trials we kept the resonance parameters fixed,
thus performing adjustment of the background parameters $b_i$
as well as the exponents $m$ and $n$ controlling  low-$Q$ behavior of the structure functions
(see Sec.~\ref{sec:model:bg}).
Then we performed the adjustment of the resonance parameters with the background fixed.
This allowed us to study correlations between the parameters
and also to locate the parameters to which the fit is most (least) sensitive.

In the course of our analysis we also observed that the fit prefers the $n\to m$ limit on the exponents describing extrapolation of the structure functions (see Sec.~\ref{sec:model:bg}).
We then fixed $n=m$ for both the transverse and the longitudinal SFs.
Note that the exponent $m$ is treated independently for $F_T$ and $F_L$.

We also performed the studies aiming to optimize the parameter $t_0=Q_0^2$,
the scale from which we start extrapolating the DIS structure functions down to $Q^2=0$
(see Sec.~\ref{sec:model:bg}).
Note that our background DIS SFs are constrained by data for $Q^2>1\gevsq$~\cite{Alekhin:2007fh,Alekhin:2008ua}.
Our fit prefers $t_0=1\gevsq$ resulting in a minimum of global $\chi^2$.
However, the low-$Q$ extrapolation starting from $t_0=1\gevsq$
following the method of Sec.~\ref{sec:model:bg}
results in $F_L<0$ for high values of $W$ and $Q^2<1\gevsq$ (effectively for $x<0.01$)
owing to a negative contribution from the derivative term in \eq{eq:ext:FL2}.
Raising the extrapolation scale $t_0$ allows us to reduce the impact of this derivative term.
We found that $t_0=2\gevsq$
results in a stable behavior of extrapolated $F_L$ at low values of $x$ and $Q^2$.
Note also that higher value of $t_0$ allows us to reduce uncertainties
in evaluating the DIS SF by \eq{eq:sf:dis} at this scale.

It should be also commented that our fit results in numerically small values
of the longitudinal parameters
$c_1(\text{Roper})$, $c_1(R_2)$, $c_2(\Delta)$, $c_2(R_2)$, $c_2(R_3)$.
On the final step we set these parameters to 0 as indicated in Table~\ref{tab:res_ampl_pars_S}.
Also, our preliminary fits prefer the numerical value of the $m_L$ exponent close to 1.
Note that for the reason of vanishing $R=F_L/F_T\to0$ at $Q^2\to0$, the value of the exponent $m_L$ must be $m_L>1$.
In the final fit trial we also fixed $m_L=1.1$,
thus leaving  49 fit parameters.

Our fit results in the minimum of $\chi^2=27785.77$ for the total number of data points $\text{NDP}=18298$ (most of the data are in the resonance region).
Thus, we have  $\chi^2/\text{NDP}\approx 1.52$.
The best fit parameters together with relative fit uncertainties are listed in Table~\ref{tab:respars}
and Tables~\ref{tab:res_ampl_pars_A} to \ref{tab:bg_part_pars}.
The values of $\chi^2$ normalized per number of data points of individual experiments
are listed in Table~\ref{tab:csdata} for all hydrogen electroproduction data sets used in the fit,
and the corresponding $\chi^2$ values for photoproduction data are given in Table~\ref{tab:photodata}.
The last column in Tables~\ref{tab:csdata} and \ref{tab:photodata} lists the values of $\chi^2$ computed for the model of Ref.~\cite{Christy:2007ve}, where applicable.
We observe significant improvement over the results of the empirical model of Ref.~\cite{Christy:2007ve}
for all studied data sets except for the cross-section data from the JLab-E94-110 experiment.

%
%
\begingroup
\squeezetable
\begin{table}[htb!]
\vspace*{-1ex}
\caption{The best fit parameters  describing the resonant contributions to the transverse helicity amplitude by \eq{eq:AMPL:T}.
The estimate of fractional parameter uncertainty is given in parentheses in percent units.
\label{tab:res_ampl_pars_A}}
\begin{ruledtabular}
\begin{tabular*}{\linewidth}{l@{\extracolsep{\fill}}llll}
{}             & $a_1\ (\mathrm{GeV}^{-1/2})$         &  $a_2\  (\mathrm{GeV}^{-5/2})$                            & $a_3\ (\mathrm{GeV}^{-2})$         & $a_4$ \\
\hline
$\Delta(1232)$ & 0.31115(0.31) &                     2.02940(0.57) & 1.67130(1.06) & 2.7600(0.41) \\
$N(1440)$      & 0.08955(4.61) &                     0.18087(1.16) & 0.23431(0.87) & 4.1173(0.35) \\
$R_1$          & 0.10677(2.08) &                     0.24897(1.62) & 0.55621(0.66) & 3.0798(0.38) \\
$R_2$          & 0.38953(0.60) & \,\hspace{-\lminus}$-$0.17962(1.88) & 0.37638(3.09) & 2.9622(1.70) \\
$R_3$          & 0.06708(5.72) &                     0.09733(6.26) & 0.27891(4.74) & 3.5372(1.42) \\
\end{tabular*}
\end{ruledtabular}
%
\caption{The best fit parameters describing the resonant contributions to the longitudinal helicity amplitude by \eq{eq:AMPL:L}. The estimate of fractional parameter uncertainty is given in parentheses in percent.}
\label{tab:res_ampl_pars_S}
\begin{ruledtabular}
\begin{tabular*}{\linewidth}{l@{\extracolsep{\fill}}lll}
{}             & $c_1\ (\mathrm{GeV}^{-1/2})$ & $c_2\ (\mathrm{GeV}^{-5/2})$ & $c_3\ (\mathrm{GeV}^{-2})$\\
\hline
$\Delta(1232)$ & 0.05029(6.72) &                     0             & 0.42522(6.40)  \\
$N(1440)$      & 0             &                     0.23847(2.62) & 1.4982(2.03)  \\
$R_1$          & 0.09198(4.33) & \,\hspace{-\lminus}$-$0.10652(5.81) & 1.0758(3.48)  \\
$R_2$          & 0             &                     0             & 0              \\
$R_3$          & 0.12027(1.68) &                     0             & 0.89367(2.72)  \\
\end{tabular*}
\end{ruledtabular}
%
\caption{The best fit parameters describing the background function by \eq{eq:B}. The estimate of fractional parameter uncertainty is given in parentheses in percent.}
\label{tab:bg_part_pars}
\begin{ruledtabular}
\begin{tabular*}{\linewidth}{l@{\extracolsep{\fill}}lll}
{}    & $b_1\ (\mathrm{GeV}^{-2 b_2})$         &  $b_2$       & $m_{T,L}$ \\
\hline
$B_T$ & 0.14453(4.19) & 3.1297(1.76) & 1.6302(0.19) \\
$B_L$ & 3.4742(2.44) & 0.54193(1.26) & 1.1 \\
\end{tabular*}
\end{ruledtabular}
\end{table}
\endgroup

\begin{figure*}[htb!]
\centering
\includegraphics[width=0.5\linewidth]{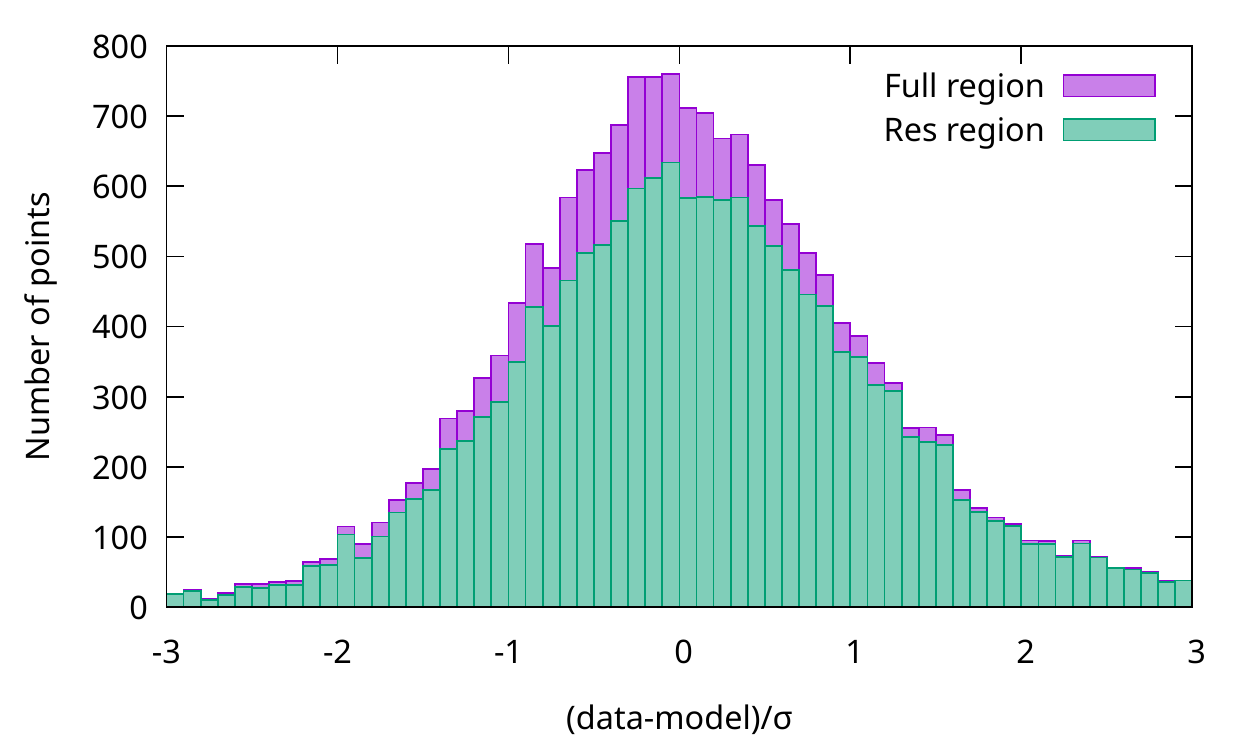}%
\includegraphics[width=0.5\linewidth]{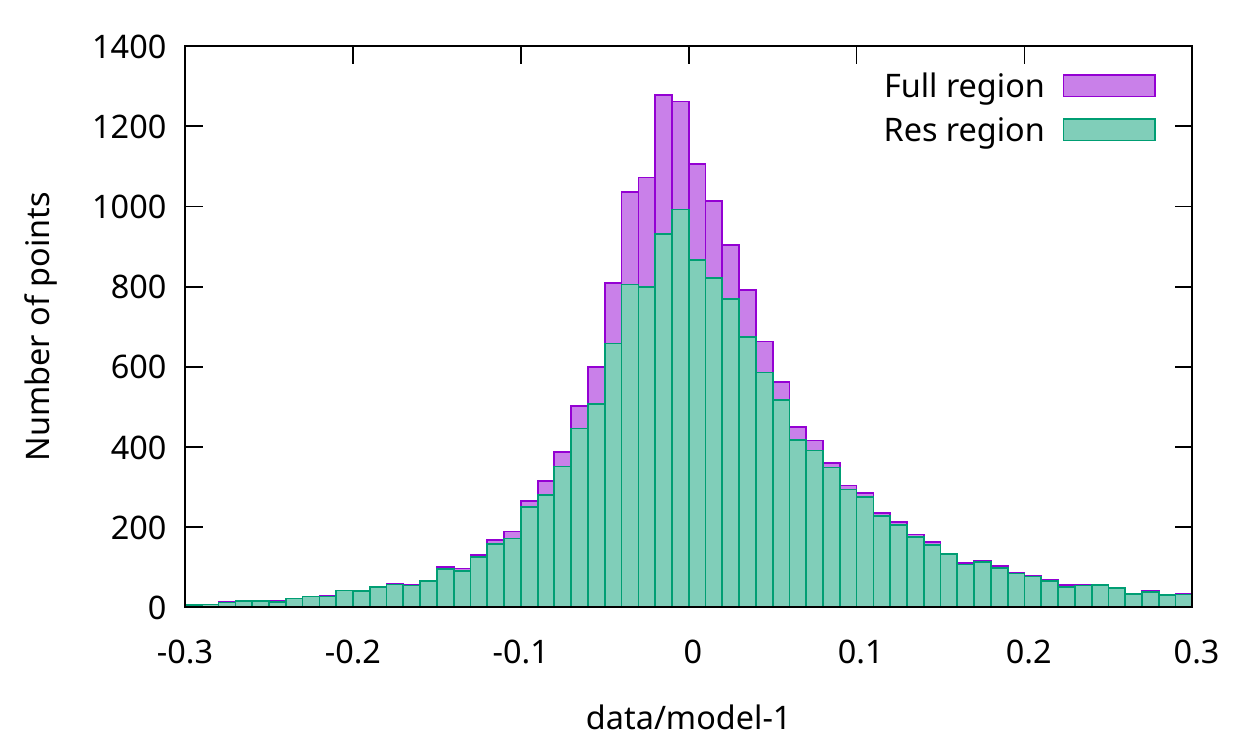}
\caption{\label{fig:CSdist}
Left panel: NDP distribution 
vs the residual  $(v_\text{dat}-v_\text{model})/\sigma_\text{dat}$ (see text).
Separate distributions are shown for the full data set and for resonance data with the legend in the figure panel.
Right panel: NDP distribution vs $v_\text{dat}/v_\text{model}-1$.}
\end{figure*}

In order to illustrate the overall quality of our fit,
in the left panel of Fig.~\ref{fig:CSdist} we show the distribution of the number of data points vs
the residual $(v_\text{dat}-v_\text{model})/\sigma_\text{dat}$,
where $v_\text{dat}$ and $\sigma_\text{dat}$ are the measured cross section value
and its experimental uncertainty
and $v_\text{model}$ is the corresponding model value.
The distribution is presented separately for the resonance region ($W<2\gev$)
and for the full set of data points.
The analysis of this residual distribution helps to understand the overall fit uncertainty
together with the shift of fit results vs data.
We found that both these residual distributions follow the normal distribution with  good accuracy,
with about 67\% of data points within $\pm 1\sigma$ interval,
and the average residual is consistent with 0.
In the right panel of Fig.~\ref{fig:CSdist} we show a similar distribution
vs. $\text{data}/\text{model}-1$ shift.
We have about 78\% of data points within $\pm10\%$ interval of  $\text{data}/\text{model}-1$.

\begin{figure*}[p]
	\centering
	\includegraphics[width=1.0\linewidth]{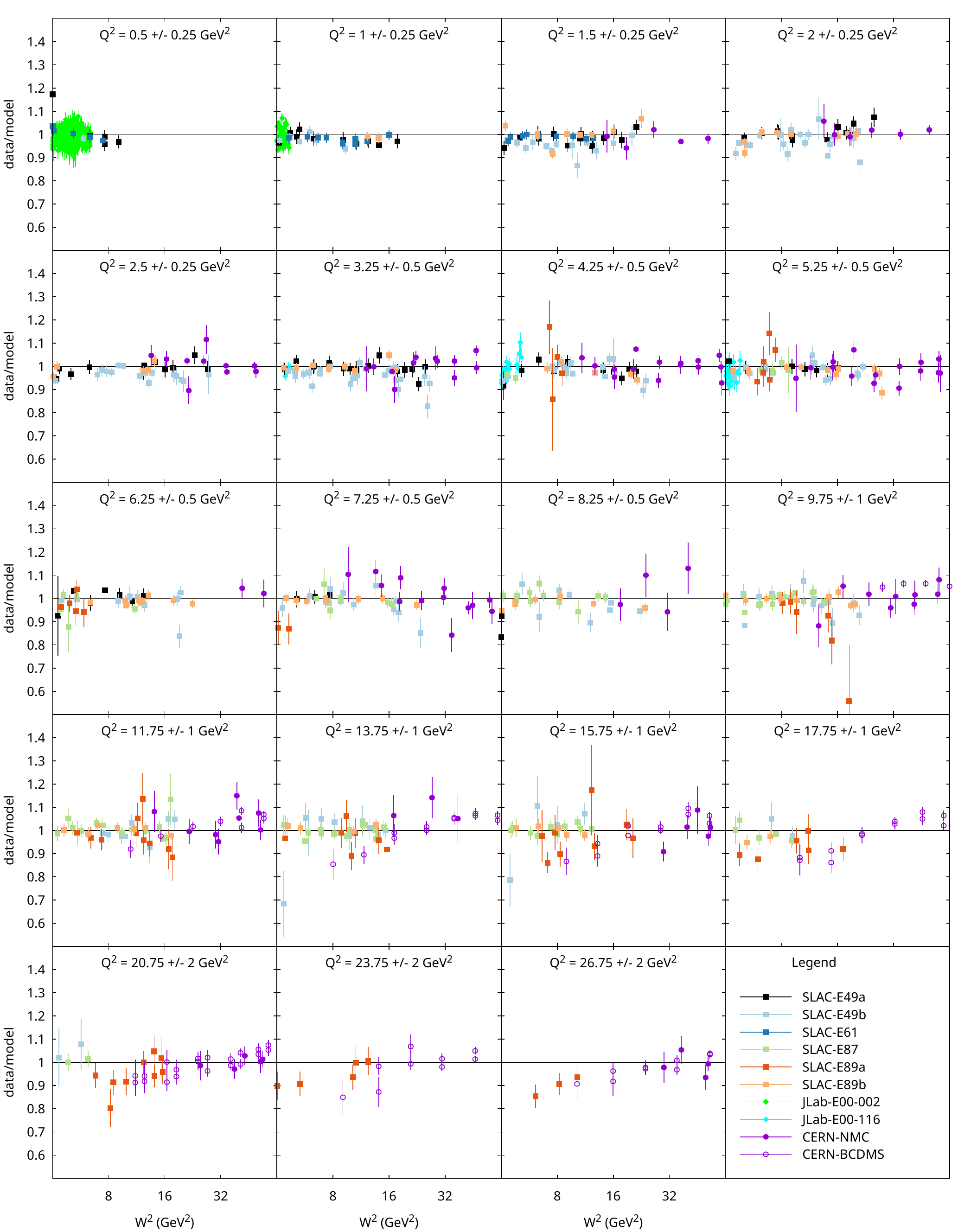}
	\caption{\label{fig:pulls_CS_DIS}
		The pulls for cross-section data of Table~\ref{tab:csdata} with $W^2>4\gevsq$.
		Data points are grouped in $Q^2$ bins indicated in the panels.
		The error bars include statistical and systematic uncertainties of data taken in quadrature.
		The legend for the data points is shown in the lower-right panel.}
\end{figure*}

\begin{figure*}[p]
	\centering
	\includegraphics[width=1.0\linewidth]{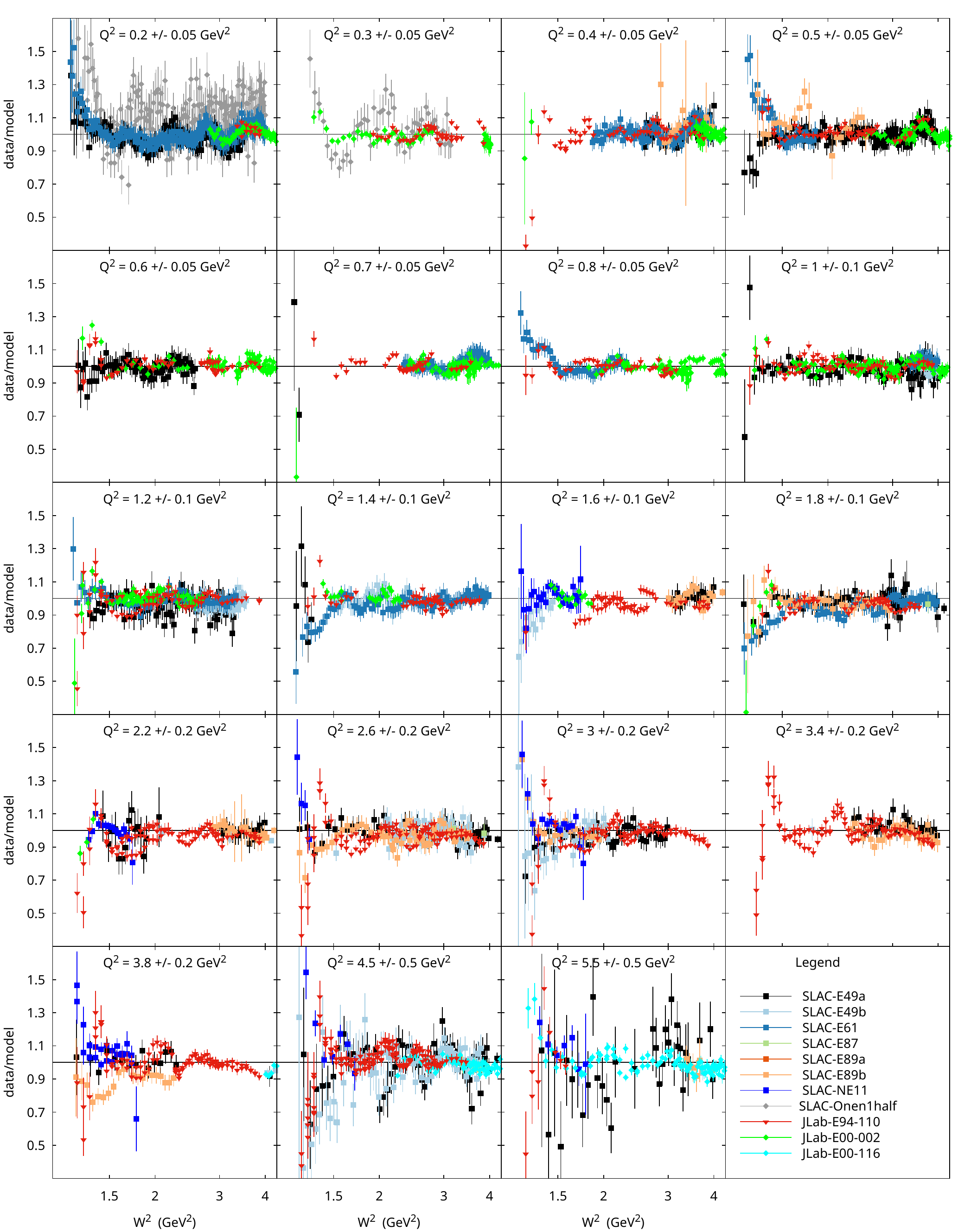}
	\caption{\label{fig:pulls_CS_RES}
		The pulls similar to Fig.~\ref{fig:pulls_CS_DIS} but focused at the resonance region $W^2 \lesssim 4\gevsq$.
		For the pulls of JLab-CLAS data see Fig.~\ref{fig:pulls_CS_CLAS}.}
\end{figure*}

\begin{figure*}[p]
	\centering
	\includegraphics[width=1.0\linewidth]{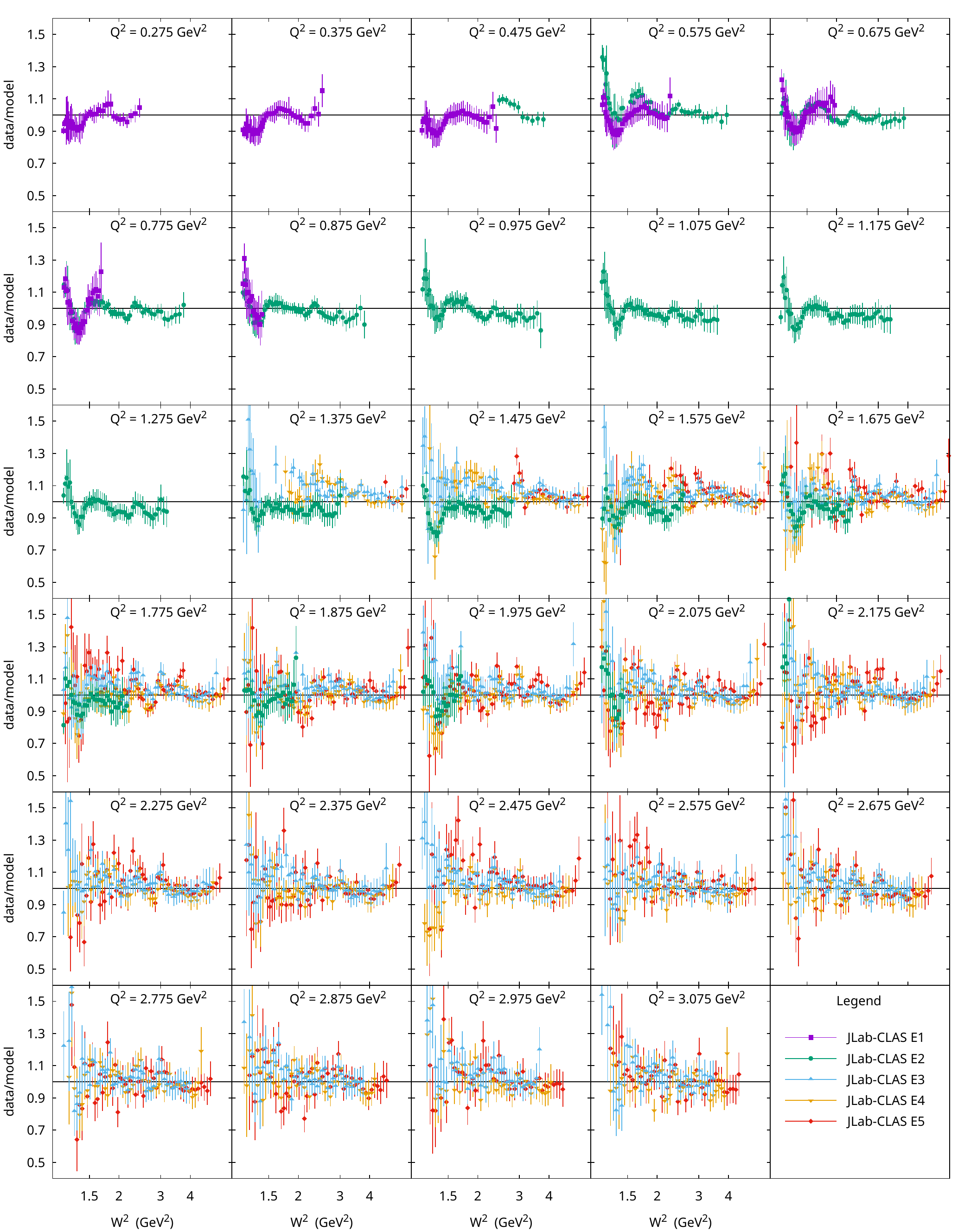}
	\caption{
		The pulls for JLab-CLAS cross-section data listed in Table~\ref{tab:csdata}.
		The panels correspond to $Q^2$ bins from JLab-CLAS measurement.
		The error bars include statistical and systematic uncertainties of data taken in quadrature.
		The legend for data points is shown in the lower-right panel.}
	\label{fig:pulls_CS_CLAS}
\end{figure*}

In Figs.~\ref{fig:pulls_CS_DIS} to \ref{fig:pulls_CS_CLAS} we show  data/model ratios (pulls)
for different electroproduction cross section data samples and for different kinematical regions.
The pulls for the electroproduction cross section data are organized in terms of a set of panels
corresponding to different $Q^2$ bins indicated in the plots.
Note the logarithmic scale in $W^2$.
Figure~\ref{fig:pulls_CS_DIS}
refers to the region $W^2>4\gevsq$ with the set of $Q^2$ bins selected to cover the region
$0.25<Q^2<30\gevsq$,
while Fig.~\ref{fig:pulls_CS_RES} is focused on the resonance region $W^2 < 4\gevsq$
and $0.25<Q^2<6\gevsq$.
For completeness, the pulls in Fig.~\ref{fig:pulls_CS_DIS} also include the DIS cross section data,
such as CERN-NMC~\cite{Arneodo:1996qe},
CERN-BCDMS~\cite{Benvenuti:1989rh} and
DESY-HERMES~\cite{Airapetian:2011nu},
which were not used in our fit but included in the global QCD analysis of
Refs.~\cite{Alekhin:2007fh,Alekhin:2008ua} whose PDFs and HT terms are used in our study.
In view of a large number of data points from
the JLab-CLAS experiment~\cite{Osipenko:2003bu,Osipenko:2003ua,Osipenko:pc,CLASDatabase},
which were presented for fixed $Q^2$ bins,
we show the comparison with JLab-CLAS data points in Fig.~\ref{fig:pulls_CS_CLAS}.

Overall, our model shows a very good agreement with data in a wide region of $Q^2$ and $W^2$.
However, we also observe rather strong fluctuations of the pulls for some data sets.
In particular, the pulls for different beam energies of JLab-CLAS cross section data
are not always fully consistent (see, e.g., the bins with $Q^2>1.3\gevsq$ in Fig.~\ref{fig:pulls_CS_CLAS}).
The fluctuations of the data/model ratio  are high in the region $W^2<1.5\gevsq$,
as can be seen in Fig.~\ref{fig:pulls_CS_RES}.
It is worth mentioning that SLAC-Onen1half cross section data~\cite{ResDatabase}
in the energy bin $E=5\gev$ and at very low values of $Q^2\sim 0.01\gevsq$
are underestimated in our model resulting in a rather high
$\chi^2$ value in Table~\ref{tab:csdata}.
However, for other energy bins of this experiment our model is in a reasonable agreement with data, as
illustrated by the cross-section plots in the Supplemental Material~\cite{supplement}.

Also, our global fit to world electroproduction and photoproduction cross section data leads to a good overall description of the total photoproduction cross section
(see Fig.~12 in the Supplemental Material~\cite{supplement}).
We found that our result for
the average transverse helicity amplitude squared
$A(Q^2)^2$ for the $\Delta(1232)$ resonance state is in a good agreement
with the results of a low-$Q$ analysis in a unitary isobar model~\cite{Drechsel:1998hk}.
For the $N(1440)$ resonance state we obtained the value of $a_1^2$ to be consistent with that
reported in the Particle Data Group review~\cite{Tanabashi:2018oca}.
However, $Q^2$ dependence of the transverse amplitude $A(Q^2)$ for the $N(1440)$ state is different from that of Ref.~\cite{Drechsel:1998hk}.

Note that the value of $a_1^2$ drives the strength of corresponding resonance peak
in the photoproduction cross section at $Q^2=0$.
However, for the third resonance peak our model is somewhat off the data.
Note in this context that, in a fit to only photoproduction data,
we achieve a good description of data in the full region with
a somewhat different set of resonance parameters,
whose values are in a tension with the results from our global fit.
In particular, the photoproduction data require a lower value of the mass of
the third effective resonance together with a significantly higher value of $a_1^2$ for this state.
This may indicate $Q^2$ dependence of Breit-Wiegner pole positions in our effective model.
This point will be addressed in future studies.

A detailed comparison of our predictions with cross-section data on both
the electroproduction differential cross section from Table~\ref{tab:csdata}
and the total photoproduction cross section from Table~\ref{tab:photodata}
can be found in the Supplemental Material~\cite{supplement}.

\section{\label{sec:discus}Discussion}

To verify our results
we compare the model predictions with data which were not used in our fit,
in particular the measurements of the structure function $F_2$ and $R=F_L/F_T$
and the DIS cross-section data in Table~\ref{tab:f2_chi2}.

We first discuss the measurements of the structure function $F_2$.
The $F_2$ data in Table~\ref{tab:f2_chi2}
include the results of Refs.~\cite{Whitlow:1990dr,Whitlow:1991uw} obtained from re-analysis of ``old" SLAC data,
the measurements
from JLab experiments~\cite{Osipenko:2003ua,Osipenko:2003bu,Liang:2004tj,Liang:thesis,Malace:2009kw,Tvaskis:2016uxm},
and we also include the $F_2$ and cross-section data from
DESY-HERMES~\cite{Airapetian:2011nu}, CERN-NMC~\cite{Arneodo:1996qe}
and CERN-BCDMS~\cite{Benvenuti:1989rh}.
To illustrate the quality of data description,
Table~\ref{tab:f2_chi2} lists the values of $\chi^2$ per one data point for each of the data set.

Note that the $F_2$ extractions from various experiments listed in Table~\ref{tab:f2_chi2}
depend on the input for $R$.
The $F_2$ measurements of Refs.~\cite{Liang:2004tj,Tvaskis:2016uxm} were based on
the Rosenbluth separation of $F_T$ and $F_L$, while
the $F_2$ extractions from Refs.~\cite{Whitlow:1990dr,Osipenko:2003ua,Malace:2009kw} used different models of $R$ constrained by data.
The $F_2$ extraction of Ref.~\cite{Arneodo:1996qe} was based on their own measurement of $R$,
while the $F_2$ extraction of Ref.~\cite{Benvenuti:1989rh} assumed $R=0$.
This may explain the significant difference in the values of $\chi^2$ in our model
for the cross-section data and $F_2$ data of Ref.~\cite{Benvenuti:1989rh} in Table~\ref{tab:f2_chi2}.

\begin{figure*}[p]
\centering
\includegraphics[width=1.0\linewidth]{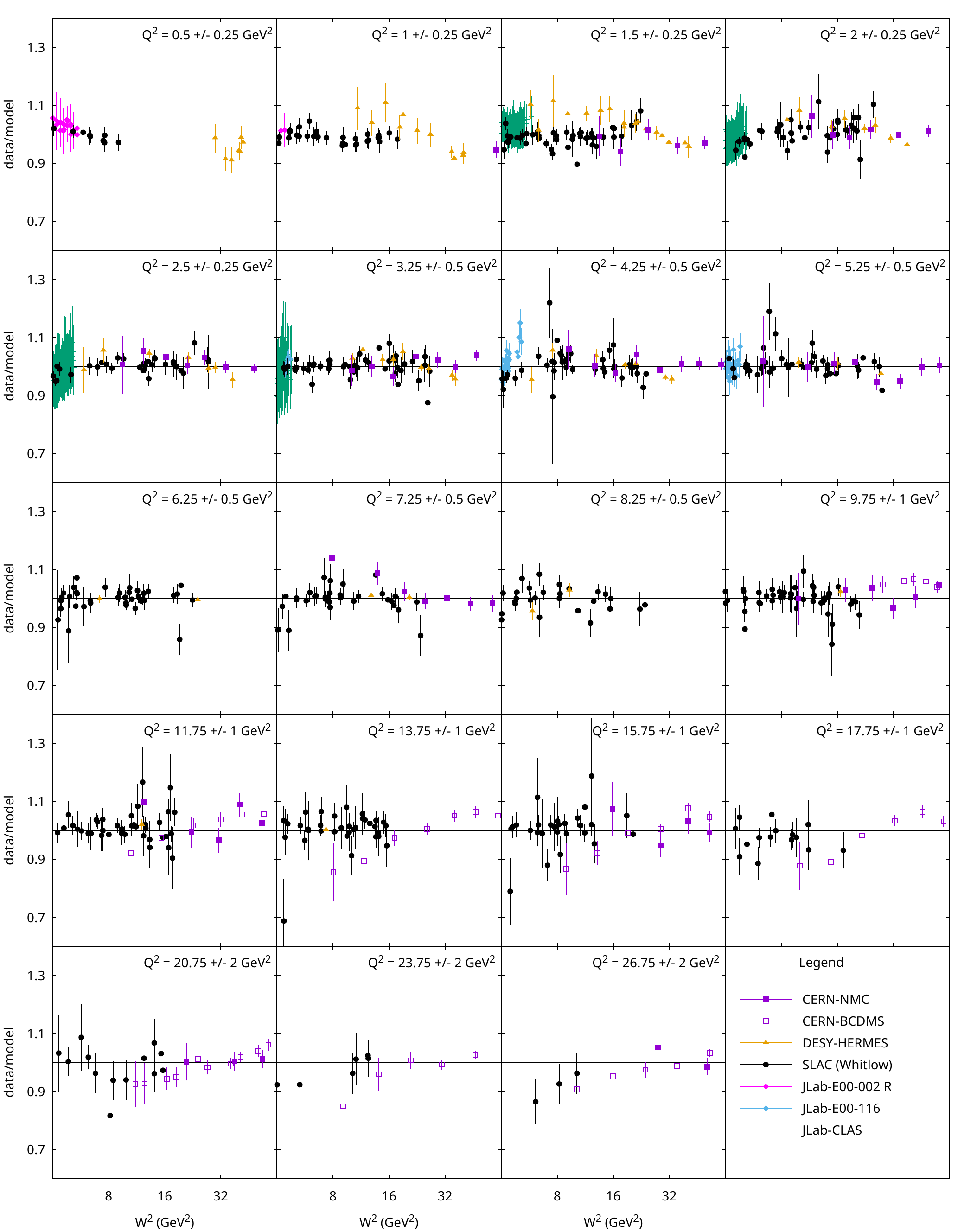}
\caption{\label{fig:pulls_F2_DIS}
The pulls for $F_2$ data in Table~\ref{tab:f2_chi2} vs $W^2$.
Shown is the region $W^2>4\gevsq$.
Data points were grouped in $Q^2$ bins indicated in the panels.
The error bars include statistical and systematic uncertainties of data taken in quadrature.
The legend for the data points is shown in the lower-right panel.}
\end{figure*}

\begin{figure*}[p]
\centering
\includegraphics[width=1.0\linewidth]{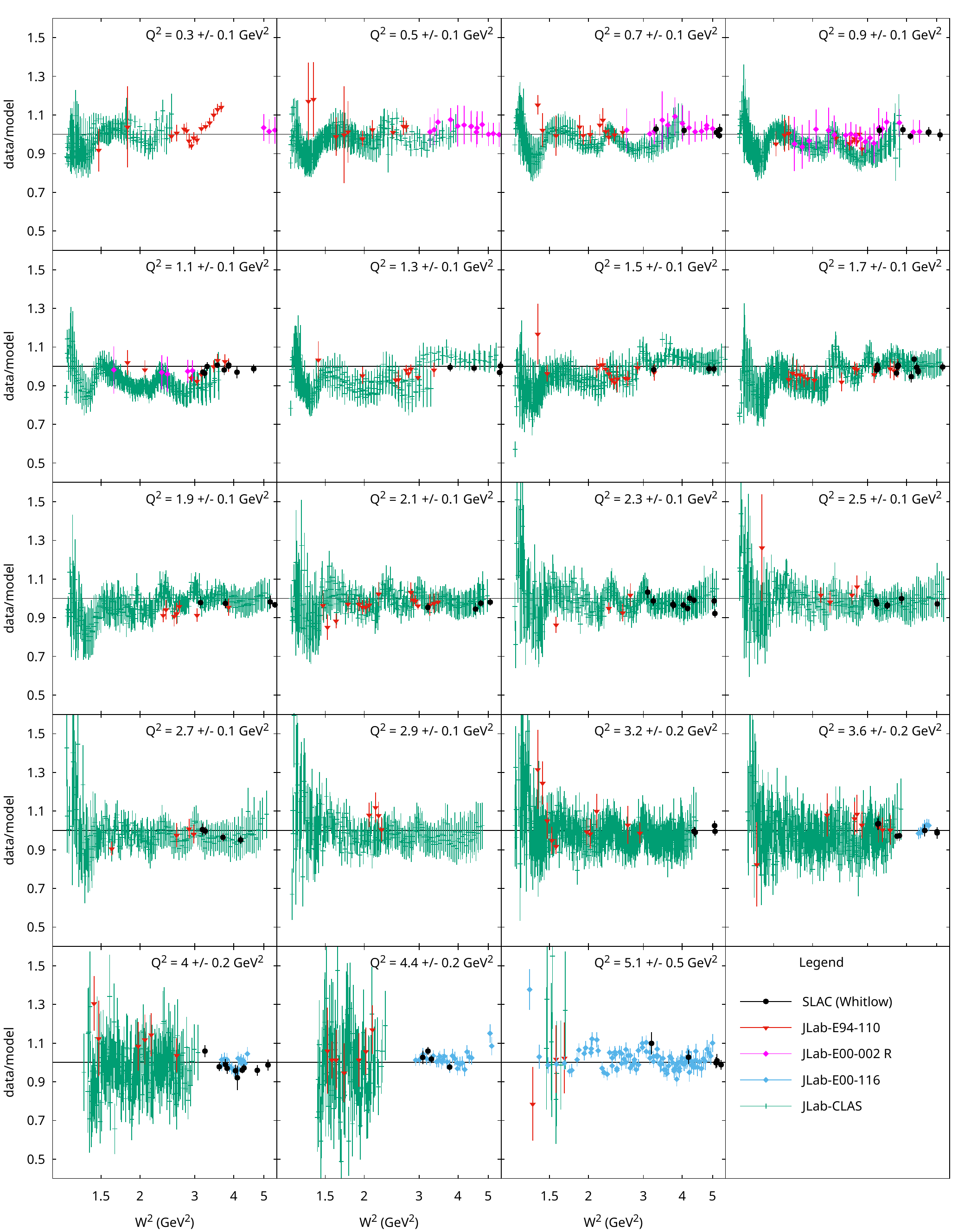}
\caption{\label{fig:pulls_F2_RES}
Similar to Fig.~\ref{fig:pulls_F2_DIS} but for the region of $W^2<5.5\gevsq$.}
\end{figure*}

\begin{figure*}[p]
\centering
\includegraphics[width=1.0\linewidth]{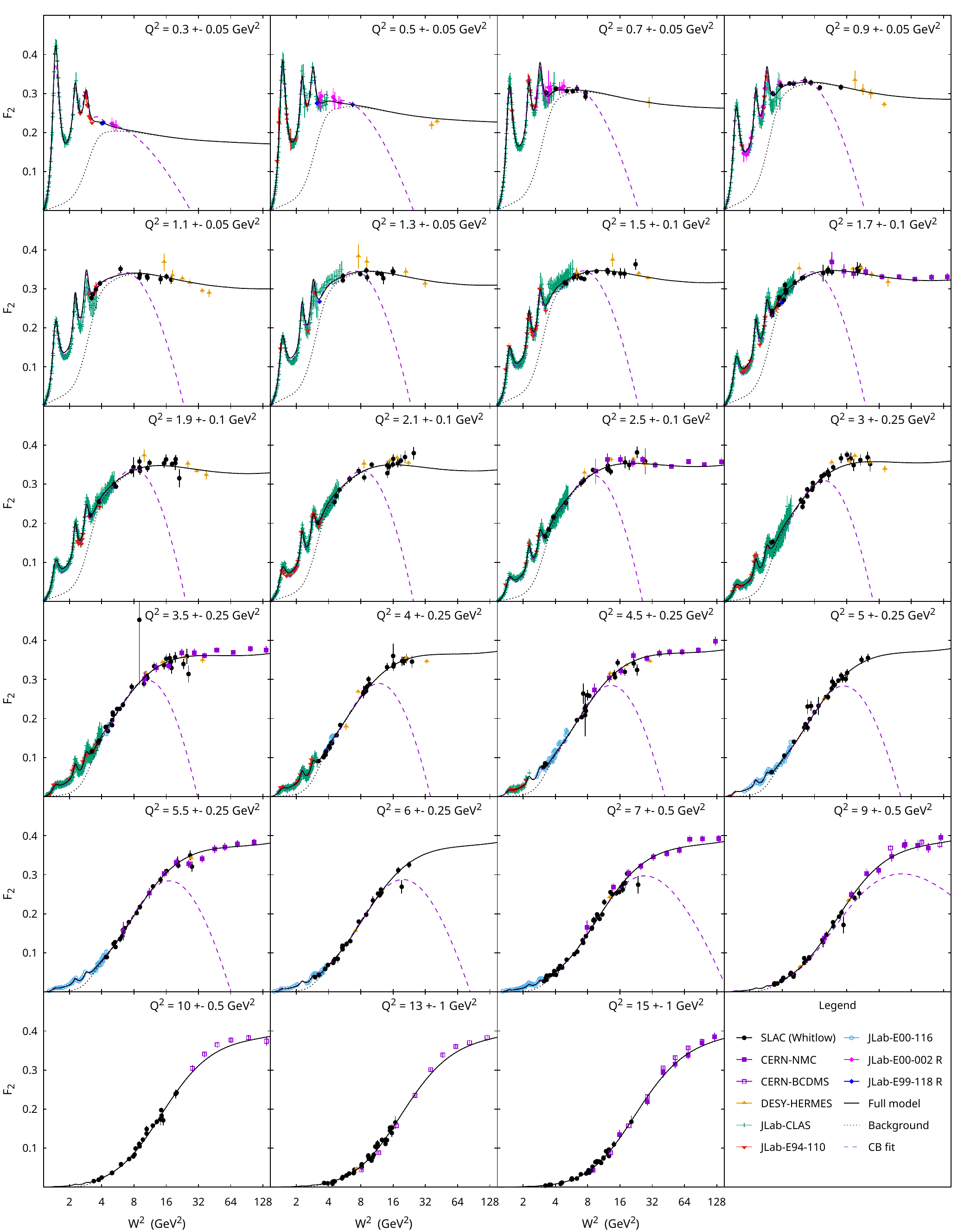}
\caption{\label{fig:F2}
Our predictions on $F_2$ (solid line) in comparison with data for $W^2<150\gevsq$ and for a number of $Q^2$ bins indicated in the panels. The curves are drawn for the central value of each of the $Q^2$ bin.
The legend for the data points is shown in the lower-right panel.
The background contribution is shown by dotted line while
the results of Ref.~\cite{Christy:2007ve} (CB) are shown by the dashed line.}
\end{figure*}

\begin{figure*}[p]
\centering
\includegraphics[width=1.0\linewidth]{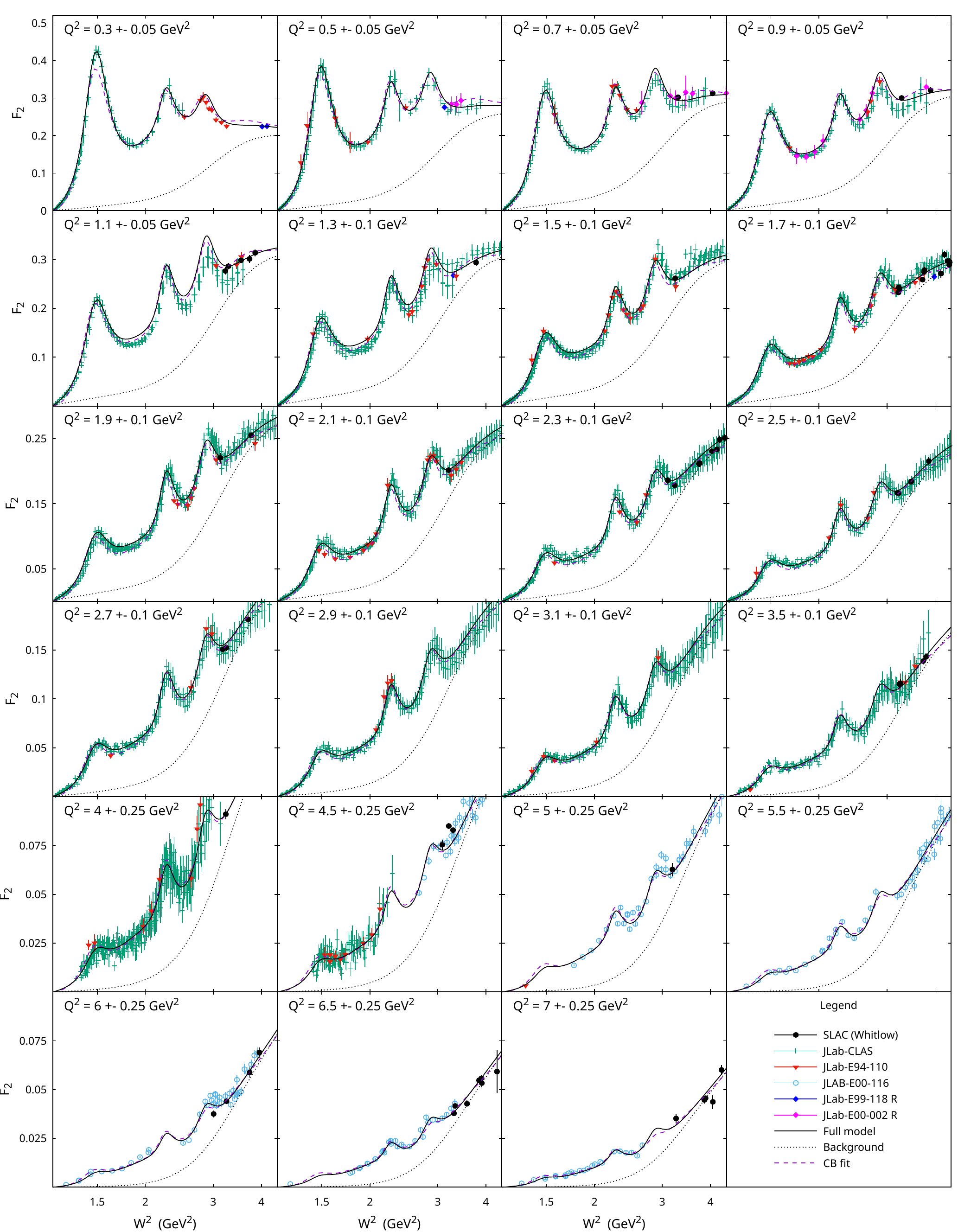}
\caption{\label{fig:F2RES}
Similar to Fig.~\ref{fig:F2} but focusing on the region $W^2 < 4.5\gevsq$ and $Q^2\lesssim 7\gevsq$.}
\end{figure*}

Our results on $F_2$ are illustrated in Figs.~\ref{fig:pulls_F2_DIS} to \ref{fig:F2RES}.
Figure~\ref{fig:pulls_F2_DIS} shows the pulls for $W^2>4\gevsq$ and $0.25<Q^2<30\gevsq$, which
are organized in the panels of $Q^2$ bins indicated in the plot.
Figure~\ref{fig:pulls_F2_RES} shows similar pulls for the resonance region of $W^2<4\gevsq$ and $0.2<Q^2<5.5\gevsq$.

In Figs.~\ref{fig:F2} and \ref{fig:F2RES} the structure function $F_2$ is shown vs $W^2$ for
a number of $Q^2$ bins, which are indicated in the panels.
The curves with our predictions are drawn for the central value of each $Q^2$ bin.
Also shown are the data points from different experiments selected in the given $Q^2$ bins.
Figure~\ref{fig:F2} covers the region up to $W^2 = 150\gevsq$ and $0.3<Q^2<16\gevsq$,
while Fig.~\ref{fig:F2RES} focuses on the resonance region
$W^2 < 4.2\gevsq$ and $0.25<Q^2<7\gevsq$.

For comparison, together with our results in Fig.~\ref{fig:F2} and \ref{fig:F2RES}
we also show the predictions from Ref.~\cite{Christy:2007ve} (the dashed curve labeled ``CB").
We observe that our predictions are consistent with the CB fit for $W^2<8\gevsq$ and $Q^2<10\gevsq$.
The inspection of $\chi^2$ values in Table~\ref{tab:f2_chi2} suggests that
our approach provides better overall description of data in the resonance region for almost all data sets
except for the JLab-E94-110 experiment.
The CB model fails for $W^2>8\gevsq$.
In contrast, our predictions can be applied in a wide range of kinematics
since at high values of $Q^2$ and $W^2$ our model merges the DIS description from
a global QCD analysis~\cite{Alekhin:2007fh}.
For completeness, in Figs.~\ref{fig:F2} and \ref{fig:F2RES} we also show the background contribution
which clearly dominates for $W^2>3.5\gevsq$ and merges with the DIS structure function at higher values of $W^2$.


%
\begin{figure*}[p]
\includegraphics[width=1.0\linewidth]{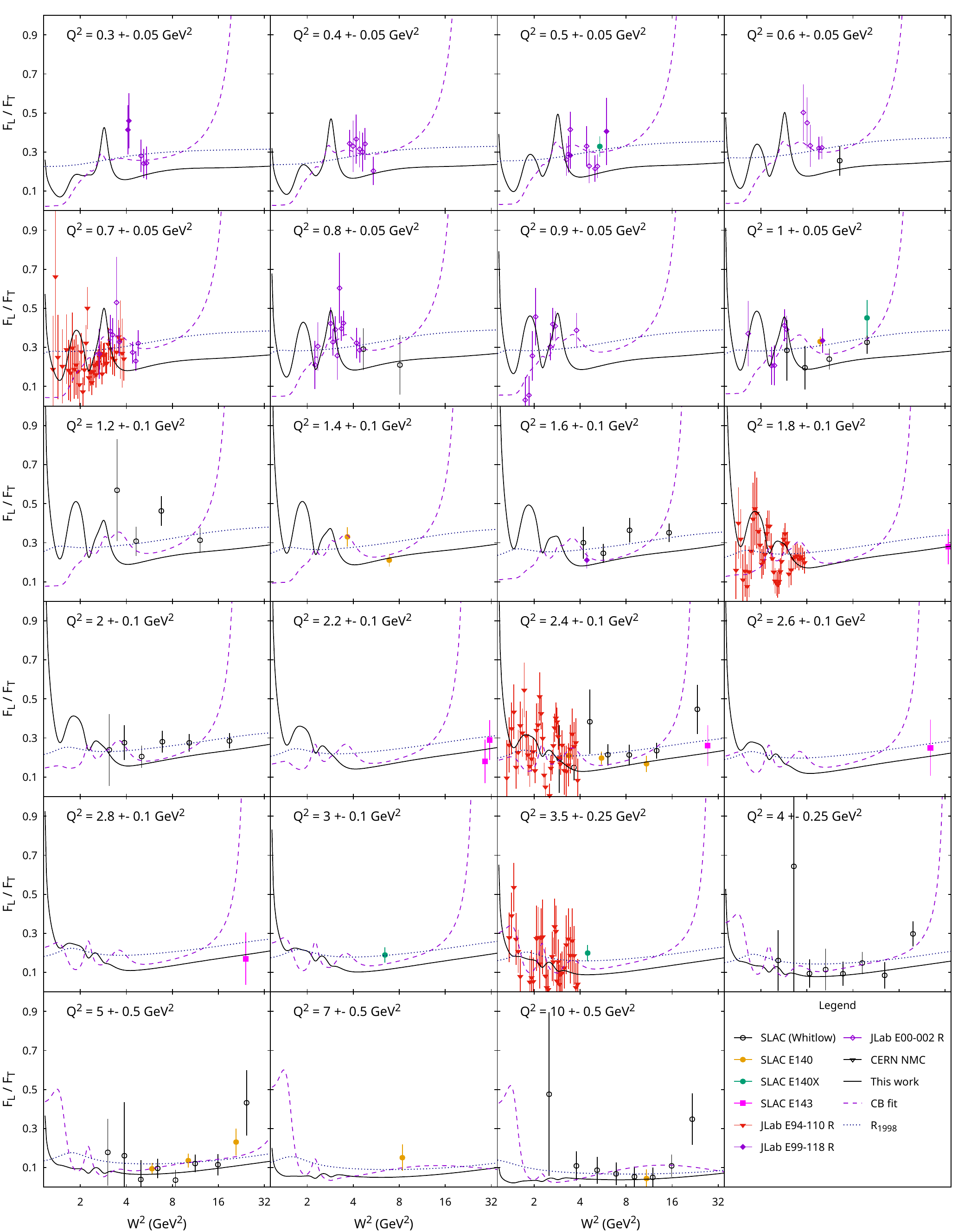}
\caption{\label{fig:R}
Our predictions (solid line) on the ratio $R=F_T$/$F_L$ vs $W^2$
in comparison with data for a number of $Q^2$ bins indicated in the panels.
The curves are drawn for the central value of each $Q^2$ bin.
Also shown are the prediction from Ref.~\cite{Christy:2007ve} (CB, dashed)
and $R_{1998}$ fit~\cite{Abe:1998ym} (dotted).
The legend is given in the lower-right panel.}
\end{figure*}

\begingroup
\begin{table*}[htb!]
\vspace{-5ex}
\caption{\label{tab:f2_chi2}
The list of $F_2$ and cross-section data sets not used in our fit.
Also shown the corresponding number of data points (NDP) and kinematics coverage.
The values of $Q^2$ and $W^2$ are in $\gevsq$ units.
The ``DIS'' label indicates data which are mostly in the DIS region
while ``RES'' labels the data samples which are mostly in the resonance region.
The last two columns are the values of $\chi^2$ normalized per NDP computed,
respectively, in our model and using parameterization of Ref.~\cite{Christy:2007ve}
(``N/A" has the same meaning as in Table~\ref{tab:csdata}).}
\begin{ruledtabular}
\begin{tabular*}{\linewidth}{l@{\extracolsep{\fill}}rcccccc}
$F_2$ data set                                               & NDP   & $Q^2_\text{min}$ &  $Q^2_\text{max}$  & $W^2_\text{min}$  & $W^2_\text{max}$ & $\chi^2$  & $\chi_\text{CB}^2$ \\
\hline
SLAC-E49a~\cite{Whitlow:1990dr} (DIS) &  117  & 0.586   &  8.067  &   3.131   &  27.24 &  0.50   & N/A \\
SLAC-E49b~\cite{Whitlow:1990dr} (DIS) &  208  & 0.663   &  20.08  &   3.007   &  27.51 &  0.72   & N/A \\
SLAC-E61 ~\cite{Whitlow:1990dr} (DIS) &   32  & 0.581   &  1.738  &   3.213   &  16.05 &  0.34   & N/A \\
SLAC-E87 ~\cite{Whitlow:1990dr} (DIS) &  109  & 3.959   &  20.41  &   3.287   &  17.16 &  0.59   & N/A \\
SLAC-E89a~\cite{Whitlow:1990dr} (DIS) &   77  & 3.645   &  30.31  &   3.303   &  20.46 &  1.01   & N/A \\
SLAC-E89b~\cite{Whitlow:1990dr} (DIS) &  118  & 0.887   &  19.18  &   3.099   &  27.78 &  0.52   & N/A \\
JLab-CLAS~\cite{Osipenko:2003ua} (RES)                & 4191  & 0.225   &  4.725  &   1.162   &  5.804 &  1.17   & 6.13\\
JLab-E94-110~\cite{Liang:thesis} (RES)                &  170  & 0.181   &  4.794  &   1.325   &  3.850 &  1.90   & 0.72\\
JLab-E00-116~\cite{Malace:thesis} (RES)               &  261  & 3.585   &  7.384  &   1.243   &  5.132 &  1.42   & 1.94\\
JLab-E00-002~\cite{Tvaskis:2016uxm} (RES)             &   54  & 0.300   &  1.000  &   1.650   &  5.419 &  0.24   & 0.05\\
CERN-NMC~\cite{Arneodo:1996qe} (DIS)                  &  157  & 0.750   &  65.00  &   6.380   &  553.9 &  1.74   & N/A \\
CERN-BCDMS~\cite{Benvenuti:1989rh} (DIS)              &  177  & 7.500   &  230.0  &   8.042   &  351.0 &  2.31   & N/A \\
DESY-HERMES~\cite{Airapetian:2011nu} (DIS)            &  80   & 0.354   &  12.78  &   5.751   &  42.17 &  0.28   & N/A \\
\hline
Cross-section data set                                               & NDP   & $Q^2_\text{min}$ &  $Q^2_\text{max}$  & $W^2_\text{min}$  & $W^2_\text{max}$ & $\chi^2$  & $\chi_\text{CB}^2$ \\
\hline
CERN-NMC~\cite{Arneodo:1996qe} (DIS)                  &  292  & 0.750   &  65.00  &   6.380   &  553.9 &  1.38   & N/A \\
CERN-BCDMS~\cite{Benvenuti:1989rh} (DIS)              &  351  & 7.500   &  230.0  &   8.042   &  351.0 &  1.15   & N/A \\
DESY-HERMES~\cite{Airapetian:2011nu} (DIS)            &  81   & 0.354   &  12.78  &   5.751   &  42.17 &  0.45   & N/A \\
\end{tabular*}
\end{ruledtabular}
\end{table*}
\endgroup
\newcommand{\gwidth}{0.46\linewidth}
\begin{figure*}[!]
\vspace{-2ex}
\centering
\includegraphics[width=\gwidth]{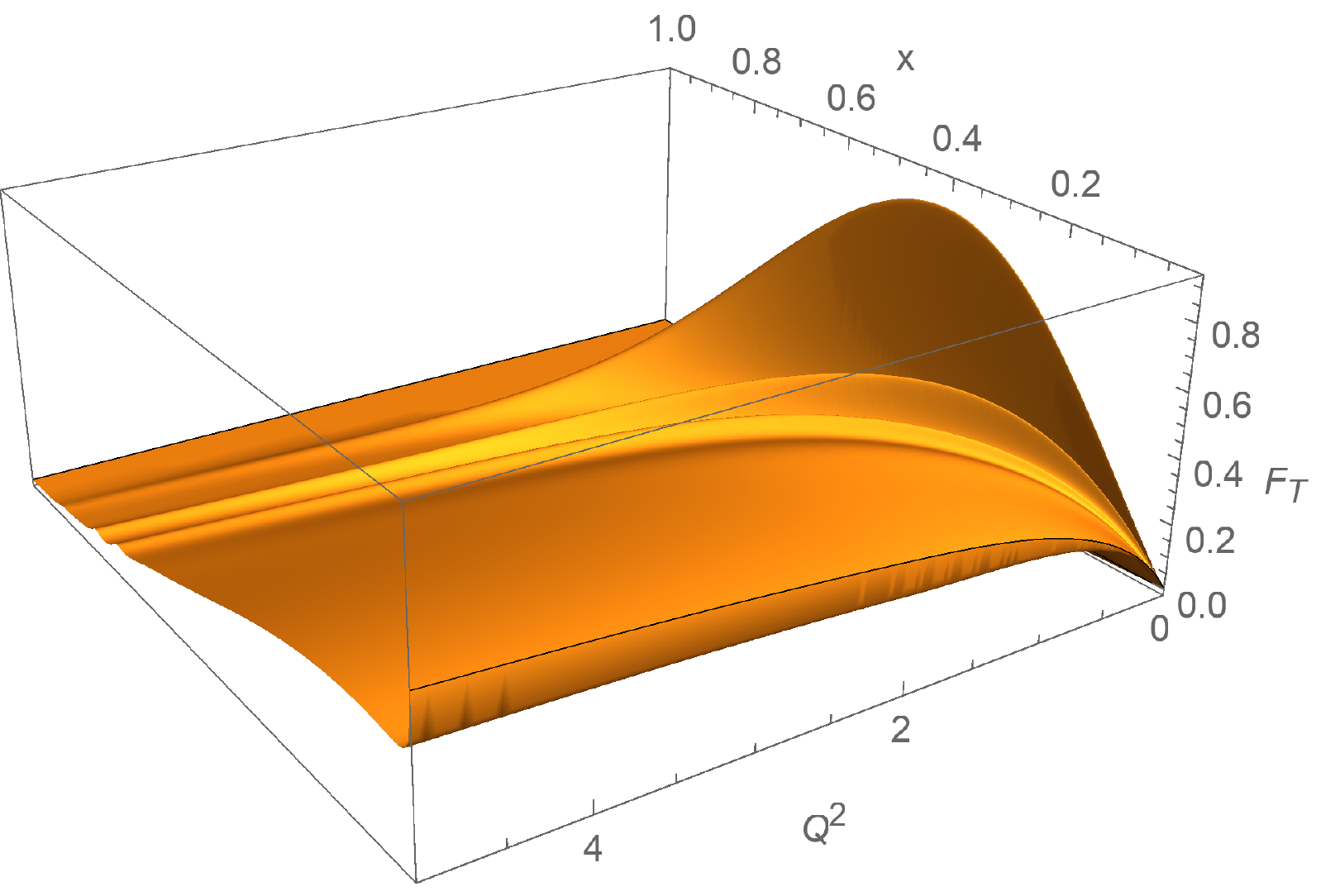}\quad\includegraphics[width=\gwidth]{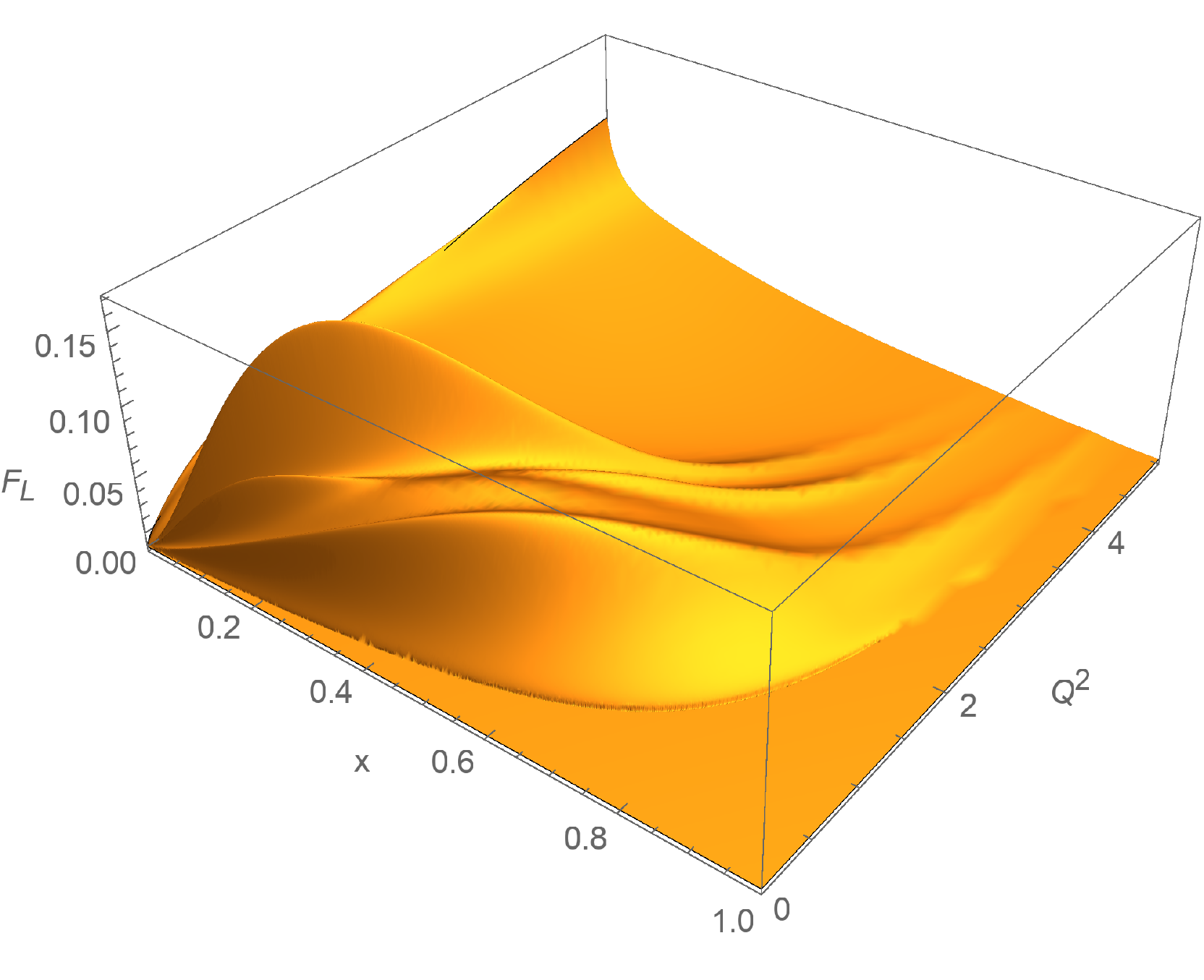}\\
\includegraphics[width=\gwidth]{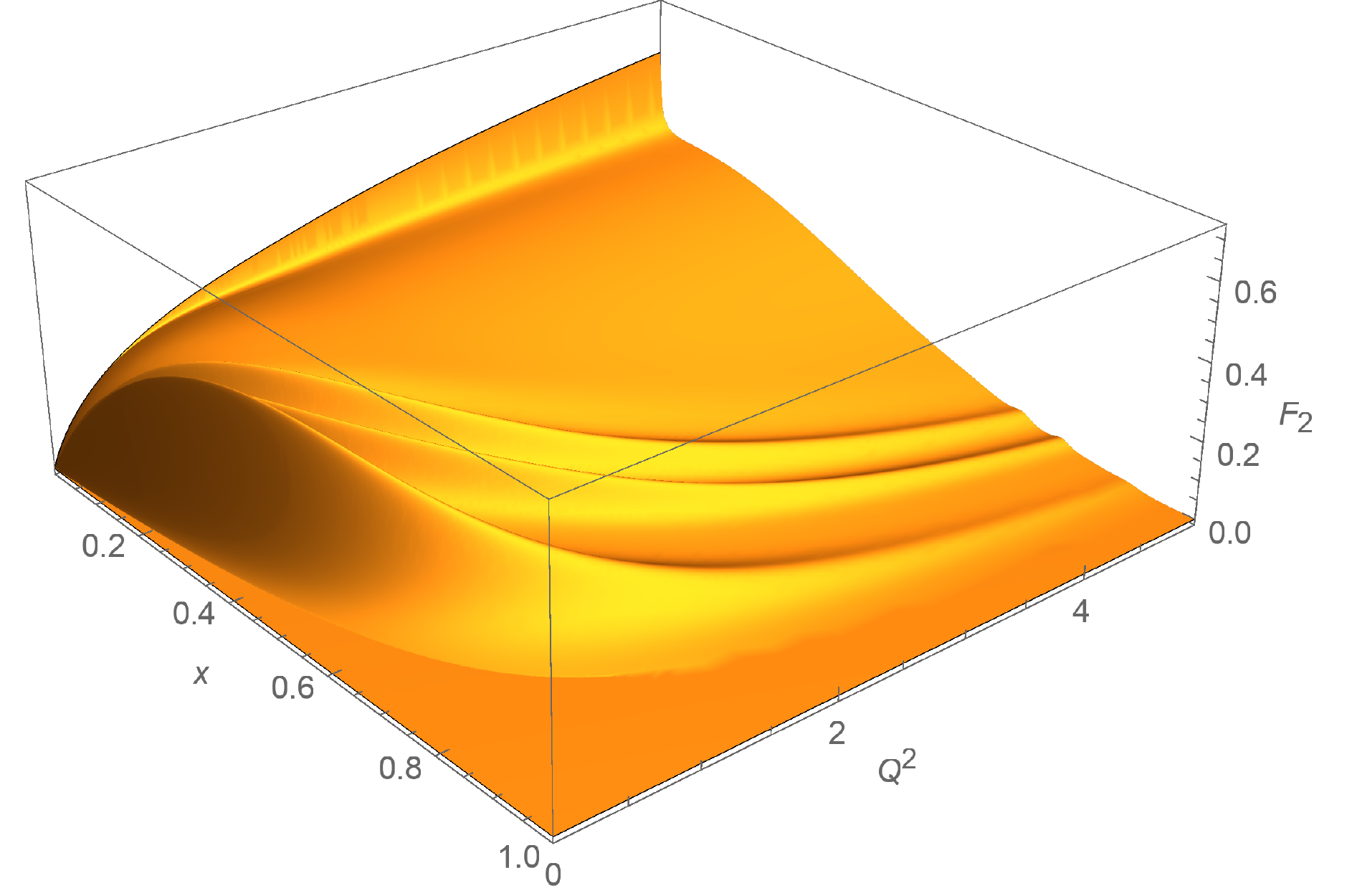}\quad\includegraphics[width=\gwidth]{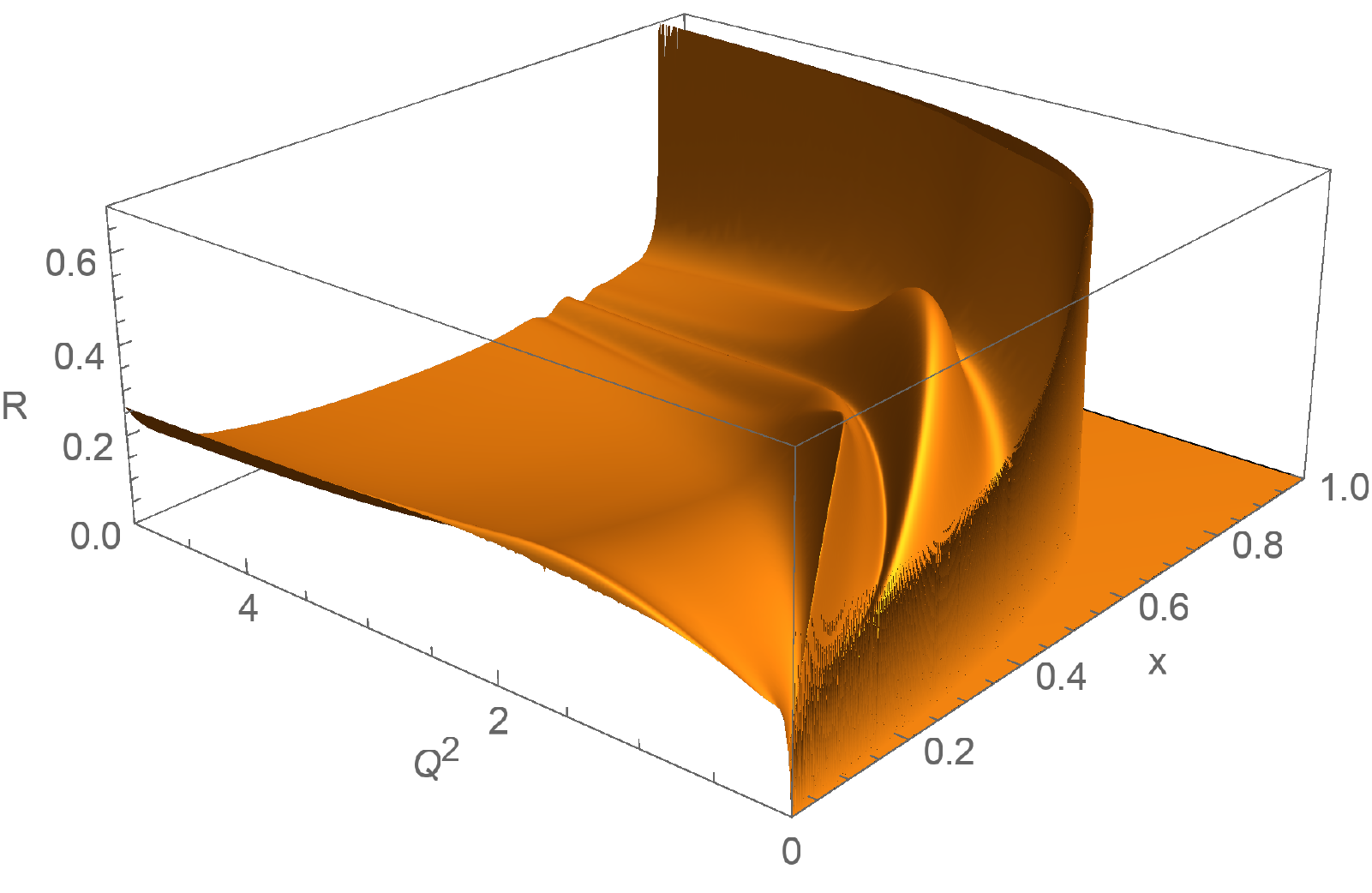}
\caption{\label{fig:sfsurface}
The structure functions $F_T$, $F_L$, $F_2$, and $R$ vs $(x,Q^2)$.}
\end{figure*}

Figure~\ref{fig:R} shows the measurements of
$R=F_L/F_T$~\cite{Whitlow:1990dr,Dasu:1993vk,Tao:1995uh,Abe:1998ym,Liang:2004tj,Tvaskis:2016uxm}
as a function of $W^2$ together with our predictions.
The plots in Fig.~\ref{fig:R} are organized in terms of panels of $Q^2$ bins covering the region from $0.25$ to $10\gevsq$.
The data points marked with the symbol ``R'' in the legend correspond to the measurements based on the Rosenbluth separation method.
Our predictions are indicated by the solid curve and for comparison we also show
the results obtained with the parametrization $R_{1998}$~\cite{Abe:1998ym}
and the CB model~\cite{Christy:2007ve}.
Note also that while our predictions are for the proton,
the data on $R$ in Fig.~\ref{fig:R} are collected for different nuclear targets
including ${}^2$H, ${}^{56}$Fe, and ${}^{197}$Au (SLAC-E140~\cite{Dasu:1993vk}),
${}^2$H and ${}^9$Be (SLAC-E140X~\cite{Tao:1995uh}), and
${}^{12}$C (SLAC-E143~\cite{Abe:1998ym}).

It should be remarked that while our model provides $R=0$ at $Q^2=0$,
the transition to this limit
occurs at a very low scale $Q^2\ll 0.1\gevsq$,
because the value of parameter $m_L$,
which drives a low-$Q$ asymptotic of background part of $F_L$, is close to 1.
If we release $m_L$, the fit prefers $m_L\to 1$ (although with significant uncertainty).
In order to have vanishing $R$ at the real photon point,
in the final fit trial we fixed the value $m_L=1.1$.

Note also an oscillating behavior of $R$ vs. $W$ in the resonance region
with $Q^2$-dependent amplitudes, as can be seen in Fig.~\ref{fig:R}.
Overall, our predictions for $R$ are in a reasonable agreement with available measurements,
although the data uncertainties are rather large.

In Fig.~\ref{fig:sfsurface} we illustrate the $x$ and $Q^2$ dependence of SFs in our model
in the region $0.001<x<1$ and $0.01<Q^2<5\gevsq$ by plotting the surfaces of $F_T$, $F_L$, $F_2$ and $R$.
We show SFs at different surface view points in order to better view
various features of resonance structures.
We clearly see that for $Q^2<2\gevsq$ the resonance structures drive SF strength
at low values of $W$
across the $x$ region while their impact rapidly decreases with $Q^2$.
The details of resonance contributions differ for various SFs.
In this context, we note a sharp $\Delta(1232)$ resonance structure on the $F_T$ surface and
a pronounced peak from the third effective resonance on the $F_L$ surface.
Also pronounced resonance structures are present in the ratio $R=F_L/F_T$.
Note that a sharp wall-like structure in $R$ at large values of $x$ 
is not because of a resonance contribution but is owed to a different $W$ dependence
of background contributions to $F_T$ and $F_L$ in a region close to the inelastic threshold.

In conclusion, we examine the duality property of our model.
The quark-hadron duality principle suggests an integral relation between the observed structure function,
which includes the resonance contributions,
and a smooth DIS structure function~\cite{Bloom:1970xb}.
We then verify the following integral relation:
\begin{equation}\label{eq:f2:duality}
 \int_{W^2_\text{th}}^{W_0^2} \ud W^2 F_2(W^2, Q^2) =
 \int_{W^2_\text{th}}^{W_0^2} \ud W^2 F_2^\text{DIS}(W^2, Q^2),
\end{equation}
where on the left we use SF of the present model and
on the right the DIS SF by \eq{eq:sf:dis} from Ref.~\cite{Alekhin:2007fh}.
The integration is taken from the pion production threshold $W^2_\text{th}=(M+m_\pi)^2$
to the boundary of the resonance region for which we take $W_0^2 = 4\gevsq$.
We found that this relation holds with rather high accuracy,
with the relative error ranging between 1 and 2\% for $1.3<Q^2<5\gevsq$.
We also studied  \eq{eq:f2:duality} for  $F_T$ and $F_L$.
We found somewhat lower accuracy of the duality relation for $F_T$,
with the relative error up to 5\% for $1<Q^2<3\gevsq$.
For higher $Q^2$ the accuracy of the duality relation for $F_T$ is on the level of that for $F_2$.
The accuracy of \eq{eq:f2:duality}  is poor for $F_L$ for $1<Q^2<3\gevsq$, where the difference between the left and right sides of \eq{eq:f2:duality} is between 15 and 25\%.
Nevertheless \eq{eq:f2:duality} for $F_L$ gradually becomes more accurate at higher $Q^2$.

\section{\label{sec:sum}Summary}

To summarize, in this paper we develop a hybrid model of the proton structure functions
applicable in a wide region of $Q^2$ and $W^2$.
In the  nucleon resonance region, $W<2\gev$, we account for contributions from
the $\Delta(1232)$ resonance, the $N(1440)$ Roper resonance, and
three more heavy effective resonances responsible for
the second and third resonance regions in the spectra.
Nonresonant background is computed in terms of DIS structure functions properly continued into a low-$Q$ and low-$W$ region.
Our extrapolation method respects the pion production threshold as well as
the $Q^2\to0$ real photon limit.
The onset of a low-$Q$ region is defined by the parameter $Q_0$,
the scale from which we start extrapolations of DIS SFs.
The value $Q_0^2=2\gevsq$ provides an optimum description of electroproduction data in our analysis,
as discussed in Sec.~\ref{sec:fit}.
The DIS region of $Q>Q_0$ and $W>2\gev$ is well described in terms of the proton PDFs
and the higher-twist terms from a global QCD analysis~\cite{Alekhin:2007fh,Alekhin:2008ua}.

The model parameters, such as resonance masses and widths,
parameters of resonance helicity amplitudes,
scale parameter for transition region,
as well as parameters responsible for extrapolation to low-$Q$ and low-$W$ values,
are adjusted from a global fit to the world data on hydrogen
electroproduction and photoproduction cross section.
This approach allows us to determine parameters of both
the transverse  and the longitudinal SFs
reproducing available cross-section data with a very good accuracy,
as illustrated in detail by the data/model ratios in  Figs.~\ref{fig:pulls_CS_DIS} to \ref{fig:pulls_CS_CLAS}.
For a detailed comparison of our model with cross-section data see the Supplemental Material~\cite{supplement}.

We verify the model performance by comparing our predictions with available measurements of the proton $F_2$ and the ratio $R=F_L/F_T$.
Figures~\ref{fig:pulls_F2_DIS} to \ref{fig:F2RES} and Table~\ref{tab:f2_chi2} show a very good overall agreement of our predictions with $F_2$ measurements.
Figure~\ref{fig:R} illustrates the model performance against available measurements of $R$.

Also, we verified that our hybrid model of structure functions is dual in the integral sense
to the underlying DIS structure functions. The duality relation, \eq{eq:f2:duality}, holds with a good accuracy for $F_2$.

Work is in progress on extending this approach to determine parameters of
the neutron structure functions from a combined set of the proton and nuclear data.
Also in progress is the generalization of this model to neutrino-nucleon scattering in the resonance and DIS transition region,
which is of primary importance for interpretation of data from current and future neutrino experiments.

Code to numerically compute the model structure functions $F_T$, $F_L$ and $F_2$ is available from the authors upon email request.

\onecolumngrid
\section{\label{sec:ackn}Acknowledgments}

We thank S.~Alekhin, A.~Kataev, and R.~Petti, for useful discussions,
and M.~Osipenko for providing the data of the JLab-CLAS Collaboration.
V.V.B. was supported by the BASIS Foundation for the Development of Theoretical Physics and Mathematics.
\twocolumngrid

\bibliography{literatura}

\newpage
\newcommand{\xsecaption}[2]{Model predictions (solid line) in comparison with {#1}
data on differential cross section $\ud^2\sigma/(\ud\Omega\ud E')$ in {#2}/(sr GeV) vs $W^2$ in GeV$^2$.
The values of the beam energy (GeV) and scattering angle (degrees) are given in the figure panels.}
\newcommand{\xsecapclas}{Model predictions (solid lines) in comparison with JLab-CLAS data on differential cross section
$\ud^2\sigma/(\ud\Omega\ud E')$ in nanobarn/(sr GeV) vs $W^2$ in GeV$^2$.
The values of $Q^2$ are given in the figure panels.
The data sets E1, E2, E3, E4, and E5 correspond to the beam energies 1.515, 2.567, 4.056, 4.247, and 4.462~GeV, respectively.}
\newcommand{\similarto}[1]{Similar to Fig.~\ref{#1}.}
\renewcommand{\gwidth}{1.0\linewidth}


\onecolumngrid
\renewcommand\thesection{\Alph{section}}
\section*{Supplemental Material\label{sec:supl}}


In the figures below we provide a detailed comparison of our model predictions with the hydrogen photoproduction
and electroproduction cross section data.
%
Figure~\ref{fig:photot} shows the total photoproduction cross section vs $W^2$.
The data points are from the experiments listed in Table~\ref{tab:photodata} and the legend is given in the figure.
The solid line corresponds to our predictions with the best fit parameters (see Sec.~\ref{sec:fit}),
while the dotted line shows the background part of the cross section.
%
In Fig.~\ref{fig:CS_E49a:1} to \ref{fig:CS_00116} we show our results (solid line) in comparison with
electroproduction cross section data from experiments listed in Table~\ref{tab:csdata}.
The figures are organized in terms of panels of given beam energy and scattering angle whose values are shown in the panels.
The data points for the resonance region ($W^2<4\gevsq$) and DIS ($W^2>4\gevsq$) are marked with different symbols/color.
The error bars in the plots are the quadrature sum of statistical and systematic errors of corresponding experiment,
the normalization uncertainties of data are not shown.

\setcounter{figure}{11}
\begin{figure*}[htb]
\centering
\includegraphics[width=\gwidth]{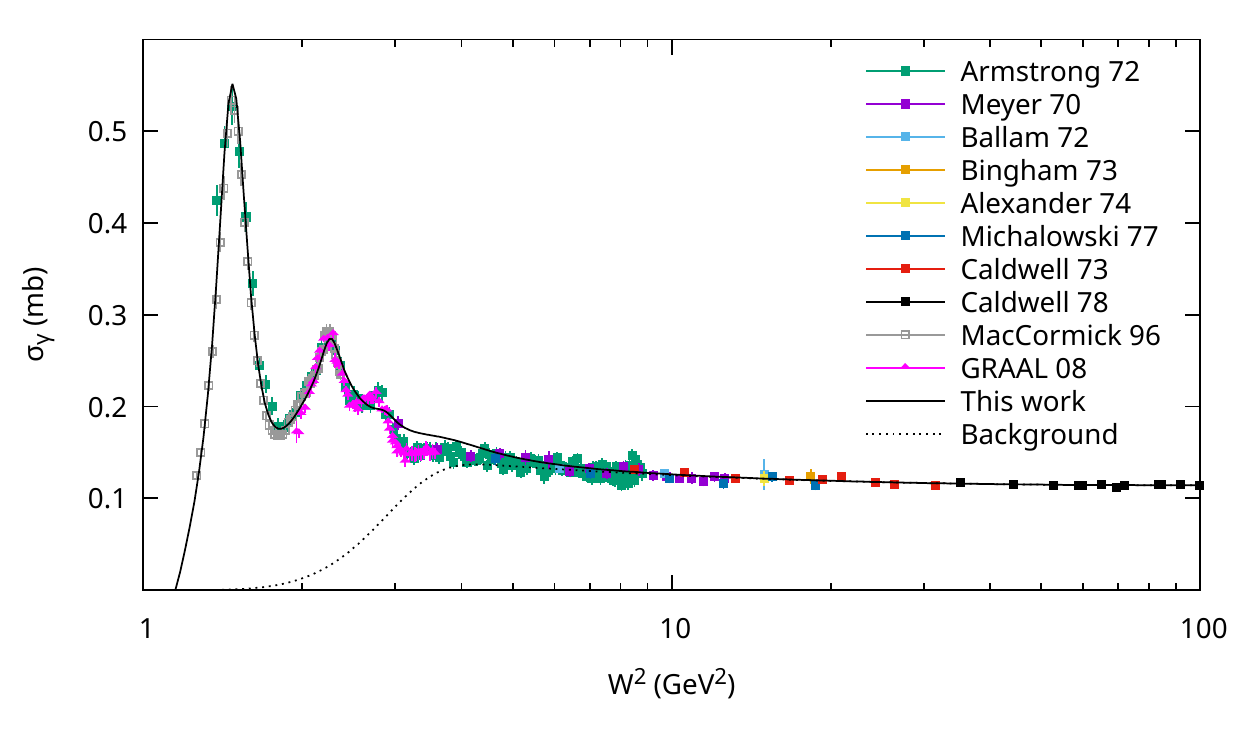}
\caption{\label{fig:photot}Model predictions (solid line) in comparison with data on the total photoproduction cross section off hydrogen vs $W^2$.
The legend of various photoproduction data is given in the figure.}
\end{figure*}

\begin{figure*}[p]
\includegraphics[width=\gwidth]{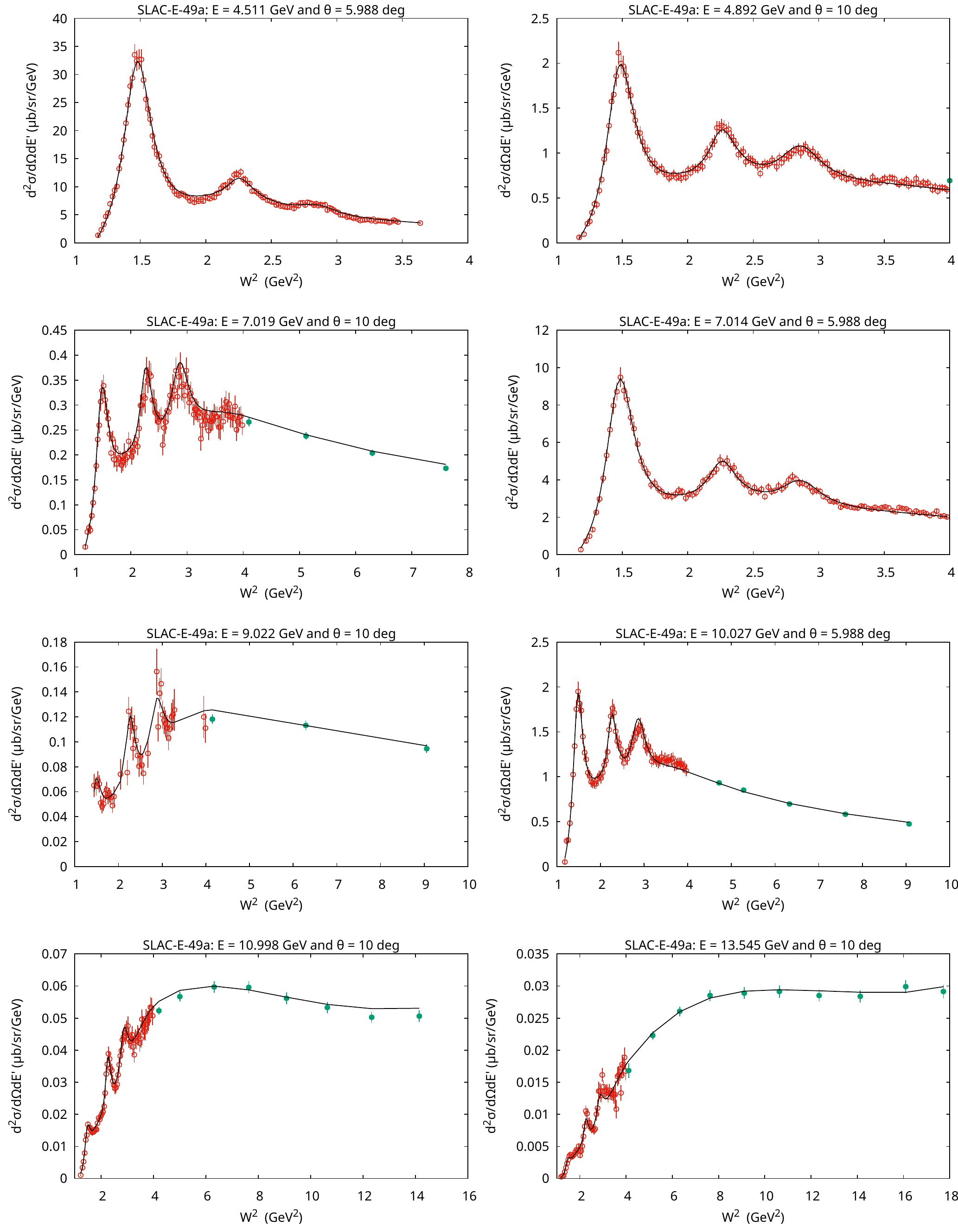}
\caption{\xsecaption{SLAC-E-49a}{$\mu$b}\protect\label{fig:CS_E49a:1}}
\end{figure*}
\begin{figure*}[p]
\includegraphics[width=\gwidth]{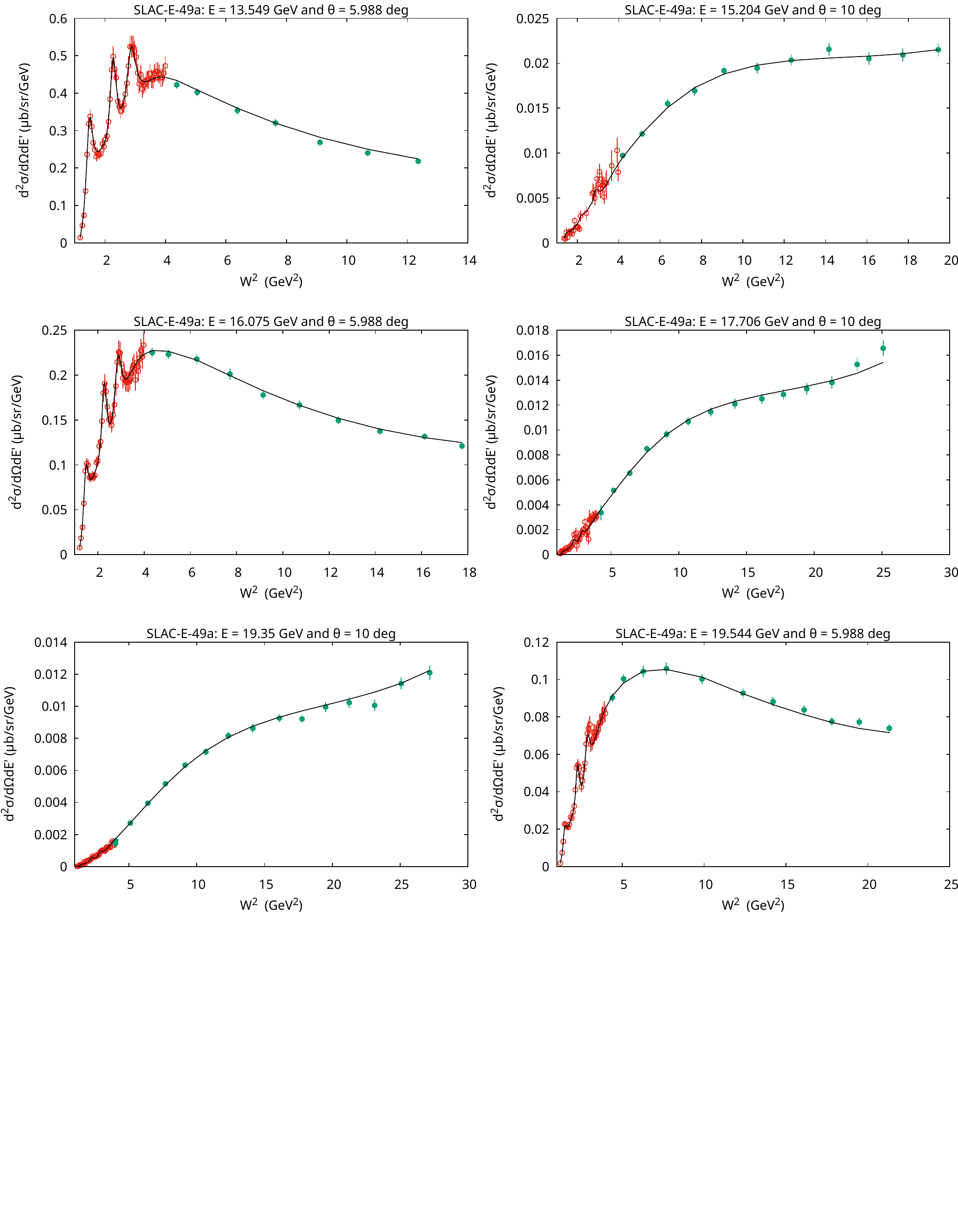}
\caption{Similar to Fig.~\ref{fig:CS_E49a:1}.\label{fig:CS_E49a:2}}
\end{figure*}

\begin{figure*}[p]
\includegraphics[width=\gwidth]{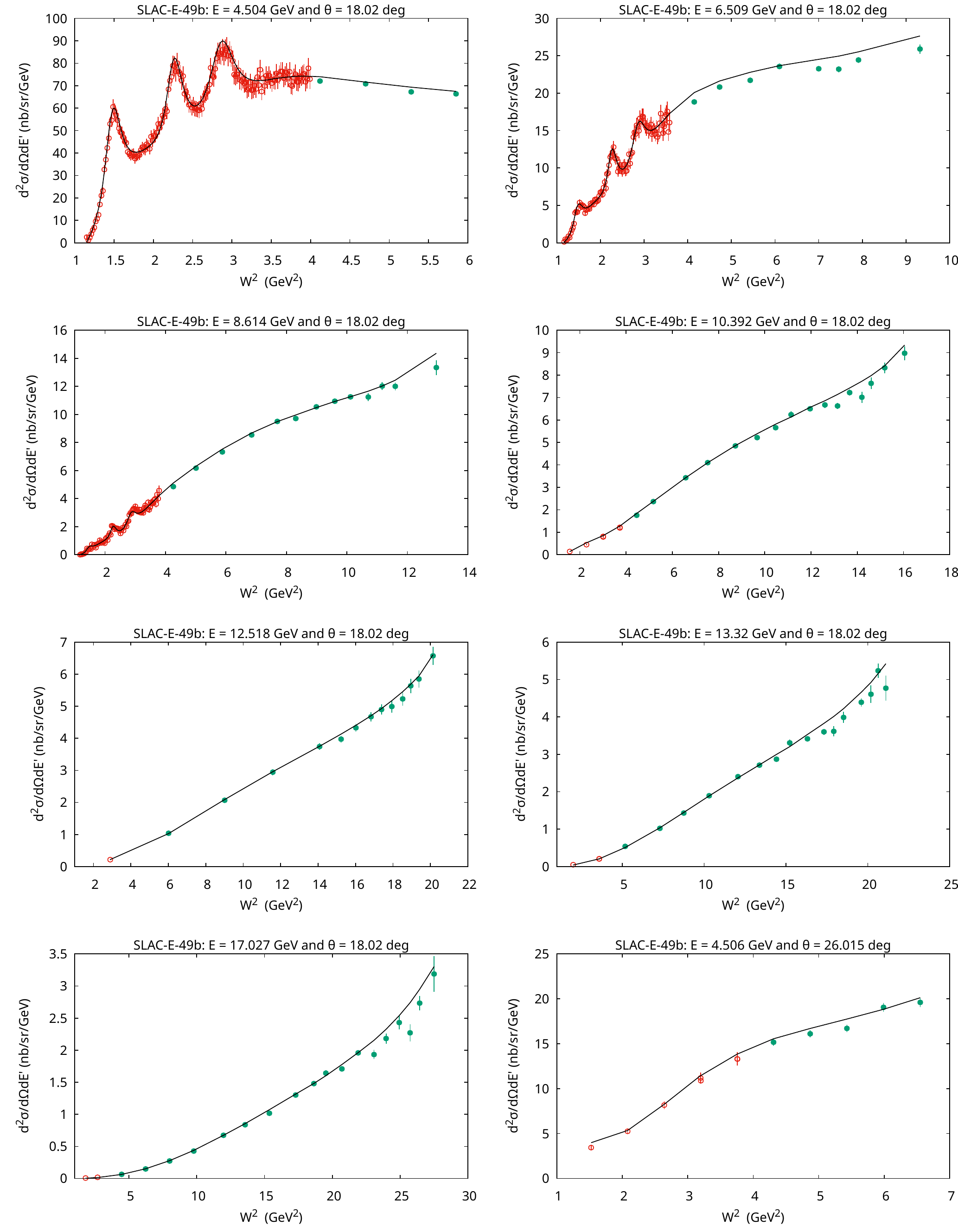}
\caption{\xsecaption{SLAC-E-49b}{nanobarn}}
\label{fig:CS_E49b:1}
\end{figure*}
\begin{figure*}[p]
\includegraphics[width=\gwidth]{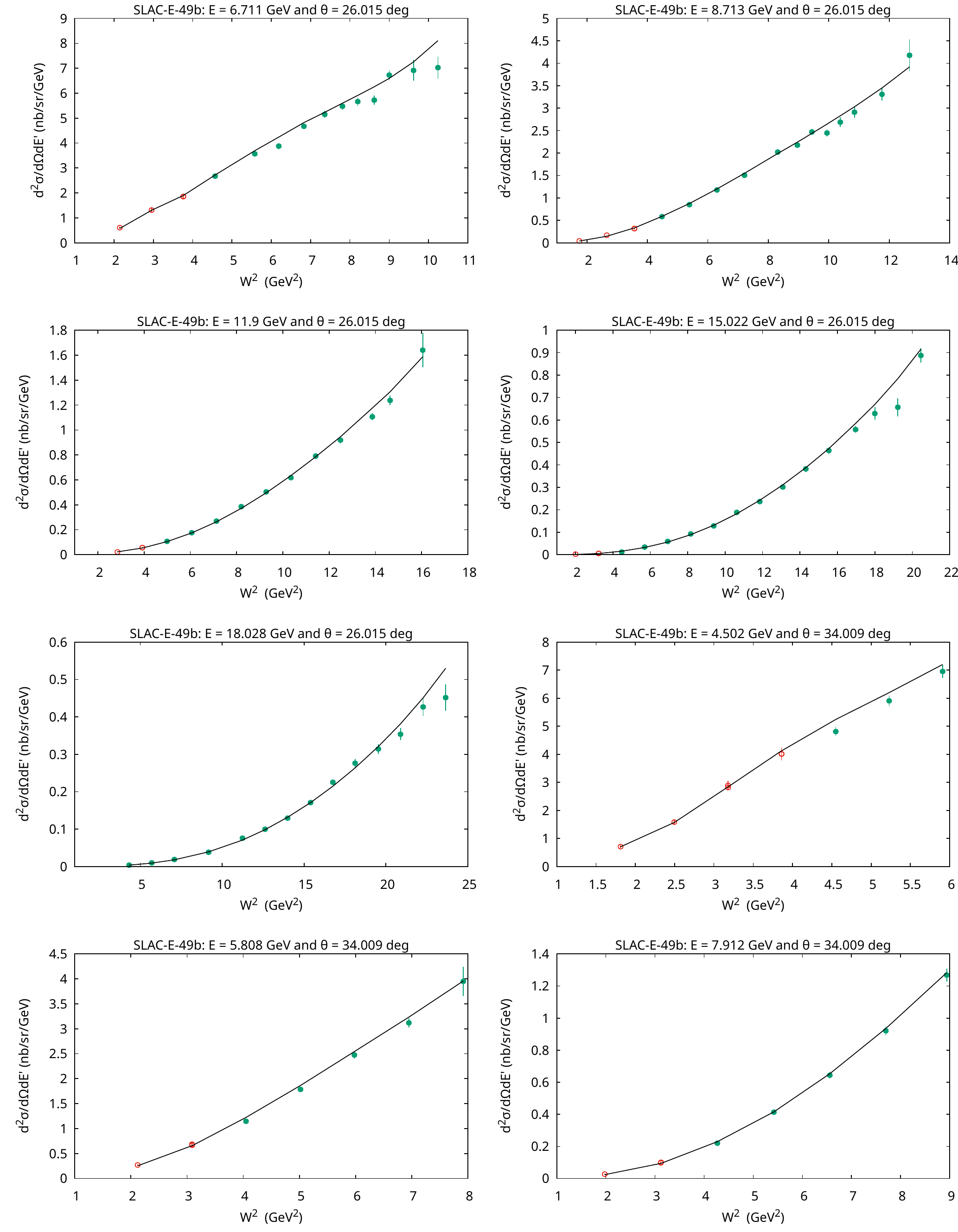}
\caption{Similar to Fig.~\ref{fig:CS_E49b:1}.\label{fig:CS_E49b:2}}
\end{figure*}

\begin{figure*}[p]
\includegraphics[width=\gwidth]{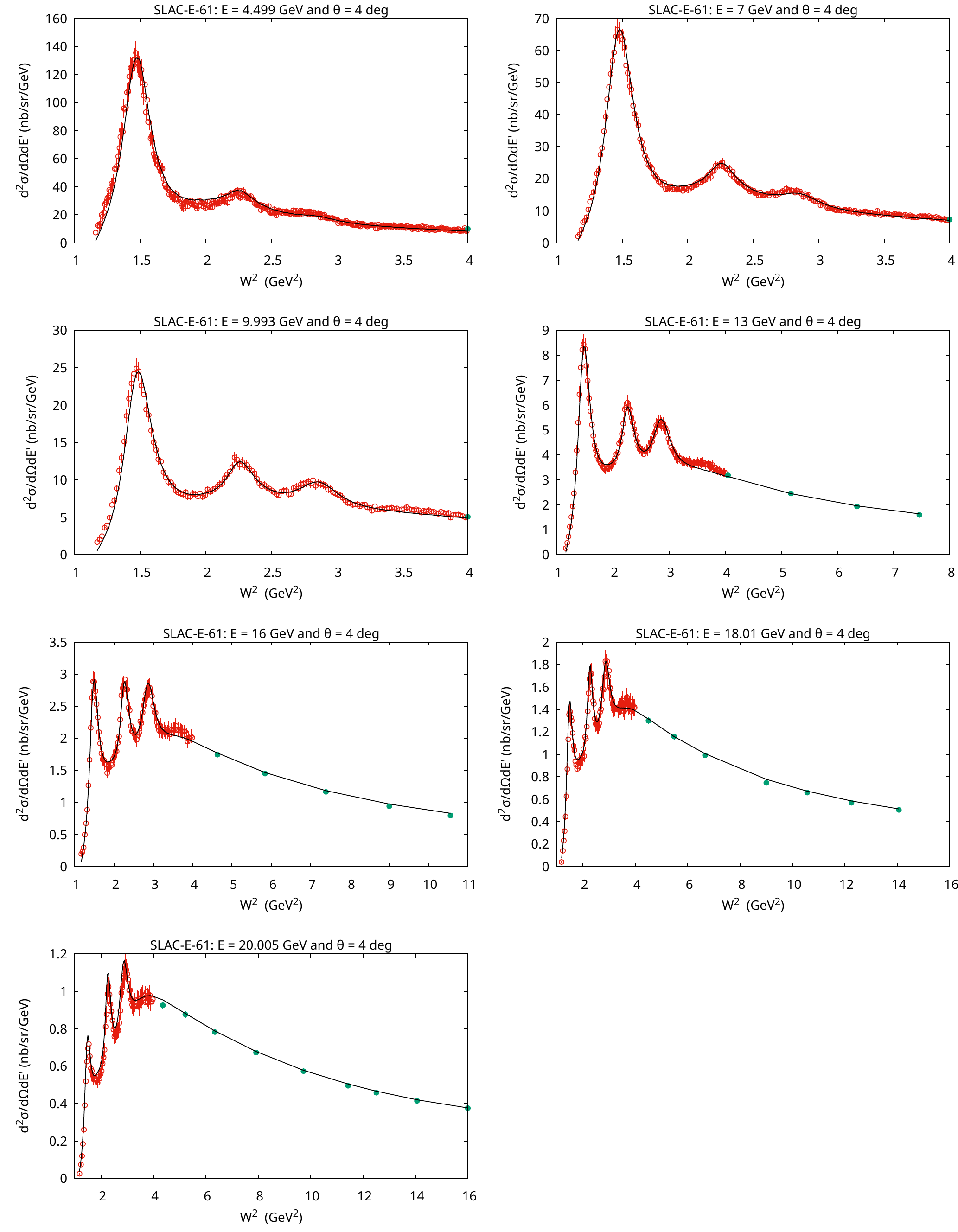}
\caption{\xsecaption{SLAC-E-61}{nanobarn}}
\label{fig:CS_E61}
\end{figure*}

\begin{figure*}[p]
\includegraphics[width=\gwidth]{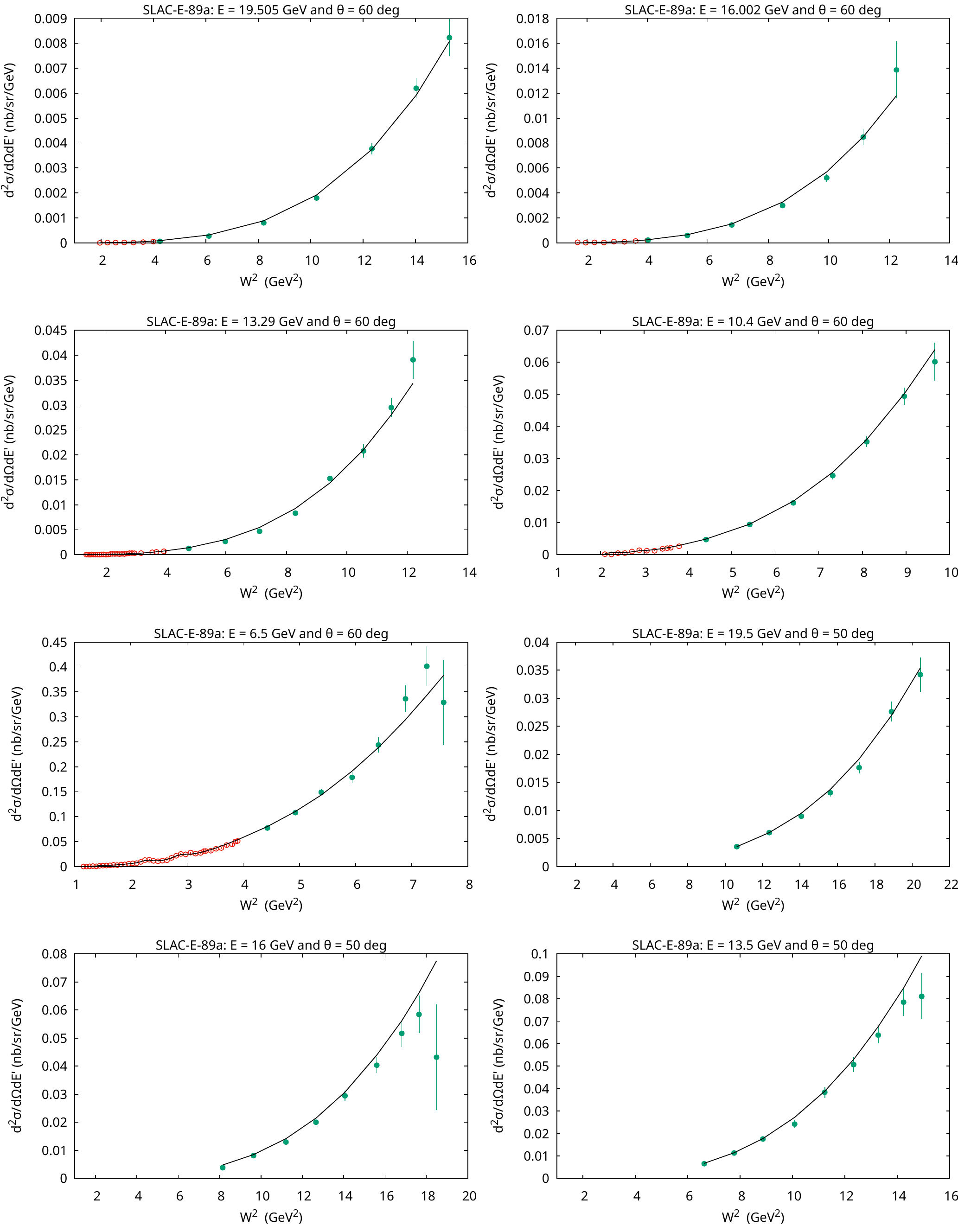}
\caption{\xsecaption{SLAC-E-89a}{nanobarn}}
\label{fig:CS_E89a}
\end{figure*}

\begin{figure*}[p]
\includegraphics[width=\gwidth]{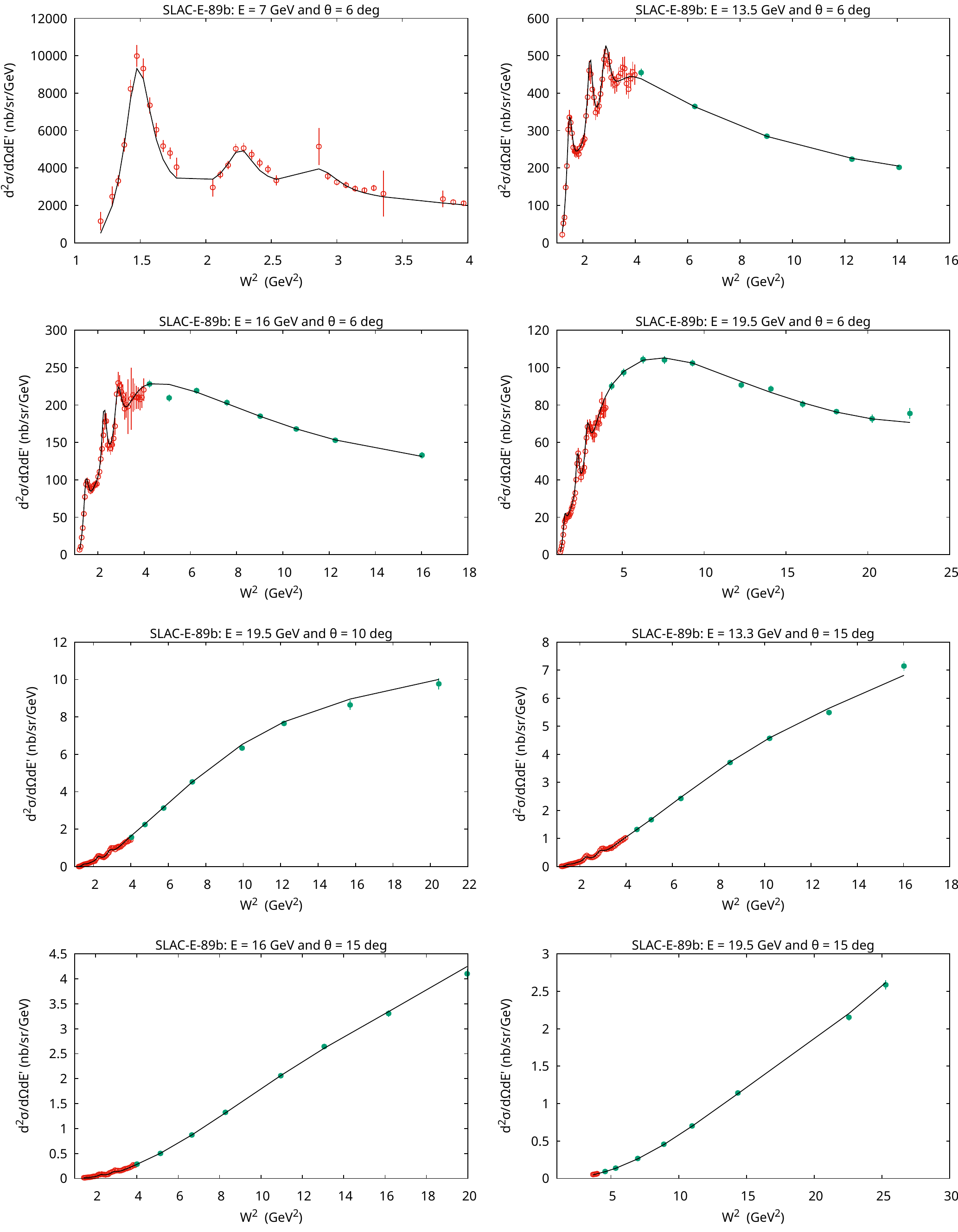}
\caption{\xsecaption{SLAC-E-89b}{nanobarn}}
\label{fig:CS_E89b:1}
\end{figure*}
\begin{figure*}[p]
\includegraphics[width=\gwidth]{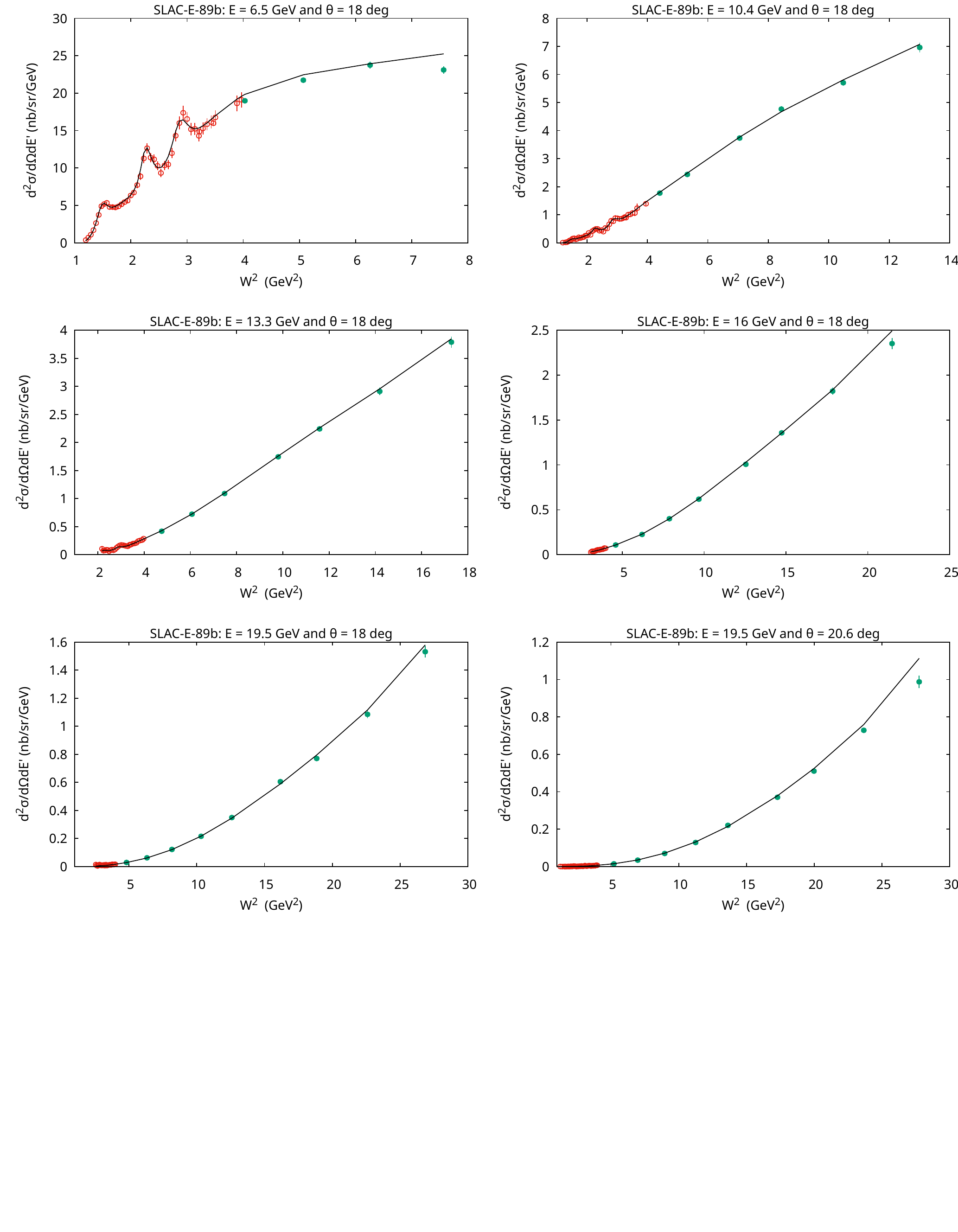}
\caption{Similar to Fig.~\ref{fig:CS_E89b:1}.\label{fig:CS_E89b:2}}
\end{figure*}

\begin{figure*}[p]
\centering
\includegraphics[width=\gwidth]{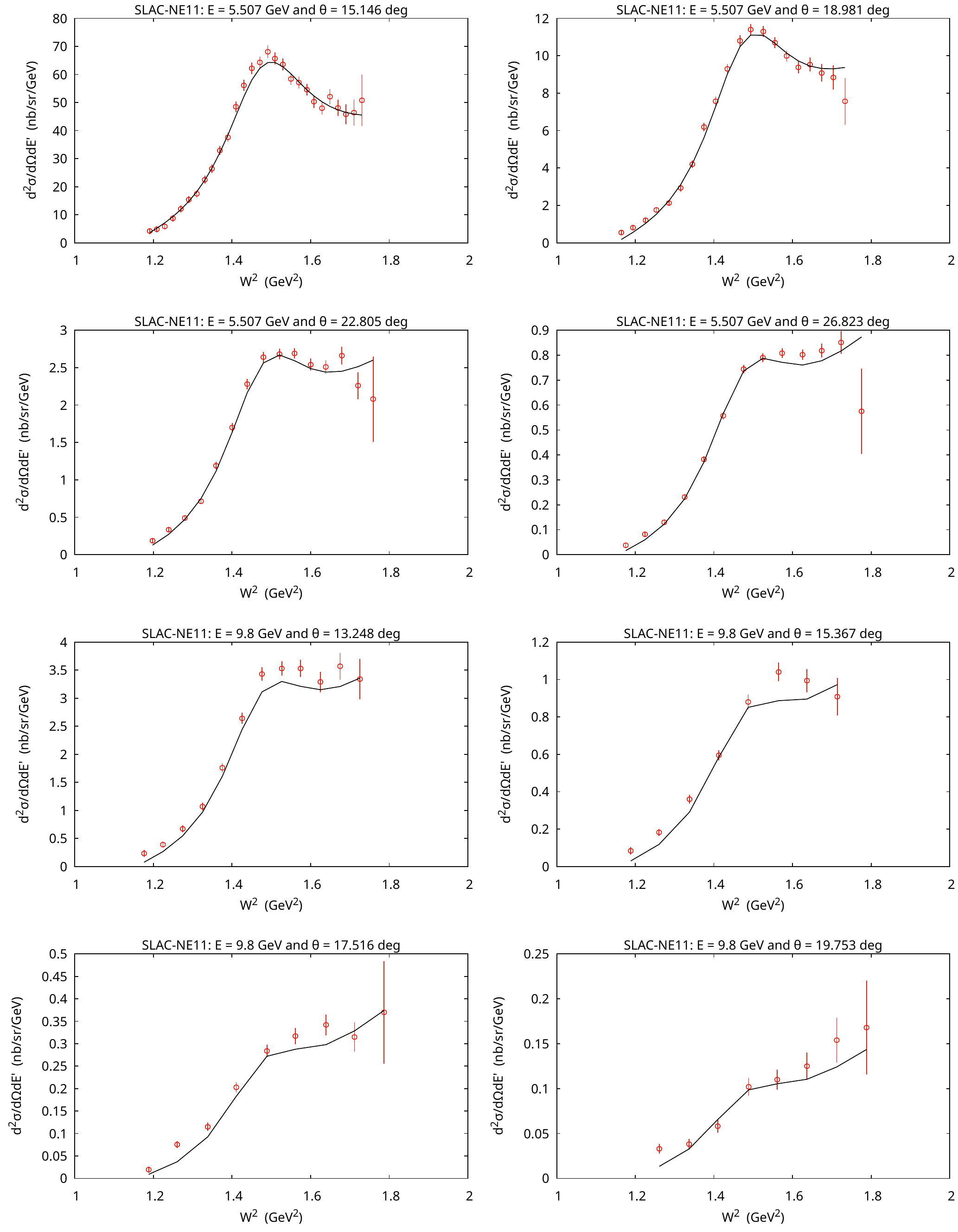}
\caption{\xsecaption{SLAC-NE11}{nanobarn}}
\label{fig:CS_NE11}
\end{figure*}

\begin{figure*}[p]
\centering
\includegraphics[width=\gwidth]{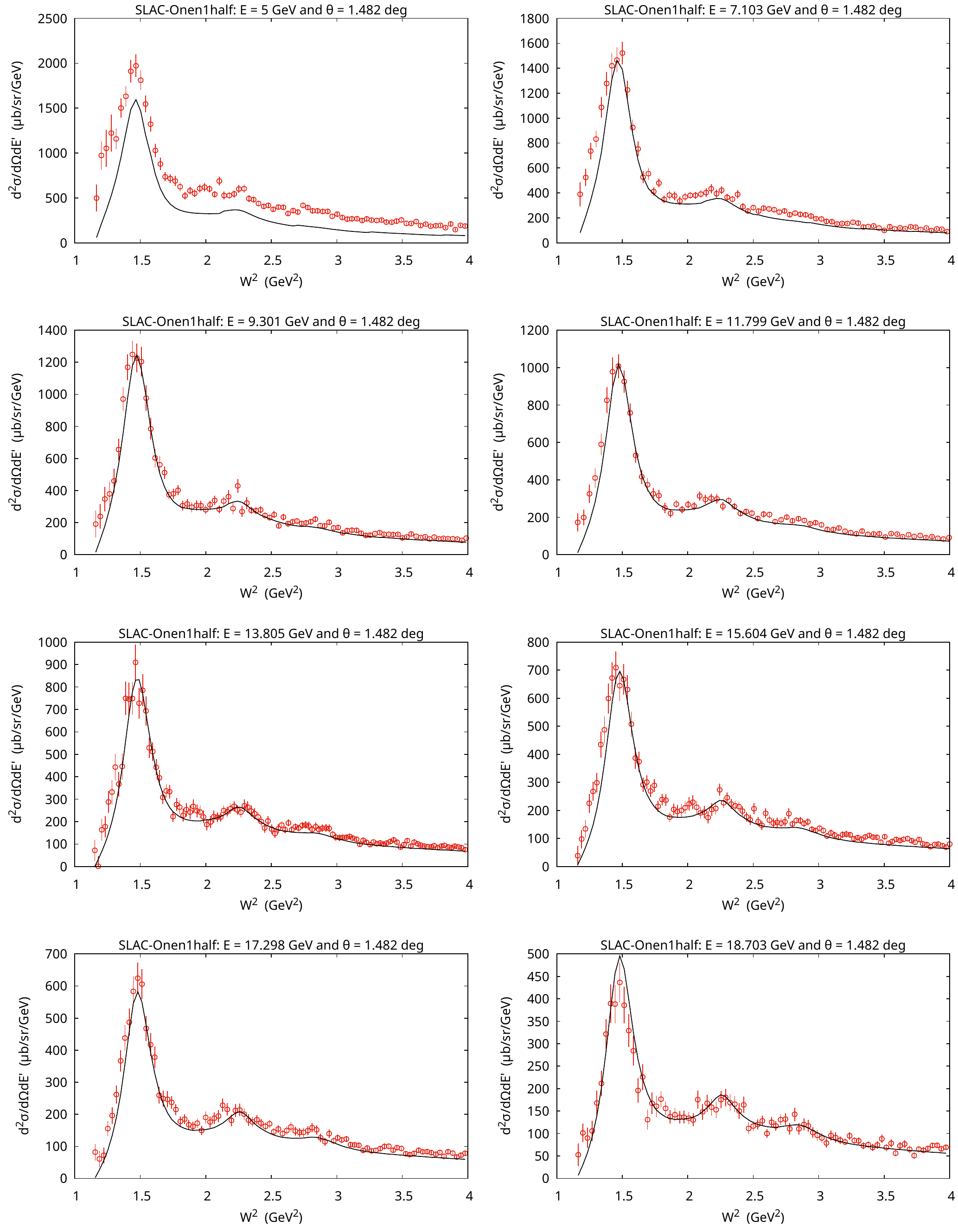}
\caption{\xsecaption{SLAC-Onen1half}{$\mu$b}}
\label{fig:CS_1n1half}
\end{figure*}

\begin{figure*}[p]
\centering
\includegraphics[trim=0 0 0 1cm,width=\gwidth]{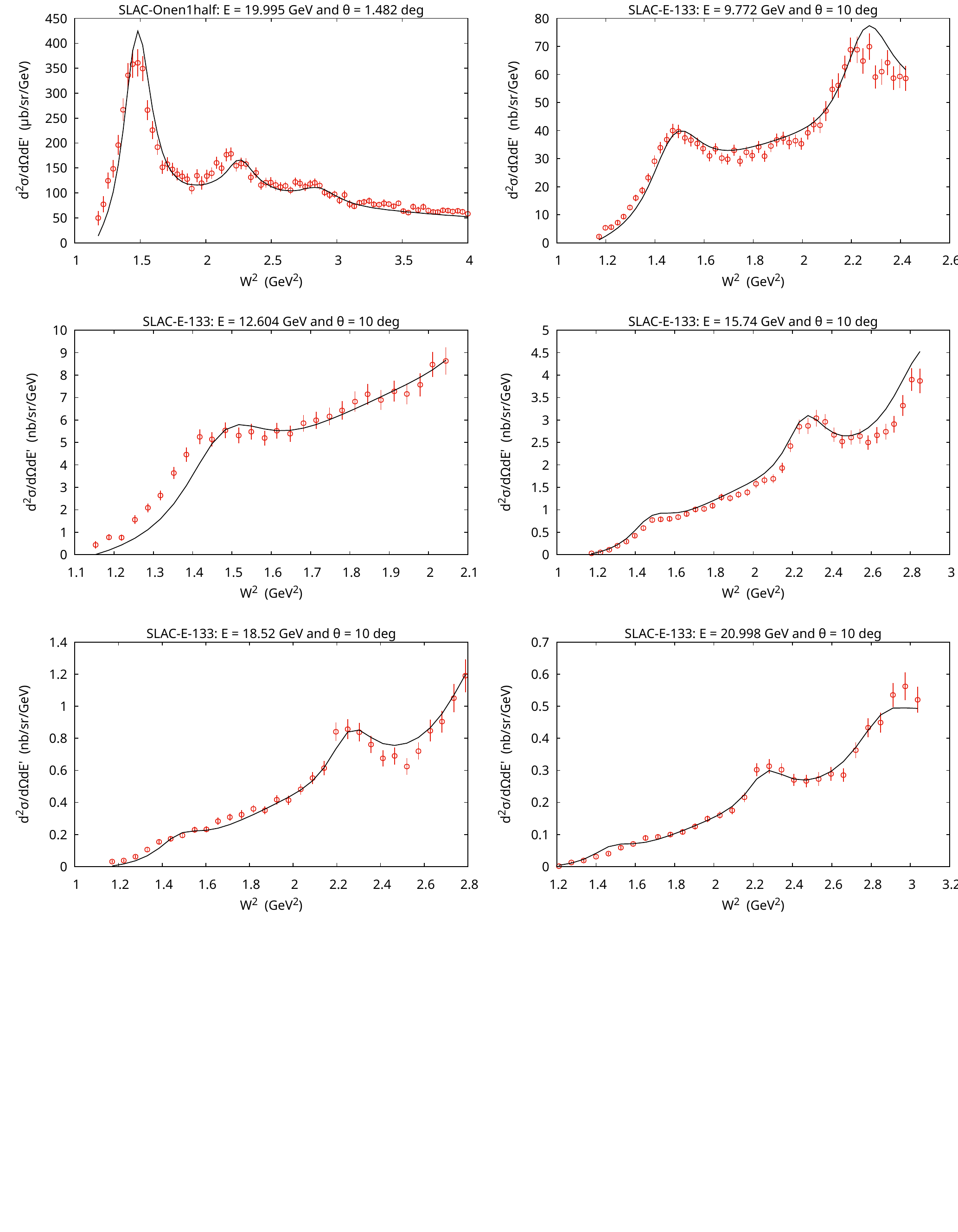}
\caption{\xsecaption{SLAC-E-133}{nanobarn}}
\label{fig:CS_E133}
\end{figure*}

\begin{figure*}[p]
\centering
\includegraphics[trim=0 0 0 1cm,width=\gwidth]{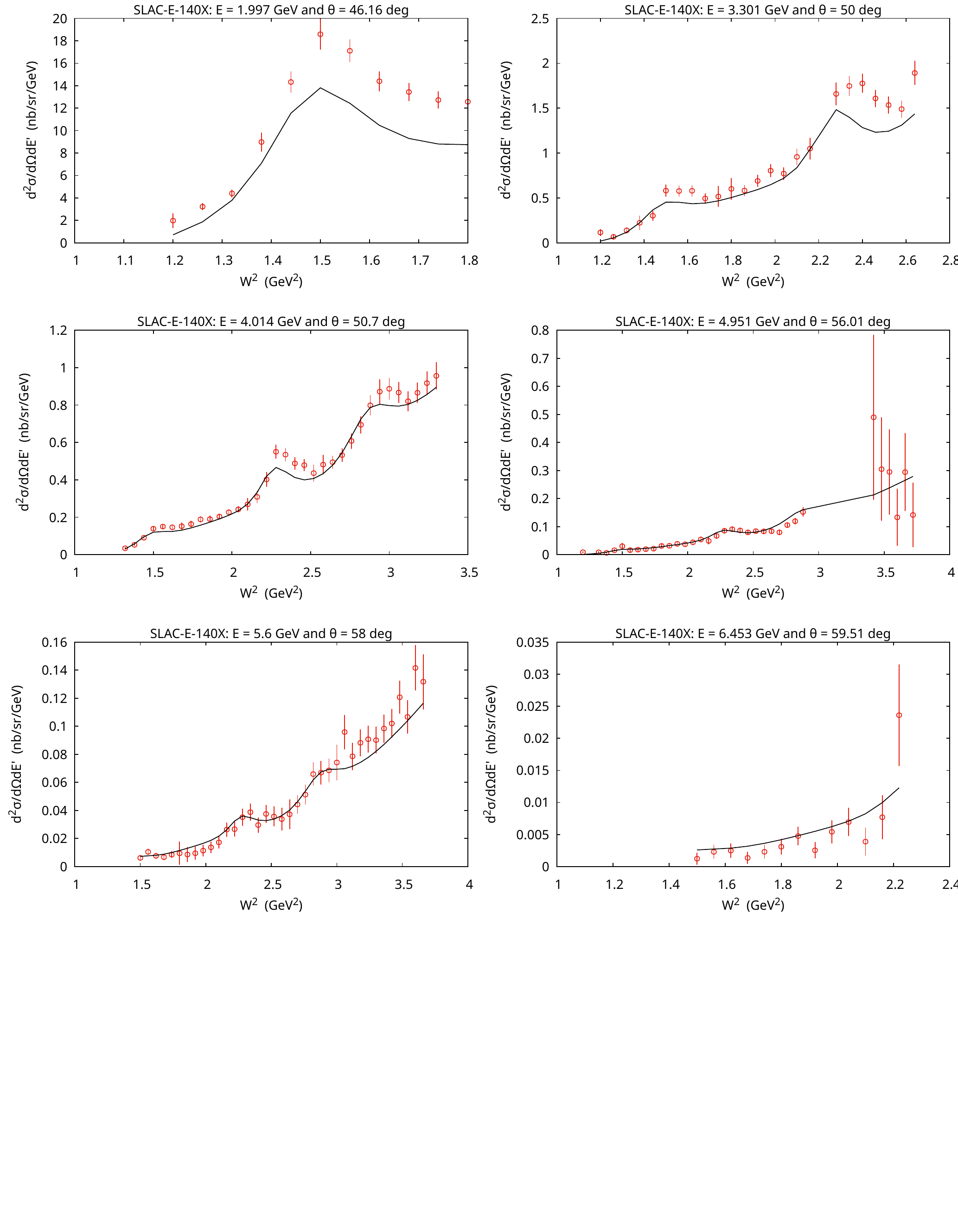}
\caption{\xsecaption{SLAC-E-140X}{nanobarn}}
\label{fig:CS_E140X}
\end{figure*}

\begin{figure*}[p]
\centering
\includegraphics[trim=0 0 0 1cm,width=\gwidth]{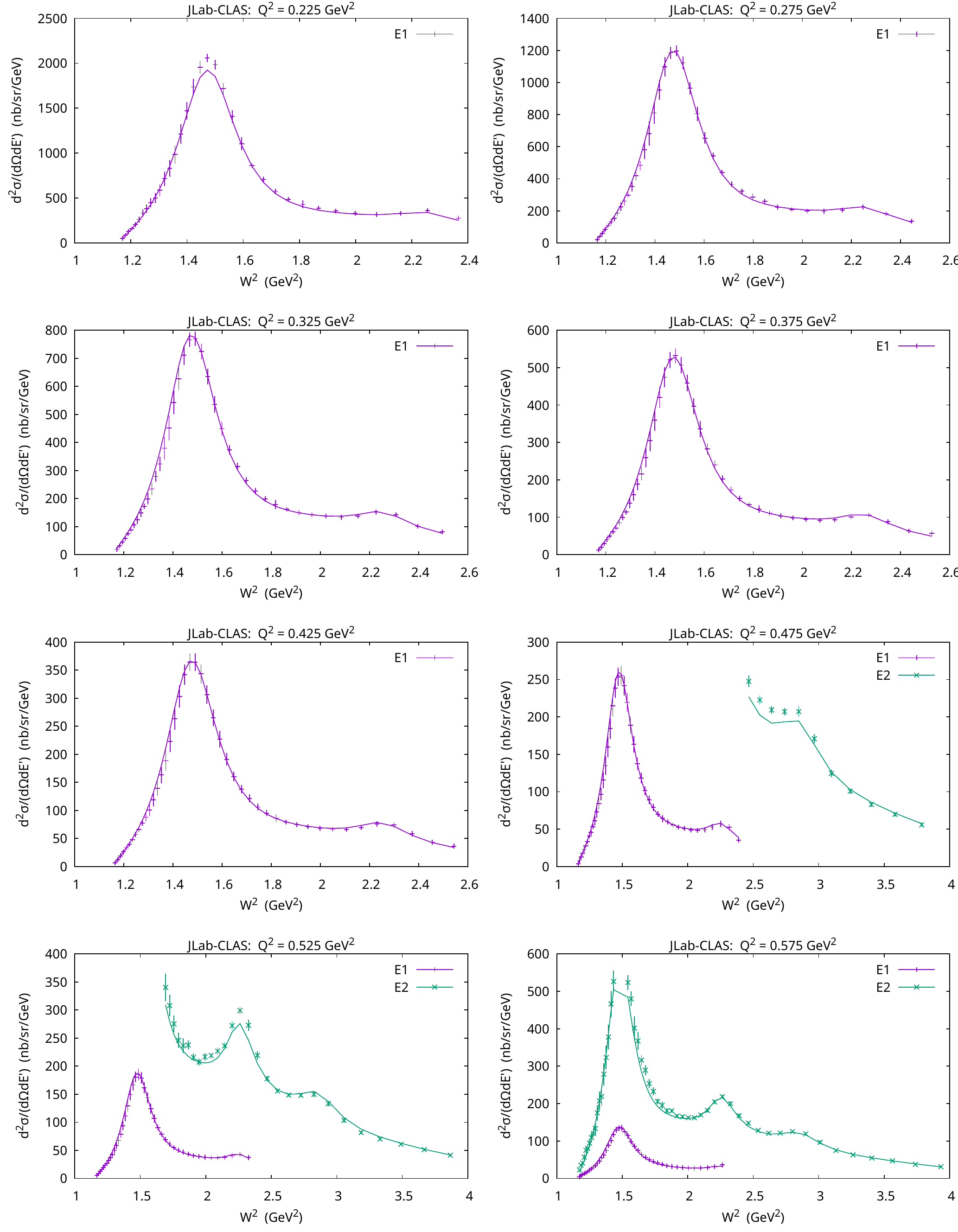}
\caption{\xsecapclas}
\label{fig:CS_CLAS:1}
\end{figure*}
\foreach \x in {2,3,4,5,6,7,8,9,10,11}
{
\begin{figure*}[p]
\includegraphics[width=\gwidth]{xsec-jlab-clas-\x.pdf}
\caption{Similar to Fig.~\ref{fig:CS_CLAS:1}.\label{fig:CS_CLAS:\x}}
\end{figure*}
}

\begin{figure*}[p]
\includegraphics[width=\gwidth]{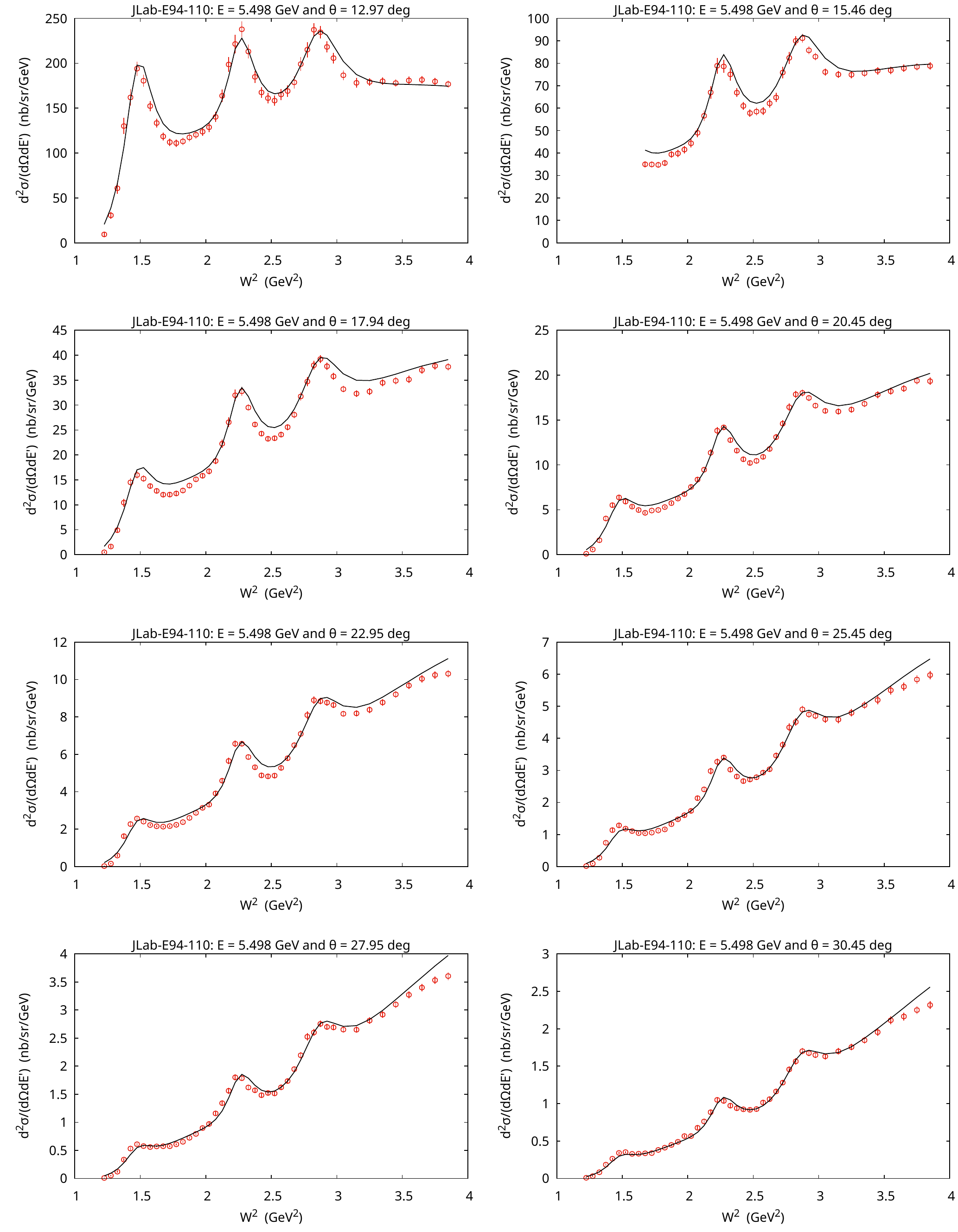}
\caption{\xsecaption{JLab-E94-110}{nanobarn}}
\label{fig:CS_94110:1}
\end{figure*}
\foreach \x in {2,3,4}
{
\begin{figure*}[p]
\includegraphics[width=\gwidth]{xsec-jlab-110-\x.pdf}
\caption{Similar to Fig.~\ref{fig:CS_94110:1}.\label{fig:CS_94110:\x}}
\end{figure*}
}

\begin{figure*}[p]
\centering
\includegraphics[width=\gwidth]{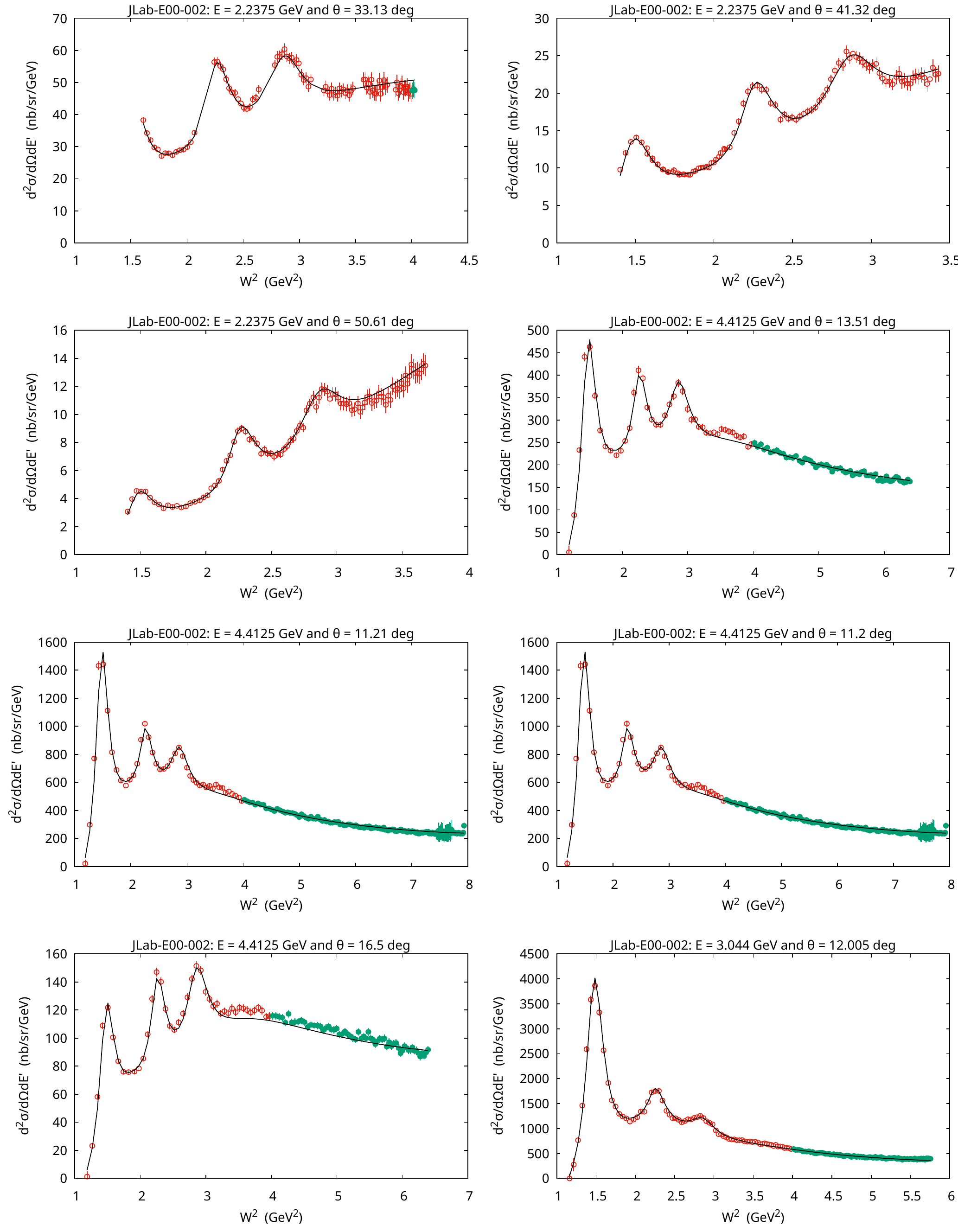}
\caption{\xsecaption{JLab-E00-002}{nanobarn}}
\label{fig:CS_00002:1}
\end{figure*}
\begin{figure*}[p]
\centering
\includegraphics[width=\gwidth]{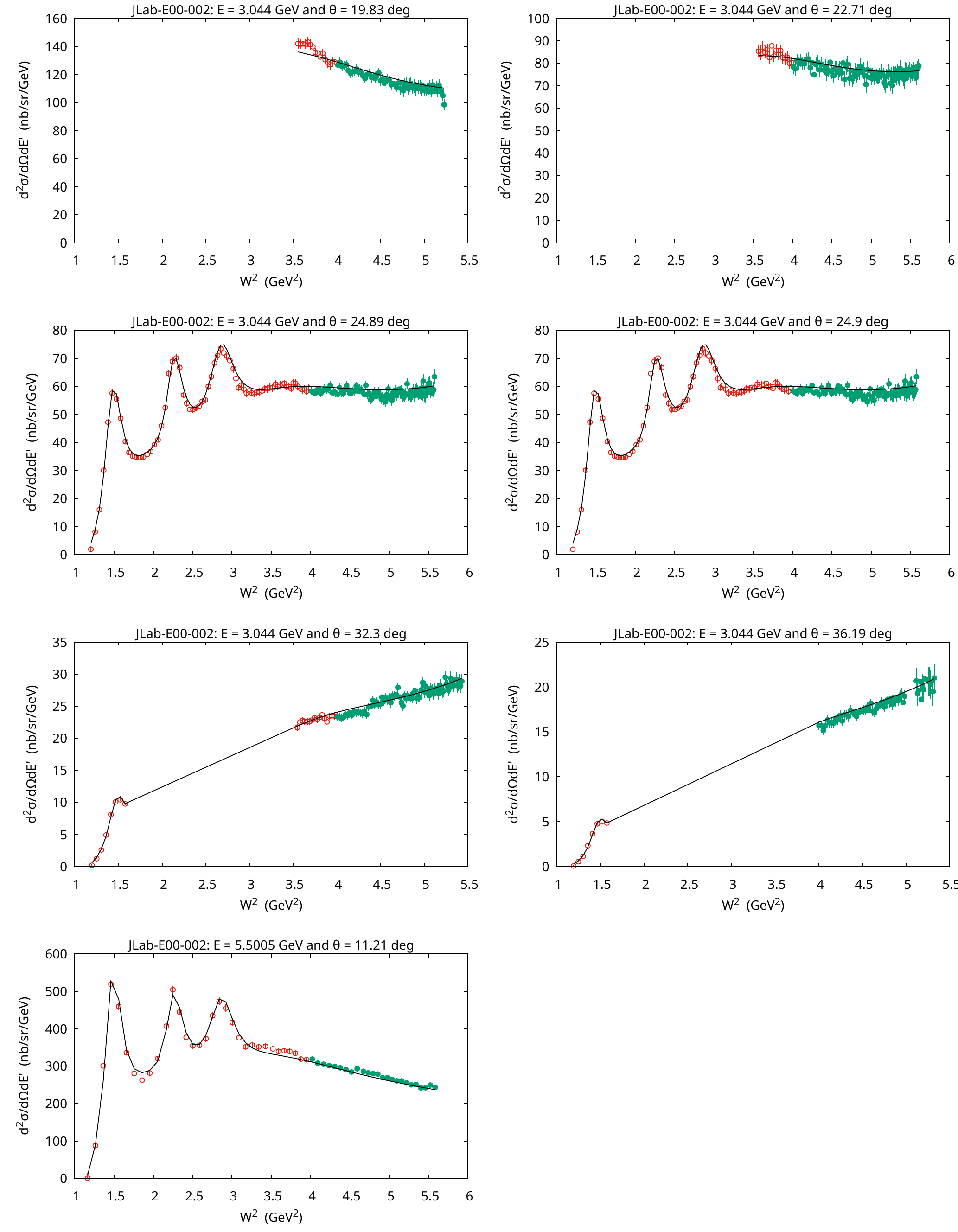}
\caption{Similar to Fig.~\ref{fig:CS_00002:1}.\label{fig:CS_00002:2}}
\end{figure*}
%
\begin{figure*}[p]
\centering
\includegraphics[width=\gwidth]{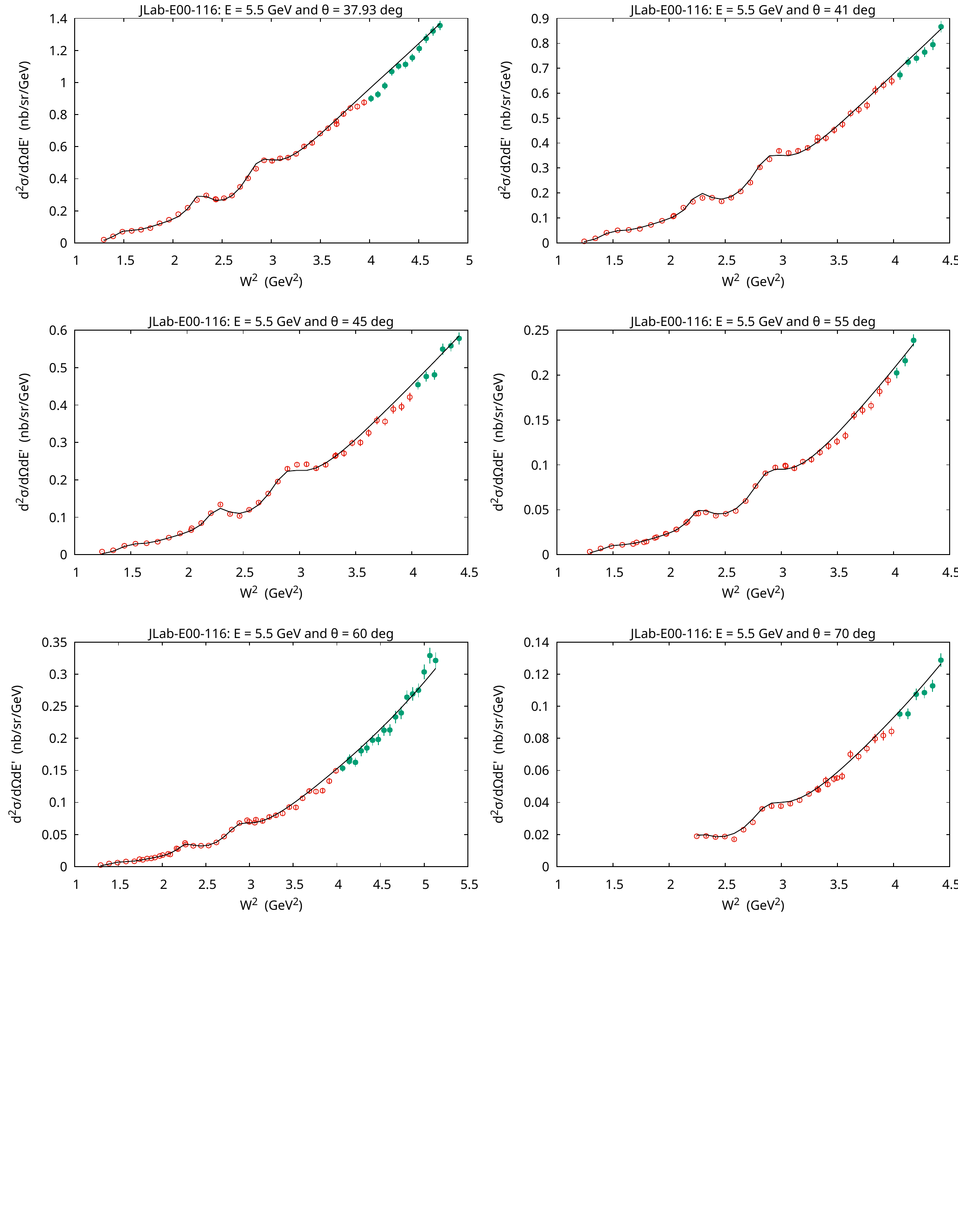}
\caption{\xsecaption{JLab-E00-116}{nanobarn}}
\label{fig:CS_00116}
\end{figure*}

\end{document}